\documentclass[%
aps,
pra,%
amsmath,amssymb,showpacs,
reprint,%
floatfix,
twocolumn,
nofootinbib,
longbibliography
]{revtex4-1}

\usepackage{amsmath,amsfonts,amssymb}
\usepackage{cases} % for subnumcases environment
\usepackage{xfrac}

\usepackage{nicefrac}

\def\doubleunderline#1{\underline{\underline{#1}}} % for matrix in double underline notation

\usepackage{empheq}
\usepackage[most]{tcolorbox}

\newtcbox{\mymath}[1][]{%
    nobeforeafter, math upper, tcbox raise base,
    enhanced, colframe=blue!30!black,
    colback=blue!30, boxrule=1pt,
    #1}

\allowdisplaybreaks[3]

\DeclareMathAlphabet\mathbfcal{OMS}{cmsy}{b}{n}

\usepackage{booktabs}

\usepackage{xcolor}
\definecolor{pltC0}{HTML}{1F77B4}
\definecolor{pltC1}{HTML}{FF7F0E}
\definecolor{pltC2}{HTML}{2CA02C}
\definecolor{pltC3}{HTML}{D62728}

\definecolor{richBlack}{HTML}{252525}
\definecolor{richLGray}{HTML}{D4D8DE}
\definecolor{richDGray}{HTML}{656565}
\definecolor{richTurq}{HTML}{46BCDE}
\definecolor{richGreen}{HTML}{52D273}
\definecolor{richRed}{HTML}{E94F64}
\definecolor{richOrange}{HTML}{E57254}
\definecolor{richYellow}{HTML}{E5C454}

\usepackage{adjustbox}

\newcommand{\dl}[1]{{#1}}

\begin{document}

    \title{\textit{\textit{Ab initio}} quantum models for thin-film x-ray cavity QED}

    \author{Dominik \surname{Lentrodt}}
    \email[]{dominik.lentrodt@mpi-hd.mpg.de}

    \author{Kilian P.~\surname{Heeg}}

    \author{Christoph H.~\surname{Keitel}}

    \author{J\"org \surname{Evers}}
    \email[]{joerg.evers@mpi-hd.mpg.de}

    \affiliation{Max-Planck-Institut f\"ur Kernphysik, Saupfercheckweg 1, 69117 Heidelberg, Germany}

    \date{\today}

    \begin{abstract}
    We develop two \textit{ab initio} quantum  approaches to thin-film x-ray cavity quantum electrodynamics with spectrally narrow x-ray resonances, such as those provided by M\"ossbauer nuclei. The first method is based on a few-mode description of the cavity, and promotes and extends existing phenomenological few-mode models to an \textit{ab initio} theory. The second approach uses analytically-known Green's functions to model the system.
    The two approaches not only enable one to \textit{ab initio} derive the effective few-level scheme representing the cavity and the nuclei in the low-excitation regime, but also  provide a direct avenue for studies at higher excitation, involving non-linear or quantum phenomena.
    The {\it ab initio} character of our approaches further enables direct optimizations of the cavity structure and thus of the photonic environment of the nuclei, to tailor the effective quantum optical level scheme towards particular applications. 
    To illustrate the power of the {\it ab initio} approaches, we extend the established quantum optical modeling to  resonant cavity layers of arbitrary thickness, which is essential to achieve quantitative agreement for cavities used in recent experiments. 
    Further, we consider multi-layer cavities featuring electromagnetically induced transparency, derive their  quantum optical few-level systems {\it ab initio}, and identify the origin of discrepancies in the modeling found previously using phenomenological approaches as arising from cavity field gradients across the resonant layers.
    \end{abstract}

    \maketitle

    \section{Introduction}
        
    In recent years, quantum optics with x-rays has become an active field of research \cite{Adams2013,Kuznetsova2017,Rohlsberger2014,PhysRevLett.96.142501,adams2012nonlinear,Adams2019}, driven by the progress in high photon intensities and beam quality at modern light sources \cite{springerreview,RevModPhys.88.015007,Shenoy2008,Adams2019}. As one promising platform, M\"ossbauer nuclei have received considerable attention. These nuclei feature transitions with exceptionally narrow linewidths, which form the basis for their broad range of applications~\cite{moessbauer_story_book,Rohlsberger2005}. From the viewpoint of quantum optics, the narrow resonances translate into favorably long lifetimes of the excitations, and thus into excellent coherence properties~\cite{Rohlsberger2014}. On the downside, the narrow linewidth renders an efficient state preparation, driving and readout of the nuclei challenging.  
   	Therefore, so far  all experiments have been restricted to the low-excitation regime \cite{Heeg2016arxiv}. Nevertheless, a rich variety of quantum optical coherence effects have already been observed, both in nuclear forward scattering \cite{Shvydko1996,Vagizov2013,Vagizov2014,Heeg2017,Heeg2020arxiv,Helisto1991,Sakshath2017,Goerttler2019} and thin-film cavity setups \cite{Rohlsberger2010,Rohlsberger2012,Heeg2013a,Heeg2015a,Heeg2015b,Haber2016a,Haber2017}. More possibilities have been theoretically suggested~\cite{Palffy2009,Liao2012,PhysRevLett.116.197402,Liao2016,PhysRevB.95.245429,Huang:17,Antonov2017,PhysRevApplied.10.014003,Zhang2019,Herkommer2020arxiv,Joshi2015}. With the first successful experiment at an x-ray free electron laser \cite{Chumakov2018}, including photon number resolved detection~\cite{Baron2000,Baron2006,Sofer2019}, this platform provides an exciting candidate for non-linear or correlated quantum effects with extreme transitions.
			
    While progress has largely been enabled by the ``source driven revolution'' \cite{Shenoy2008} in the x-ray regime, it is likely that it will not be sufficient for the establishment of more general x-ray quantum optical methods including non-linear and correlated quantum phenomena. Instead, the development of novel control techniques, target optimizations as well as novel physical platforms will be paramount~\cite{Heeg2016arxiv}. In particular, suitable photonic environments for the nuclei enable one to engineer enhanced couplings between x-rays and nuclei, to implement more versatile nuclear level schemes, and to simulate otherwise unavailable coherent control fields.

\begin{figure}[t]
		\includegraphics[width=1.0\columnwidth]{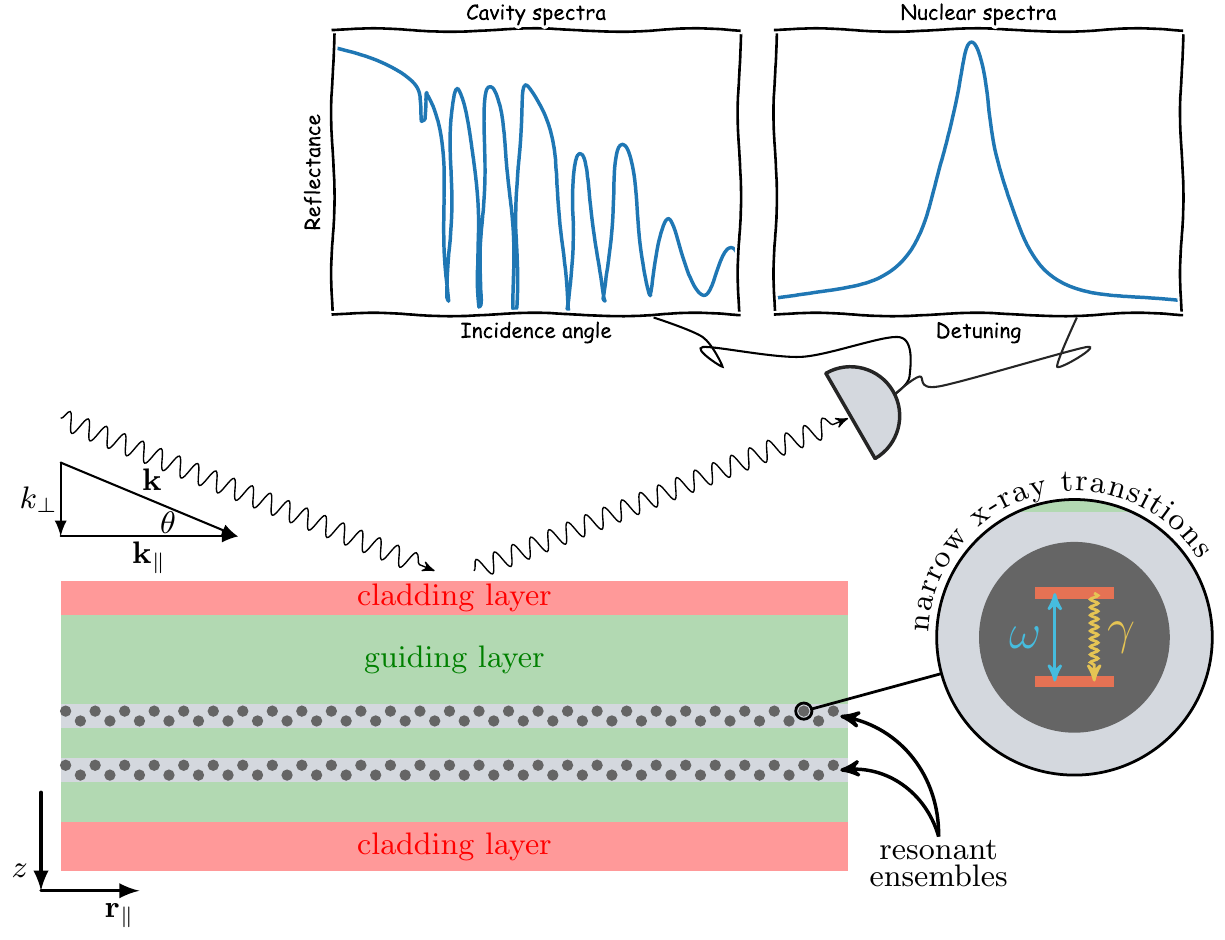}
		\caption{(Color online) Sketch of  a typical setting in thin-film x-ray cavity QED. A side-on view of a thin-film layer cavity, typically consisting of cladding layers (red) and a guiding layer (green) doped with thin resonant layers (gray), is shown. In the example in the figure, the resonant layers contain atoms or nuclei (black) featuring ultra-narrow transitions in the hard x-ray range, such as those provided by M\"ossbauer nuclei (see zoom, for the example of $^{57}$Fe, $\hbar\omega\approx14.4~$keV and $\gamma\approx4.7~$neV). The system is probed spectroscopically at grazing incidence (wave vector $\mathbf{k}$ with parallel component $\textbf{k}_\parallel$, perpendicular component $k_\perp$ and incidence angle $\theta$) by x-radiation. Typical observables include off-resonant cavity reflection spectra as a function of incidence angle and resonant nuclear spectra (sketched plots).}
		\label{fig::sketch}
	\end{figure}

    In this context, thin-film cavities with resonant nuclei embedded in the guiding layers are particularly promising~\cite{Rohlsberger2005,Rohlsberger2014}.  Such cavities (see Fig.~\ref{fig::sketch} for an example) have already facilitated the experimental observation of a multitude of quantum optical phenomena, including the collective Lamb shift \cite{Rohlsberger2010}, an electromagnetically-induced transparency (EIT) mechanism without an externally applied control field \cite{Rohlsberger2012}, and spontaneously generated nuclear coherences \cite{Heeg2013a}. Certain schemes have been found to provide access to new nuclear and x-ray observables, such as interferometric phase detection via Fano resonances \cite{Heeg2015a}. While the coupling between x-rays and single nuclei is weak compared to the rate of cavity leakage, the observation of collective strong coupling \cite{Haber2016a} and  Rabi oscillations between nuclear ensembles~\cite{Haber2017} have recently been reported, and the experimental realization of slow light \cite{Heeg2015b} further hints at the possibility of quantum systems with strong non-linearities \cite{Fleischhauer2005} even at low light levels \cite{Harris1999}. A key factor contributing to the success of the x-ray cavities is the fact that in the low-excitation regime, the total system of cavity and nuclear ensembles can equivalently be reinterpreted as an effective nuclear few-level scheme, with properties going beyond what is available in the bare nuclei~\cite{Rohlsberger2012,Heeg2013b,Heeg2015c,Longo2016}. 
    We note that while thin-film cavities have become particularly popular in the nuclear resonance scattering community, they also facilitate x-ray quantum optics  with electronic resonances~\cite{Haber2019}.

    Together, these advances establish the field of nuclear cavity quantum electrodynamics (QED) with hard x-rays, opening various avenues to connect to the impressive quantum optical toolkit in the optical and microwave regime that is available throughout various platforms of resonator  \cite{Reiserer2015,Berman1994,Haroche2013,Ritsch2013,Carusotto2013,Xiang2013,Schoelkopf2008} and waveguide \cite{Lodahl2015,Chang2018} QED.

    On the theoretical side, the nuclear resonant scattering community has largely employed semi-classical or mean-field methods based on perturbative scattering theory for the various transitions \cite{Afanasev1965,Kagan1967,Hannon1969,Hannon1999,Belyakov1975}, including variants of linear dispersion theory \cite{Born1980,Zhu1990} known as the layer \cite{Rohlsberger1999,Rohlsberger2005,Sturhahn2000} or Parratt's \cite{Parratt1954} formalism, Shvyd'ko's time and space picture \cite{Shvydko1999} as well as Maxwell-Bloch equation treatments \cite{Shvydko1999,Liao2012}.
    These approaches enable one to accurately model the related experiments, which  so far essentially operate in the  weak-excitation regime.  However, they are not well suited to interpret the resulting spectra in terms of quantum optical phenomena. Motivated thereby, a recently developed quantum model for the nuclei-cavity dynamics based on the input-output formalism \cite{Heeg2013b,Heeg2015c} has crucially provided the possibility to interpret the observed features quantum optically. By expressing the dynamical scattering process in terms of effective quantum optical few-level schemes, this theoretical approach has enabled the identification of various phenomena in the linear regime \cite{Heeg2013a,Heeg2015c,Heeg2015a,Haber2017,Heeg2015b}. Beyond that, it may be able to serve as a predictive theory for non-linear and quantum phenomena at higher intensities \cite{Heeg2016arxiv}.

    Despite the practical success of the phenomenological input-output model for thin-film x-ray cavities with M\"ossbauer nuclei \cite{Heeg2013b,Heeg2015c}, there are various open questions connected to its applicability. In the multi-mode and multi-ensemble case \cite{Heeg2015c}, which is for example crucial to describe the EIT phenomenon observed in \cite{Rohlsberger2012}, two heuristic extensions of the model, including a dispersion phase and an envelope factor, are required to fit the cavity spectra. While the resulting model is able to reproduce the main EIT phenomenon, quantitative and qualitative deviations are found when the spectra are scanned against incidence angle \cite{Heeg2015c}. In the linear regime, where the spectral observables can be modeled rigorously and with excellent empirical agreement by established semi-classical and scattering theory methods \cite{Hannon1999,Rohlsberger1999,Rohlsberger2005,Sturhahn2000,Shvydko2000}, these issues mainly limit the interpretation in terms of quantum optical phenomena. Beyond semi-classics and at higher excitation, however, these observations cast doubts on to what extent such phenomenological quantum models can be used as a predictive tool at larger driving field intensities, where no experimental results are available at the moment.

	\begin{figure*}[t]
		\includegraphics[width=0.8\textwidth]{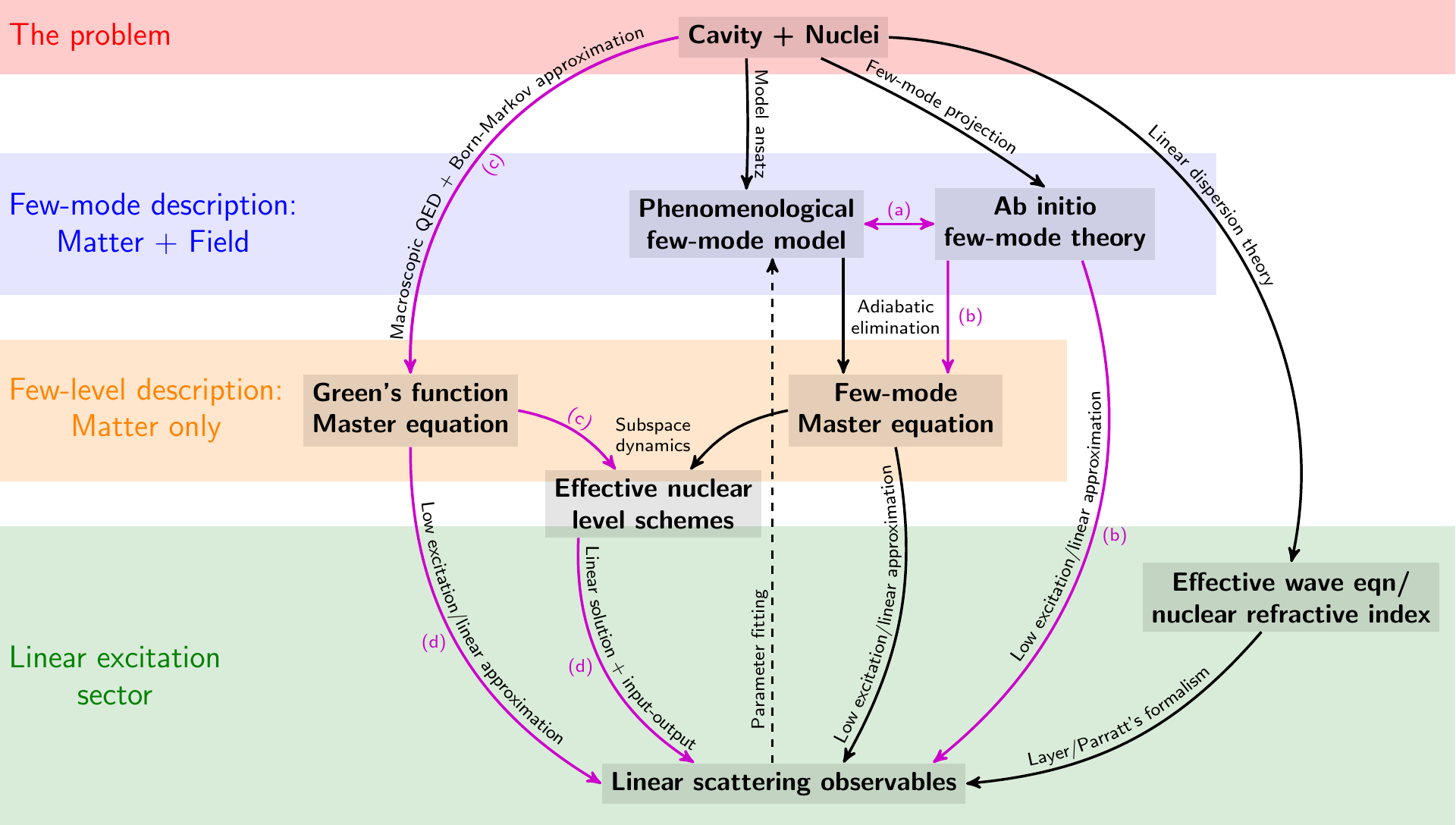}
		\caption{(Color online) Overview of theoretical approaches to thin-film x-ray cavity QED with M\"ossbauer nuclei, including the standard semi-classical layer formalism \cite{Rohlsberger2005}, which is based on a variant of linear dispersion theory \cite{Parratt1954,Zhu1990,Born1980}, and a successful phenomenological few-mode model \cite{Heeg2013b,Heeg2015c}, which is based on the quantum input-output formalism of cavity QED \cite{Gardiner2004}. In this paper, we present two additional \textit{ab initio}  quantum approaches to the problem, including an application of the \textit{ab initio}  few-mode theory \cite{Lentrodt2020} and of established Green's function techniques \cite{Gruner1996,Dung2002,Scheel2008,Asenjo-Garcia2017a} to the layer geometry and the nuclear quantum optics setting. Besides completing the connections between the theories illustrated in this diagram and showing which approximations are involved in each case, the main results presented in this paper are indicated by the labeled magenta arrows. (a) We establish the connection of the successful phenomenological model to \textit{ab initio}  theory. In the process, we propose an improved \textit{ab initio}  version of the model (Sec.~\ref{sec::aifmt}). (b) Based on this \textit{ab initio}  model, we derive general scattering solutions and effective level schemes for multi-mode multi-ensemble systems in closed form (Sec.~\ref{sec::aifmt_GeneralSolution}). (c) We show how the Green's function technique can be used as a numerically efficient method for calculating effective nuclear level schemes (Sec.~\ref{sec::Green_lowExc_levScheme}). (d) We demonstrate the connection of the Green's function approach to the layer formalism when calculating linear scattering observables and show how to reconstruct scattering observables once the nuclear level scheme dynamics are solved (Sec.~\ref{sec::Green_linDisp} and \ref{sec::Green_reconstructScatt}).}
		\label{fig::overview}
	\end{figure*}
	
    Here, we develop two \textit{ab initio} quantum approaches to thin-film x-ray cavity QED. Throughout the analysis, we focus on cavities containing M\"ossbauer nuclei, but the general results equally well apply to cavities with electronic resonances.   Our results not only resolve the previous discrepancies between the quantum optical model and the semi-classical descriptions, but also establish new theoretical tools that can be applied in the fully quantum sector and at higher intensities. The {\it ab initio} character of our approaches further enables direct optimizations of the cavity structure and thus of the photonic environment of the nuclei, by providing a predictive theory in which the involved approximations are well under control. As one application, the theory opens an avenue towards the design of effective quantum optical level structures  including the control fields realized by the  cavity. 
    
    Our first approach constitutes an \textit{ab initio} version of the phenomenological few-mode model \cite{Heeg2013b,Heeg2015c}, based on our recent general framework for few-mode approaches to quantum scattering problems~\cite{Lentrodt2020}. It removes the need for heuristic extensions and to fit the  quantum optical parameters. Instead, the {\it ab initio} approach directly determines  the parameters from the cavity geometry, and improves the prediction of scattering spectra to essentially perfect agreement. We further provide insight into the approximations involved in the model, providing a solid foundation for its application in the linear sector and opening the path for applications beyond this regime. At the same time, this progress promotes the quantum optical approach of constructing few-level models for x-ray cavity QED systems to an essentially exact and theoretically well-founded method. 
    
    Our second approach is based on a well-known Green's function technique \cite{Gruner1996,Dung2002,Scheel2008,Asenjo-Garcia2017a}, which in our case allows to derive a Markovian master equation for the nuclei coupling via the cavity environment directly from the cavity geometry. This method provides an alternative to the \textit{ab initio}  few-mode approach, involving different approximations. Most importantly, the Green's function formalism provides a numerically efficient way to calculate effective nuclear level schemes, which is crucial for optimizations. \dl{We note that after the submission of this work, another approach using Green's functions for nuclear quantum optics in x-ray cavities was brought to our attention \cite{Kong2020_private}.}
    
    Next to these general results, the examples we use to verify the two {\it ab initio} methods and to showcase their capabilities are chosen in such a way that they at the same time substantially advance the understanding of nuclear cavity QED, mainly in three respects. First, we demonstrate that the {\it ab initio} methods are capable of modeling the cavity-nuclei system not only in the vicinity of a chosen resonance, but across the entire range of probing photon frequencies and incidence angles. This opens perspectives for new applications beyond the current focus on individual cavity resonances. Second, we generalize the presently established quantum optical  approaches to the practically relevant case of arbitrarily thick resonant layers. This is mandatory to achieve quantitative agreement for standard cavities used in recent experiments.  Interestingly, we find that in the thick-layer limit the system is no longer equivalent to an effective nuclear few-level scheme, but rather to a set of self-coupled continuum ensembles. Third, we resolve previously not understood discrepancies in the quantum optical modelling of  multi-layer cavities featuring EIT spectra, and show that they originate from spatial field gradients across the resonant layers.  
    
    Finally, we provide a  clarifying guide and overview of the existing and here presented theoretical approaches to thin-film x-ray cavity QED in Fig.~\ref{fig::overview}. \dl{This diagram further illustrates the complementary nature of our two \textit{ab initio} approaches. The first few-mode theory rigorously justifies and extends the phenomenological models and completes connections between existing approaches, while the second approach based on Green's functions provides a computationally highly efficient new avenue to the nuclear scattering problem. Both approaches are applicable at weak coupling, but offer different advantages in more extreme regimes, as we discuss in detail.}
    
    The paper is structured as follows. Sec.~\ref{sec::few} introduces and discusses the few-mode {\it ab initio} approach. As a starting point and for later reference, we summarize the main features of the phenomenological quantum optical model as presented in \cite{Heeg2013b,Heeg2015c} in  Sec.~\ref{sec::recap}. In Sec.~\ref{sec::aifmt}, we develop the corresponding \textit{ab initio}  theory. The approach is based on our recently developed method \cite{Lentrodt2020} connecting the input-output formalism to quantum potential scattering theory. In Sec.~\ref{sec::aifmt_GeneralSolution}, we provide a general solution for linear scattering observables of the resulting \textit{ab initio} model. Section~\ref{sec::adiabatic_lineShape} disentangles the effects of the empty cavity and the contributions of the nuclei to the observables, which is exemplified in Sec.~\ref{sec:fanocavity} for well-studied cavities featuring Fano resonances. The effective nuclear level scheme represented by the cavity is derived in Sec.~\ref{sec::eff_scheme} by eliminating the cavity degrees of freedom. In Sec.~\ref{sec::aifmt_SimpleEx}, we verify  and illustrate the method for an analytically solvable example geometry\dl{, providing analytical formulas for all observables, and explicitly explain the computational algorithm for better accessibility to our method}. As an application, we generalize thin-film x-ray cavity QED to cavities containing thick resonant layers in Sec.~\ref{sec::aifmt_ThickLayers}.
    
    The following Sec.~\ref{sec::Green} develops and discusses the Green's function approach as a second \textit{ab initio}  method. After introducing the relevant concepts of Green's functions and macroscopic QED in Sec.~\ref{sec:macroQED}, we use the linear dispersion theory to derive an effective wave equation in Sec.~\ref{sec::Green_linDisp} that demonstrates the relation of this quantum theory to the established semi-classical approaches \cite{Rohlsberger1999,Rohlsberger2005}. 
    Section~\ref{sec:greenMaster} derives the nuclear Master equation using the Green's-function method.  As a main result and advantage of this method, we show how to derive effective nuclear level schemes in Sec.~\ref{sec::Green_lowExc_levScheme}, providing a numerically efficient method. Sections~\ref{sec:solgreen}-\ref{sec:efficiency} discuss the general solution in the linear sector and further analyze the numerical efficiency. \dl{To increase the accessibility to our method, Sec.~\ref{sec::Green_calc} provides a practical guide to  calculations using the Green's function approach.}
    Finally, in Sec.~\ref{sec::Green_applications}, we apply the Green's-function approach to multi-layer cavity structures in the EIT configuration, as studied in \cite{Rohlsberger2012,Heeg2015c}, and identify field gradients across the layers of finite thickness as the origin of previously not understood inaccuracies of the phenomenological quantum optical models~\cite{Heeg2015c}. We further provide the first \textit{ab initio}  calculation of the effective nuclear level scheme in the cavity environment. Our method removes the need for an ambiguous fitting procedure and opens design opportunities for more complex cavity geometries. 
    
    Finally, Sec.~\ref{sec::conclusion} discusses and summarizes our results. Details on derivations and parts of the calculations, as well as an in-depth practical comparison of the phenomenological and {\it ab initio} few-mode model are provided in the Appendices.
	
	\begin{figure}[t]
		\includegraphics[width=\columnwidth]{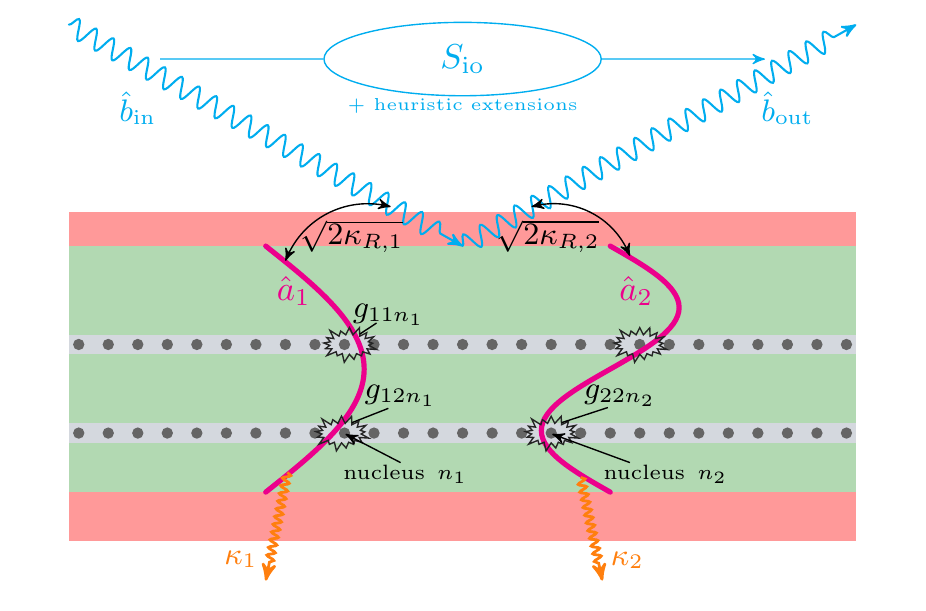}
		\caption{(Color online) Schematic illustration of the phenomenological quantum optical model for cavity QED with M\"ossbauer nuclei, indicating the relevant processes and their coupling parameters.  As an example, two of the system modes (magenta lines, $\hat{a}_\lambda$ for mode $\lambda\in\{1,2\}$) coupled to the input-drive $\hat{b}_\textrm{in}$ and output field $\hat{b}_\textrm{out}$ (cyan wiggly lines) are shown. The coupling constants in the model include the system-bath coupling ($\sqrt{2\kappa_{R,\lambda}}$), the cavity mode decay constants ($\kappa_\lambda$) and the mode-nucleus coupling ($g_{\lambda ln}$ for nucleus $n$ in ensemble/layer $l$). 
		See also Table~\ref{tab::pheno_vs_aifmt} for the functional dependencies of each model parameter.
		The input-output scattering matrix $S_\textrm{io}$ describes the scattering process between bath modes via the cavity interacting with the nuclei, and is assumed to provide the full scattering information for x-rays being reflected of the system. For certain cavities, heuristic extensions of the model are necessary to provide more accurate descriptions of the cavity and nuclear response (see text).}
		\label{fig::sketch-pheno}
	\end{figure}	
	\begin{figure}[t]
		\includegraphics[width=\columnwidth]{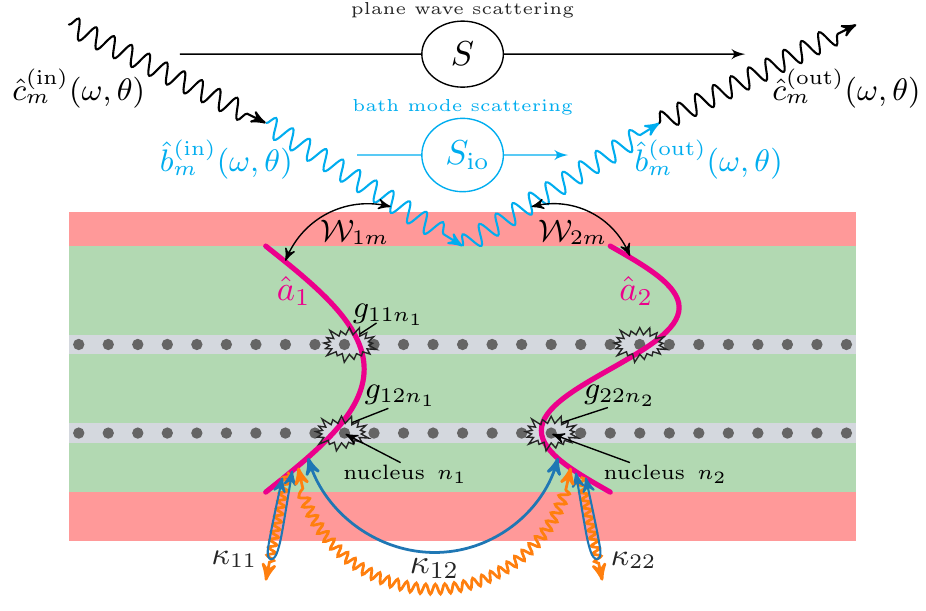}
		\caption{(Color online) Schematic illustration of the \textit{ab initio} few-mode model for cavity QED with M\"ossbauer nuclei. In comparison to the phenomenological model depicted in Fig.~\ref{fig::sketch-pheno}, the modes feature additional Lamb shifts (blue arrows at $\kappa_{11}$, $\kappa_{22}$) and cross-mode couplings ($\kappa_{12}$). Instead of a single drive, the bath is now an angle-dependent continuum with system-bath coupling constant $\mathcal{W}_{\lambda m}$ (corresponding to $\sqrt{2\kappa_{R,\lambda}}$ in the phenomenological model). The input-output matrix $S_\textrm{io}$ still contains the bath mode scattering information, however, to obtain the plane wave scattering matrix $S$, one has to employ an additional background scattering contribution (see text).  Again, Table \ref{tab::pheno_vs_aifmt} lists the functional dependencies of each model parameter.}
		\label{fig::sketch-aifmt}
	\end{figure}

	\section{\textit{Ab initio} few-mode approach to x-ray cavity QED}\label{sec::few}
	\subsection{Recap of the phenomenological model}\label{sec::recap}
	 In this section, we briefly summarize the essential features of the phenomenological quantum optical model for thin-film cavities, which are doped with layers of M\"ossbauer nuclei and probed spectroscopically in grazing incidence by hard x-radiation \cite{Heeg2013a,Rohlsberger2005}, as illustrated in Fig.~\ref{fig::sketch}. Such setups have been studied in a multitude of experiments, featuring a variety of different cavities (see \cite{Rohlsberger2010,Rohlsberger2012,Heeg2013a,Heeg2015a,Heeg2015b,Haber2016a,Haber2017,Haber2019} for the most recent quantum optics experiments).
	 For simplicity, we restrict ourselves to the unpolarized and unmagnetized case, which already features the essential ingredients. With polarization and magnetization included, one only obtains additional coupled equations \cite{Heeg2013b,Heeg2015c}. 
	
	The phenomenological model is given by the following equation for the density operator $\rho$ characterizing the cavity and the nuclei~\cite{Heeg2015c},
	\begin{align}\label{eq::recap_Master}
	   \dot{\rho} = -i[H, \rho] + \mathcal{L}_\textrm{cav}[\rho] +  \mathcal{L}_\textrm{SE}[\rho]\,.
	 \end{align}
	The Hamiltonian contribution,
	\begin{align}\label{eq::HE_H}
		H = H_\textrm{cav} + H_\textrm{nuc} + H_\textrm{int} + H_\textrm{drive} \,,
	\end{align}
	consists of a cavity part $H_\textrm{cav}$, a nuclear part $H_\textrm{nuc}$, their interaction $H_\textrm{int}$ and an external driving term $H_\textrm{drive}$.  The two Lindblad terms describe the incoherent cavity decay and the spontaneous emission of the nuclei. The model is illustrated in Fig.~\ref{fig::sketch-pheno}. We work in natural units with $\hbar = c = 1$.
	
	The dynamics of the empty cavity in the interaction picture are given by \cite{Heeg2015c}
	\begin{subequations}\label{eq:pheno-cavity}
	\begin{align}
		H_\textrm{cav} &= \sum_{\substack{\lambda \in \textrm{modes}}} \Delta_{C,\lambda}(\theta, \omega) \hat{a}^\dagger_\lambda \hat{a}_\lambda \,, \label{eq::HE_H_cav} \\
		H_\textrm{drive} &= \sum_{\substack{\lambda \in \textrm{modes}}} i\sqrt{2\kappa_{R,\lambda}} \hat{b}_\textrm{in}\hat{a}_{\lambda}^\dagger + h.c. \,, \label{eq::HE_H_drive} \\
		\Delta_{C,\lambda}(\theta, \omega) &=  \omega_{C,\lambda}(\theta, \omega) - \omega \nonumber
		\\
		& =  \sqrt{\omega^2\cos^2(\theta) + \omega_\textrm{nuc}^2\sin^2(\theta_\lambda)} - \omega \,, \label{eq::HE_CavDet_functionality}
	\end{align}
	\end{subequations}
	where $\hat{a}_\lambda$ is the bosonic cavity mode operator with index $\lambda$, $\hat{b}_\textrm{in}$ is an operator characterizing the driving field and $\sqrt{2\kappa_{R,\lambda}}$ is the drive coupling to mode $\lambda$. $\Delta_C$ is the cavity mode detuning relative to the external field with frequency $\omega$. It depends on the incidence angle $\theta$ and is parametrized in terms of the resonance angles $\theta_\lambda$ of the modes at the nuclear resonance frequency $\omega_\textrm{nuc}$, which is used as a reference. The detuning dependence on the external driving frequency can usually be neglected \cite{Heeg2013a} since the nuclear line width is typically orders of magnitude smaller than the cavity resonance width (see also Sec.~\ref{sec::adiabatic_lineShape}, where this property is illustrated).

	Interactions with nuclei in the cavity are included via the Hamiltonian contributions \cite{Heeg2015c}
	\begin{subequations}
	\label{eq:pheno-nuclei}
	\begin{align}
		H_\textrm{nuc} &= \sum_{\substack{l \in \textrm{ensembles}\\n \in 1, 2\dots N_l}} \frac{\omega_{\textrm{nuc},l}}{2} \hat{\sigma}^z_{ln} \,, \label{eq::HE_nuc} \\
		H_\textrm{int} &= \sum_{\substack{\lambda \in \textrm{modes}\\l \in \textrm{ensembles}\\n \in 1, 2\dots N_l}} g^{\mathstrut}_{\lambda l n} \hat{a}^{\mathstrut}_\lambda \hat{\sigma}^+_{ln} + h.c.\,, \label{eq::HE_int}
	\end{align}
	\end{subequations}
	\dl{where $\hat{\sigma}^z = |e\rangle\langle e| -  |g\rangle\langle g|$, $\hat{\sigma}^+ = |e\rangle\langle g|$, $\hat{\sigma}^- = |g\rangle\langle e|$ are the Pauli operators of a single nucleus}, and $ |e\rangle$ ($|g\rangle$) is the corresponding excited (ground) state. Further,  $\omega_{\textrm{nuc},l}$ is the nuclear transition frequency of ensemble $l$. The indices are chosen such that $l$ corresponds to the different resonant layers or ensembles and $n$ runs over all the $N_l$ nuclei within one such layer.  $g_{\lambda l n}$ is the coupling constant of a nucleus to the cavity mode. The coupling  is assumed to have the form $g_{\lambda l n} = g_{\lambda l} e^{i\phi_n}$ \cite{Heeg2015c}, including a position dependent phase $\phi_n$ due to propagation along the waveguide. We note that since the prefactor is independent of the nucleus index $n$ within one layer $l$, the above form includes the thin-layer approximation. That is it is assumed that the mode profile does not vary significantly across a layer that is considered as one ensemble, such that describing thick resonant layers requires the inclusion of multiple sub-ensembles dividing the thick layer into thinner ones (see Secs.~\ref{sec::aifmt_ThickLayers}, \ref{sec::Green_applications} for an illustration and further details).
	
	In addition to these Hamiltonian terms, the model features the incoherent contributions~\cite{Heeg2015c}
	\begin{subequations}\label{eq:pheno-inc}
	\begin{align}
		\mathcal{L}_\textrm{cav}[\rho] &= \sum_{\substack{\lambda \in \textrm{modes}}} \kappa_\lambda (2\hat{a}_{\lambda}\rho\hat{a}_{\lambda}^\dagger - \{\hat{a}_\lambda^\dagger \hat{a}_{\lambda},\,\rho \})\,,\\
		\mathcal{L}_\textrm{SE}[\rho] &= \sum_{\substack{l \in \textrm{ensembles}\\n \in 1, 2\dots N_l}} \frac{\gamma}{2} (2\hat{\sigma}^-_{ln}\rho\hat{\sigma}_{ln}^+ - \{\hat{\sigma}_{ln}^+\hat{\sigma}_{ln}^-,\,\rho \}) \,, \label{eq::pheno-inc_SE}
	\end{align}
	\end{subequations}
	where $\mathcal{L}_\textrm{cav}[\rho]$ describes the cavity decay and $\mathcal{L}_\textrm{SE}[\rho]$ the single nucleus incoherent decay. Here, $\rho$ is the density matrix of the system and $\{\cdot,\cdot\}$ is the anti-commutator.
	\dl{We note that these terms describe inelastic processes only, while the  elastic scattering is fully included via the interaction with the cavity modes.} 
	
	In order to calculate output operators from the driving input and the cavity dynamics, the input-output relation \cite{Heeg2015c}
	\begin{align}\label{eq::HE_io}
		\hat{b}_\textrm{out} = -\hat{b}_\textrm{in} + \sum_{\substack{\lambda \in \textrm{modes}}} \sqrt{2\kappa_{R,\lambda}} \hat{a}_\lambda \,,
	\end{align}
	is invoked. Spectroscopic observables can then be calculated from the output operators, such as the reflection coefficient given by \cite{Heeg2015c}
	\begin{align}
	 r = \frac{\langle \hat{b}_\textrm{out} \rangle}{\langle \hat{b}_\textrm{in} \rangle} = -1 + \sum_{\lambda}\sqrt{2\kappa_{R,\lambda}}\: \frac{ \langle\hat{a}_\lambda\rangle}{\langle \hat{b}_\textrm{in} \rangle}\,.
	 \end{align}
	 Within the adiabatic approximation \cite{Heeg2013b,Heeg2015c}, the mode operator dynamics are given by \cite{Heeg2015c}
	 \begin{align}\label{eq:a-pheno}
	  \hat{a}_\lambda = \frac{\sqrt{2\kappa_{R,\lambda}}\hat{b}_\textrm{in} - i\sum_{ln} g_{\lambda ln}\hat{\sigma}^-_{ln} }{\kappa_\lambda + i\Delta_{C,\lambda}}\,.
	 \end{align}
	 The nuclear dynamics can then be obtained from a nuclei-only Master equation obtained by adiabatically eliminating the cavity modes in Eq.~\eqref{eq::recap_Master}.
	
	The phenomenological few-mode model summarized so far has been employed for the analysis of various experiments in x-ray cavity QED \cite{Heeg2013a,Heeg2015a,Heeg2015b,Haber2017,Haber2019}, providing a basis for the quantum mechanical interpretation of the system. 
	
	However, for certain cavities, it was found that heuristic extensions to the model are necessary. These include a dispersion as well as an envelope factor resulting in the modified reflection coefficient \cite{Heeg2015c}
	\begin{align}
		r = \frac{\langle \hat{b}_\textrm{out} \rangle}{\langle \hat{b}_\textrm{in} \rangle} = r_\textrm{env}(\theta) \left[-r_\textrm{disp} + \sum_{\lambda}\sqrt{2\kappa_{R,\lambda}} \frac{\langle\hat{a}_\lambda\rangle}{\langle \hat{b}_\textrm{in} \rangle}\right] \,,
	\end{align}
	where $r_\textrm{disp} = |r_\textrm{disp}| e^{i\phi_C}$ is the dispersion factor that is introduced to account for a dispersion phase $\phi_C$ as well as for far off-resonant modes  and $r_\textrm{env}$ is the envelope factor that is introduced to account for total reflection behavior at grazing incidence.
	
	While with these heuristic extensions included, good agreement with the layer formalism and experiment is found in many cases (see e.g.~\cite{Heeg2013b,Heeg2013a,Heeg2015a,Heeg2015b,Haber2017,Haber2019}), there are still quantitative and even qualitative differences for important setups, such as the EIT cavity \cite{Heeg2015c} that is investigated experimentally in \cite{Rohlsberger2012}. In addition, the heuristic extensions are neither justified nor derived from the quantum mechanical input-output theory \cite{Gardiner1985,Gardiner2004}, restricting the predictive power of the model. In our analysis below, we will remove these restrictions, clarify the relations between different theoretical approaches, and prove an accurate complete description of the system without the need for heuristic extensions  (see Fig.~\ref{fig::overview} for an overview).
	
	\dl{We note that our paper does not invalidate the phenomenological model, which is intrinsically consistent and which we show to be a good description in many cases. The main improvements of the \textit{ab initio} theory developed in the following are the extension to the cases of overlapping modes and highly leaky cavities, the unambiguous computation of the quantum couplings in the effective level scheme and the justification of approximations required to obtain the description setting the ground for a predictive theory beyond the linear regime.}

    \subsection{\textit{Ab initio} few-mode theory for thin-film x-ray cavities}\label{sec::aifmt}
    
    \begin{table}
    	\begin{center}
    		\begin{tabular}{ c | c | c }
    			\hline
    			& Pheno model & \textit{Ab initio} model \\
    			\hline
    			Mode resonance & $\Delta_{C,\lambda}(\omega, \theta)$ & $\tilde{\Delta}_{C,\lambda}(\omega, \theta) $ \\
    			Cavity loss & $\kappa_\lambda$ & $\kappa_{\lambda \lambda'}(\omega, \theta)$ \\
    			Bath/drive coupling & $i\sqrt{2\kappa_{R/T,\lambda}}$ & $\mathcal{W}_{\lambda m}(\omega, \theta)$ \\
    			Mode-nucleus coupling & $g_{\lambda l}$ & $g_{\lambda l}(\theta)$ \\
    			Background scattering & - & $S_\textrm{bg}(\omega, \theta)$ \\
    			Heuristic extensions & $r_\textrm{disp}$, $r_\textrm{env}(\theta)$ & -
    			\\
    			\hline
    		\end{tabular}
    	\end{center}
    	\caption{Overview of the differences between the phenomenological model for thin-film x-ray cavities \cite{Heeg2013b,Heeg2015c} and its \textit{ab initio} counterpart developed in this paper. The quantum optical parameters (left) and their functional dependencies in each case (middle and right) are shown.}\label{tab::pheno_vs_aifmt}
    \end{table}
    
    In this section, we present an \textit{ab initio} version of the phenomenological model for thin-film x-ray cavities, based on our recently developed general formalism for few-mode theories~\cite{Lentrodt2020}.
    The \textit{ab initio}  few-mode theory for the given problem can be written in close analogy to the phenomenological version summarized in Sec.~\ref{sec::recap}. The main differences are the functional dependencies of the coupling constants and the treatment of the input-output relation as illustrated in Figs.~\ref{fig::sketch-pheno}, \ref{fig::sketch-aifmt} and Table~\ref{tab::pheno_vs_aifmt}. Here, we outline these differences and define the structure of the \textit{ab initio}  theory. A detailed derivation is given in Appendix~\ref{sec::aifmt_DetailedDerivation}, \dl{together with a functional summary of the concept of the \textit{ab initio} few-mode theory. For readers who are interested in how to apply the method in practice, we refer to Sec.~\ref{sec::aifmt_SimpleEx}, where an explicit example system including analytic formulas for the required coupling constants is presented.}
    
    The Hamiltonian governing the dynamics is analogous to the phenomenological version in Eq.~(\ref{eq::HE_H}),
    \begin{align}
      H &= H_\textrm{cav} + H_\textrm{nuc} + H_\textrm{int} + H_\textrm{drive} +  H_\textrm{field}\,.
    \end{align}
	The empty-cavity dynamics are governed by
	\begin{subequations}\label{eq:aifmt-cavity}
	\begin{align}
		H_\textrm{cav} &= \sum_{\lambda \in \textrm{modes}} \omega^{\mathstrut}_\lambda(\theta) \hat{a}_\lambda^\dagger \hat{a}_\lambda^{\mathstrut}\,, \\[2ex]
		H_\textrm{drive} &= \sum_{\substack{m \in \textrm{channels}\\\lambda \in \textrm{modes}}} \int d\omega \mathcal{W}^{\mathstrut}_{\lambda m}(\omega, \theta) \hat{b}_m^{\mathstrut}(\omega)\hat{a}_\lambda^{\dagger} \nonumber \\[1ex]
		& \qquad\qquad \qquad + h.c.\,,\\[2ex]
		H_\textrm{field} &= \sum_{m \in \textrm{channels}} \int d\omega  \tilde{\omega}(\omega, \theta) \hat{b}_m^{\dagger}(\omega)\hat{b}_m^{\mathstrut}(\omega)\,,
	\end{align}
	\end{subequations}
	which generalizes Eqs.~(\ref{eq:pheno-cavity}). \dl{The cavity mode detuning to a selected bath mode is then given by 
	\begin{align}
		\tilde{\Delta}_{C,\lambda}(\omega, \theta) = \omega_\lambda(\theta) - \tilde{\omega}(\omega, \theta) \,, \label{eq::aifmt_mode-detuning}
	\end{align}
	which replaces $\Delta_{C,\lambda}(\omega, \theta)$ in Eq.~(\ref{eq::HE_CavDet_functionality}) from the phenomenological approach. We note that in the {\it ab initio} formulation, no explicit driving term is specified on the level of the Hamiltonian}. Rather, the coupling to all bath modes $\hat{b}_m(\omega)$ is included, with their free evolution term $H_\textrm{field}$. \dl{The external driving field is then specified through expectation values or correlation functions of the input operators corresponding to the bath modes within the input-output formalism (see Eq.~\eqref{eq::eff_ioEOM} below).} For this reason, we also do not work in an interaction picture here, but with the bare frequencies such as $\tilde{\omega}(\omega, \theta)$, which is an effective bath mode frequency at a given incidence angle or parallel wave vector (see Appendix~\ref{sec::aifmt_DetailedDerivation} for details). For notational brevity, we omit the parametric angular dependence of the mode operators.
	
	We note that the angular and frequency dependence of the effective mode detuning $\tilde{\Delta}_{C,\lambda}(\omega, \theta)$ may differ from the one assumed in the phenomenological model by $\Delta_{C,\lambda}(\omega, \theta)$ in Eq.~\eqref{eq::HE_CavDet_functionality} (refer to Sec.~\ref{sec::aifmt_SimpleEx} and Fig.~\ref{fig::benchmark1} for an example). Similarly, $\mathcal{W}_{\lambda m}(\omega, \theta)$ extends the in-coupling rate $\sqrt{2\kappa_{R/T,\lambda}}$, where $m$ is an index for the external channels, such as reflection and transmission denoted by $R/T$ in the phenomenological model \cite{Heeg2013a}.
	
	Interactions with the nuclei are included via the contributions
	\begin{subequations}
	\begin{align}
		H_\textrm{nuc} &= \sum_{\substack{l \in \textrm{ensembles}\\n \in 1, 2\dots N_l}} \frac{\omega_{\textrm{nuc},l}}{2}\: \hat{\sigma}^z_{ln} \,,\\
		H_\textrm{int} &= \sum_{\substack{\lambda \in \textrm{modes}\\l \in \textrm{ensembles}\\n \in 1, 2\dots N_l}} g^{\mathstrut}_{\lambda l}(\theta) \: \hat{a}^{\mathstrut}_\lambda \: \hat{\sigma}^+_{ln} + h.c.\,. 
	\end{align}
	\end{subequations}
	which are essentially identical to their phenomenological counterparts Eqs.~\eqref{eq:pheno-nuclei}, only now the mode-nucleus coupling is dependent on the incidence angle. 
	
	The incoherent nuclear decay corresponding to Eq.~(\ref{eq:pheno-inc}) is still given by
	 \begin{align}\label{eq:aifmt-inc}
	  \mathcal{L}_\textrm{SE}[\rho] &= \sum_{\substack{l \in \textrm{ensembles}\\n \in 1, 2\dots N_l}} \frac{\gamma}{2} (2\hat{\sigma}^-_{ln}\rho\hat{\sigma}_{ln}^+ - \{\hat{\sigma}_{ln}^+\hat{\sigma}_{ln}^-,\,\rho \}) \,.
	 \end{align}
	 \dl{As in the phenomenological model, elastic scattering is included via the interaction with the cavity modes.
	 
	 The density operator fulfills a generalization of the Master equation Eq.~\eqref{eq::recap_Master}, such that also the radiative bath degrees of freedom are now included, and the Hamiltonian and Lindbladian are modified to comprise Eqs.~(\ref{eq:aifmt-cavity}-\ref{eq:aifmt-inc}). The decay constant in the spontaneous emission term is the natural linewidth $\gamma$, as in the phenomenological case Eq.~\eqref{eq::pheno-inc_SE} (see Appendix \ref{sec::aifmt_DetailedDerivation_eff1D} for details on the decay constant).}
	 
	The cavity decay term in the phenomenological model $\mathcal{L}_\textrm{cav}[\rho]$, however, has no direct counterpart in the {\it ab initio} few-mode theory, since the frequency dependence of the cavity-bath couplings induces non-Markovian dynamics of the cavity modes \cite{Hackenbroich2002,Viviescas2003,Zhang2013}. In particular for cavities with overlapping modes \cite{Lentrodt2020}, which are realized in standard x-ray cavities as we will show in Sec.~\ref{sec::aifmt_SimpleEx}, this effect is not negligible and no Markovian Lindblad term for the cavity modes can be obtained. We note, however, that when tracing out the full cavity as a structured bath for the nuclei, the Markov approximation is applicable for many cavities due to the narrow line width of the nuclear response, which is the basis for the Green's function approach presented in Sec.~\ref{sec::Green}.
	
	In the {\it ab initio} few-mode approach, instead of a Lindblad term for the cavity decay, one can write a frequency domain equation for the cavity mode operators \cite{Lentrodt2020}
	\begin{align}
		\hat{a}_\lambda(\omega) &= \sum_{\lambda'} \mathcal{D}^{-1}_{\lambda \lambda'}(\omega, \theta) \big[2\pi \sum_m \mathcal{W}^{\mathstrut}_{\lambda'm}(\omega, \theta) \hat{b}^{\mathrm{(in)}}_m(\omega) \nonumber
		\\
		&\qquad+ \sum_{ln} g^{*\mathstrut}_{\lambda l}\hat{\sigma}_{ln}^-(\omega)\big] \,, \label{eq::eff_cavityEOM}\\
		\mathcal{D}_{\lambda \lambda'}(\omega, \theta) &= -\tilde{\Delta}_{C,\lambda}(\omega, \theta)\delta_{\lambda \lambda'} + i\kappa_{\lambda \lambda'}(\omega, \theta)\,,
	\end{align}
	which is a direct generalization of Eq.~\eqref{eq:a-pheno} in the phenomenological approach going beyond the adiabatic and Markov approximations, with the effective mode detuning $\tilde{\Delta}_{C,\lambda}(\omega, \theta)$ defined in Eq.~\eqref{eq::aifmt_mode-detuning}.
	
	In comparison to the phenomenological model version in Eq.~(\ref{eq:a-pheno}), the cavity loss rate is now replaced by a frequency and incidence-angle dependent cavity loss matrix $\kappa_{\lambda \lambda'}(\omega, \theta)$, corresponding to $-i\Gamma'_{\lambda \lambda'}(\omega, \theta)$ in \cite{Lentrodt2020}. It has a matrix structure, as it includes cross-mode coupling terms, going beyond the single-mode cavity decay rates $\kappa_\lambda$ in the phenomenological model as illustrated in Figs.~\ref{fig::sketch-pheno}, \ref{fig::sketch-aifmt}.
	
	In order to calculate observables, the exact input-output theory from \cite{Lentrodt2020} allows one to obtain the frequency-domain input-output relation for the input [output] bath operators $\hat{b}^{\mathrm{(in)}}_m(\omega)$ [$\hat{b}^{\mathrm{(out)}}_m(\omega)$] at a given incidence angle,
	\begin{equation}
		\hat{b}^{\mathrm{(out)}}_m(\omega)   = \hat{b}^{\mathrm{(in)}}_m(\omega) -i\sum_\lambda \mathcal{W}^*_{\lambda m}(\omega, \theta) \hat{a}^{\mathstrut}_\lambda(\omega)\,,\label{eq::eff_ioEOM}
	\end{equation}
	generalizing Eq.~(\ref{eq::HE_io}) in the phenomenological model. In addition to the parametric dependencies of the input-output coupling, solving the input-output relation Eq.~\eqref{eq::eff_ioEOM} does not yield the full scattering information, as shown in \cite{Lentrodt2020}. Instead, an additional background scattering contribution $S_\textrm{bg}(\omega, \theta)$ is necessary to translate the bath-mode scattering into plane-wave scattering (refer to Sec.~\ref{sec::aifmt_SimpleEx} and Fig.~\ref{fig::benchmark2_rocking2D} for an example). This additional contribution is absent in the phenomenological model, but crucial to reproduce the response not only in the vicinity of the studied resonance, but also away from it. Most notably, it accounts for the fact that the empty-cavity response is treated exactly within the \textit{ab initio} few-mode approach \cite{Lentrodt2020} independent of the number of included cavity modes.

	In summary, we see that the differences between the {\it ab initio} and the phenomenological version of the model are a modified frequency and angular dependence of the coupling constants, as well as the cavity decay rate becoming a matrix introducing cross-mode decay terms. We also note that the parameters may in general be complex quantities and a background scattering matrix should be included for accuracy over a large frequency range. The differences between the models are summarized in Table~\ref{tab::pheno_vs_aifmt} and illustrated in Figs.~\ref{fig::sketch-pheno}, \ref{fig::sketch-aifmt}.

    \subsection{General solution in the linear regime}\label{sec::aifmt_GeneralSolution}
    In this section, we provide a general solution to the multi-mode multi-ensemble \textit{ab initio} few-mode theory for x-ray cavities with M\"ossbauer nuclei, which is outlined in the previous section, in the linear regime. By writing the equations in Heisenberg-Langevin form, we find a closed-form solution for the linear scattering observables  of systems containing an arbitrary number of modes and layers or ensembles. By employing the exact input-output formalism \cite{Lentrodt2020}, multi-mode coupling, overlapping modes, non-Markovian, open system and scattering effects are included in the solution method.

    For simplicity, we again consider a single polarization and unmagnetized samples. The method, however, can straightforwardly be applied to the full Hamiltonian of x-ray cavities with thin film M\"ossbauer nuclei, including polarization and magnetization \cite{Heeg2013b}.

    \subsubsection{Heisenberg-Langevin equations of motion}
    The Heisenberg-Langevin equations of motion for the nuclear operators defined by the \textit{ab initio}  model in Sec.~\ref{sec::aifmt} read \cite{Gardiner2004,Kirton2017,Heeg2015c}
    \begin{subequations}\label{eq::effFM}
        \begin{align}
        \frac{d}{dt}{\hat{\sigma}}_{ln}^-(t) &= -(i\omega_{\textrm{nuc},l} + \frac{\gamma}{2})\hat{\sigma}_{ln}^-(t) \nonumber \\
        & \qquad + i \hat{\sigma}_{ln}^z(t)\sum_{\lambda} \hat{a}^{\mathstrut}_\lambda(t) g^{\mathstrut}_{l\lambda},
        \\
        \frac{d}{dt}{\hat{\sigma}}_{ln}^z(t) &= -2i \hat{\sigma}_{ln}^+(t) \sum_{\lambda} \hat{a}^{\mathstrut}_\lambda(t) g^{\mathstrut}_{l\lambda} + h.c. \nonumber \\
        &- \gamma(\hat{\sigma}^z_{ln} + 1)\,.\label{equ::eff_atomEOM3} 
        \end{align}
    \end{subequations}
    We note that in this form, the equations are particularly simple, in the sense that the individual ensembles only interact with each other via the cavity modes. The many-particle aspect apparent from the nuclear index $n$ can be treated by introducing collective operators \cite{Heeg2013a,Heeg2015c} or by working with single particle operators and employing the permutation symmetry of the model \cite{Kirton2017,Shammah2019}. The latter can be employed since our Liouvillian is invariant under the relabeling $(l,n)\rightarrow (l,n')$ for nuclei within each ensemble $l$, which is the case due to the coupling and decay constants being independent of $n$ within the single parallel wave vector approximation. We adopt the permutation symmetry in the following, which then allows us to drop the index $n$ on expectation values of single nucleus operators in Eq.~\eqref{eq::effFM} and to reduce the corresponding sums over nuclei within one ensemble.

    The resulting equation for the cavity mode expectation values is given by Eq.~\eqref{eq::eff_cavityEOM} and now simplifies to \cite{Lentrodt2020}
    \begin{align}
    	\langle\hat{a}_\lambda(\omega)\rangle = \sum_{\lambda'} \mathcal{D}^{-1}_{\lambda \lambda'}(\omega) \big[&2\pi \sum_m \mathcal{W}^{\mathstrut}_{\lambda'm}(\omega) \langle\hat{b}^{\mathrm{(in)}}_m(\omega)\rangle \nonumber
    	\\
    	&+ \sum_{l} N^{\mathstrut}_l g^{*\mathstrut}_{l\lambda'} \langle\hat{\sigma}_l^-(\omega)\rangle\big] \,, \label{eq::eff_cavityEOM2}
    \end{align}
    where we have dropped the parametric angle dependencies for brevity. We note that the number of nuclei $N_l$ in ensemble $l$ appears explicitly in the equations due to the use of the permutation symmetry \cite{Kirton2017}.
    
    The dynamical equations~\eqref{eq::eff_cavityEOM2} are completed by the input-output relation Eq.~\eqref{eq::eff_ioEOM}.

    \subsubsection{Solution of the multi-layer multi-mode equations in the linear regime}
    In \cite{Heeg2015c}, the linear equations of motion are solved for the phenomenological model within the adiabatic elimination approximation for the special case of a two-layer EIT configuration, which is notably investigated experimentally in \cite{Rohlsberger2012}. Here, we solve the above {\it ab initio} equations in the linear regime without the need for adiabatic elimination, a Markov approximation or a specific mode-ensemble configuration. The adiabatic approximation will, however, be employed a posteriori to gain insight into the final solution and the resulting nuclear line shape.

    The crucial approximation to obtain linear spectral observables is the weak excitation approximation, which can conveniently be performed in the Heisenberg-Langevin approach by setting $\langle\hat{\sigma}^z_l\rangle \approx -1$ \cite{Waks2010} for all ensembles $l$ (see Sec.~\ref{sec::Green_linDisp} for further details). This results in the nuclear lowering operator equation of motion becoming linear and independent of the other nuclear operators. Physically, this way of solving the equations of motion corresponds to eliminating the nuclei instead of the modes, which can be done without further approximations in the linear regime (see also Sec.~\ref{sec::Green_linDisp}). An alternative way of performing this approximation is the Holstein-Primakoff method \cite{Holstein1940,DeBernardis2018}. \dl{We note that due to this approximation, it is not necessary to trace out the bath modes or the radiative bath of the nuclei. Instead, the linear operator equations of motion can be solved directly.}
	
    Within this low-excitation approximation, one obtains three linearly coupled matrix equations in the frequency domain, one for the nuclear lowering operators, one for the cavity modes and one given by the input-output relation. These equations can be solved straightforwardly by linear algebra. For the nuclear lowering operators we obtain
    \begin{align}\label{eq::eom_lower_freqSol}
	    \left (\omega - \omega_{\textrm{nuc},l} + i\frac{\gamma}{2}\right)\langle\hat{\sigma}_l^-(\omega)\rangle = \sum_{\lambda}g_{l\lambda}\langle\hat{a}_\lambda(\omega)\rangle \,.
    \end{align}
    Substituting Eq.~\eqref{eq::eom_lower_freqSol} into Eq.~\eqref{eq::eff_cavityEOM}, we find
     \begin{align}
    	\langle\hat{a}_\lambda(\omega)\rangle &= 2\pi \sum_{\lambda'} \mathcal{D}^{-1}_{\lambda \lambda'}(\omega)  \sum_m \mathcal{W}^{\mathstrut}_{\lambda'm}(\omega) \langle\hat{b}^{\mathrm{(in)}}_m(\omega)\rangle  \,, \label{eq:awithoutsigma}\\
          \mathcal{D}_{\lambda \lambda'}(\omega) &\rightarrow \mathcal{D}^{(\textrm{int})}_{\lambda \lambda'}(\omega) =  \mathcal{D}_{\lambda \lambda'}(\omega) - \sum_l \frac{N^{\mathstrut}_l g^{*\mathstrut}_{l\lambda} g^{\mathstrut}_{l\lambda'}}{\omega - \omega_{\textrm{nuc},l} + i\frac{\gamma}{2}}\,.\label{eq::mod_prop}
    \end{align}
    Thus, the nuclear ensembles effectively modify the cavity propagator matrix $\doubleunderline{\mathcal{D}}$. Finally, inserting Eq.~\eqref{eq:awithoutsigma} into the input-output relation Eq.~\eqref{eq::eff_ioEOM}, we obtain the scattering solution
    \begin{align}
    & \langle\hat{b}^{\mathrm{(out)}}_m(\omega)\rangle - \langle\hat{b}^{\mathrm{(in)}}_m(\omega)\rangle = -2\pi i \nonumber
    \\
    &\quad \times \sum_{\lambda,\lambda',m'} \mathcal{W}^*_{\lambda m}(\omega) [\mathcal{D}^{(\textrm{int})}]^{-1}_{\lambda \lambda'}(\omega) \mathcal{W}_{\lambda' m'}(\omega)\langle\hat{b}^{\mathrm{(in)}}_{m'}(\omega)\rangle.\label{eq::eff_ioEOM_subs}
    \end{align}
    The scattering matrix, defined in matrix notation by \cite{Lentrodt2020}
    \begin{equation}
    \langle\underline{\hat{b}}^{\mathrm{(out)}}(\omega)\rangle = \doubleunderline{S}_\mathrm{io}(\omega) \langle\underline{\hat{b}}^{\mathrm{(in)}}(\omega)\rangle,
    \end{equation}
    thus reads
    \begin{align}
        \label{eq::scatt_int}
        \doubleunderline{S}_{\mathrm{io}}(\omega) &= \doubleunderline{\mathbb{I}} - 2\pi i \doubleunderline{\mathcal{W}}^\dagger(\omega) \doubleunderline{\mathcal{D}}_{(\textrm{int})}^{-1}(\omega) \doubleunderline{\mathcal{W}}(\omega)\,.
    \end{align}
    As noted before, the input-output scattering is complemented by a background scattering contribution $\doubleunderline{S}_\textrm{bg}(\omega)$ \cite{Lentrodt2020}, which translates the bath-mode scattering into plane wave scattering via
        \begin{align} \label{eq::Sfull_product}
    \doubleunderline{S}(\omega) = \doubleunderline{S}_{\mathrm{bg}}(\omega)\doubleunderline{S}_{\mathrm{io}}(\omega) \,.
    \end{align}
    Since the background scattering is independent of the nuclear ensembles, it can be computed for the empty cavity and multiplied with the input-output solution without additional effort when solving the interacting system \cite{Lentrodt2020}.

    These results provide a general solution for linear scattering observables in the \textit{ab initio}  version of the phenomenological few-mode model for thin-film x-ray cavities with M\"ossbauer nuclei.

    \subsection{Separating the contributions of cavity and nuclei}\label{sec::adiabatic_lineShape}
    While the above Eq.~(\ref{eq::scatt_int}) together with the background scattering contribution constitutes a complete solution of the problem, it does not allow to straightforwardly separate the individual contributions of the cavity and the nuclei to the scattering process. To establish this separation, we note that the relevant frequency scale of the nuclei (e.g., $\sim$neV, quality factor of $Q\sim10^{12}$ for ${}^{57}$Fe, see also Fig.~\ref{fig::sketch}) is typically orders of magnitude smaller than that of the cavity ($\sim$keV).
    
    In order to isolate the nuclear contribution, we approximate the frequency response of the cavity as constant on the scale of the nuclear response. If $\omega_{\textrm{nuc}}$ is any one of the nuclear transition frequencies, we can then write
    \begin{align}\label{eq::adiabatic_spectralLevel1}
    \mathcal{W}_{\lambda m}(\omega) &\approx \mathcal{W}_{\lambda m}(\omega_{\textrm{nuc}})\nonumber
    \\
    &\equiv \mathcal{W}_{\lambda m}\,,
    \\
    \label{eq::adiabatic_spectralLevel2}
    \mathcal{D}_{\lambda \lambda'}(\omega)^{\mathstrut} &\approx (\omega_{\textrm{nuc}}-\omega_\lambda)\delta_{\lambda \lambda'} + i\kappa_{\lambda \lambda'}(\omega_{\textrm{nuc}})\nonumber
    \\
    &\equiv \mathcal{D}_{\lambda \lambda'}\,.
    \end{align}    
    That is in Eq.~\eqref{eq::scatt_int}, the only relevant frequency dependence on the nuclear scale is the $\omega$ in the nuclear addition to $\doubleunderline{\mathcal{D}}^{(\textrm{int})}$ in Eq.~\eqref{eq::mod_prop}. This approximation essentially amounts to performing the adiabatic approximation \cite{Heeg2013b,Heeg2015c} a posteriori on a spectral level and can also be considered as a type of Markov approximation \cite{Breuer2002_BOOK,Gardiner2004,Viviescas2003}, since it relies on the cavity's frequency response being flat on the scale of the nuclear response. We refer to Sec.~\ref{sec::aifmt_SimpleEx} and Fig.~\ref{fig::couplings} for an explicit demonstration of the validity of this form of the approximation. In the experiments so far \cite{Heeg2013a,Heeg2015a,Heeg2015b,Rohlsberger2010,Rohlsberger2012,Haber2017}, the approximation was found to be possible due to the narrow nuclear response. We note, however, that when strong coupling effects are present \cite{Haber2017} or for broader electronic resonances \cite{Haber2019}, the approximation may break down and it may not be possible to separate the cavity and the nuclear response.

    We can now separate the nuclear contribution by employing the Woodbury matrix identity \cite{Golub1996,Domcke1983} to write the inverse of the modified cavity propagator as
    \begin{align}
    &\doubleunderline{\mathcal{D}}^{-1}_{(\textrm{int})}(\omega) =  (\doubleunderline{\mathcal{D}} + \doubleunderline{g}^\dagger \doubleunderline{\Lambda}(\omega) \doubleunderline{g} )^{-1} \nonumber
    \\
    &\quad= \doubleunderline{\mathcal{D}}^{-1} - \doubleunderline{\mathcal{D}}^{-1} \doubleunderline{g}^\dagger (\doubleunderline{\Lambda}^{-1}(\omega) + \doubleunderline{g}\, \doubleunderline{\mathcal{D}}^{-1}\doubleunderline{g}^\dagger)^{-1} \doubleunderline{g}\,\doubleunderline{\mathcal{D}}^{-1} \,,\label{eq:wood-decomposition}
    \end{align}
    where the first summand is the (inverse) empty cavity propagator $\doubleunderline{\mathcal{D}}^{-1}$, such that the second term amounts to the nuclear contributions. 
    The matrices appearing in Eq.~\eqref{eq:wood-decomposition} are defined by their components as
    \begin{subequations}\label{eq:dmatrices}
    \begin{align}
    (\doubleunderline{g})_{l\lambda} = \tilde{g}_{\lambda l}  &= \sqrt{N^{\mathstrut}_l} g_{\lambda l} \,,
    \\
    (\doubleunderline{g}^\dagger)_{\lambda l} = \tilde{g}^*_{\lambda l} &= \sqrt{N^{\mathstrut}_l} g^*_{\lambda l}  \,,
    \\
    (\doubleunderline{\Lambda}(\omega))_{l l'} = \Lambda(\omega)_{l l'} &=  -\frac{1}{\omega - \omega_{\textrm{nuc},l} + i\frac{\gamma}{2}}\delta_{ll'}  \,,
    \\
    (\doubleunderline{\Lambda}^{-1}(\omega))_{l l'} = \Lambda^{-1}(\omega)_{l l'} &= -(\omega - \omega_{\textrm{nuc},l} + i\frac{\gamma}{2})\delta_{ll'}  \,,
    \end{align}
    \end{subequations}
    where $\tilde{g}_{\lambda l}$ are the rescaled coupling constants of the collective states \cite{Heeg2015c} (see also Appendix \ref{app::couplingsDef}). Inserting Eq.~\eqref{eq:wood-decomposition} into Eq.~\eqref{eq::scatt_int}, the scattering matrix then reads
    \begin{align}
        \label{eq::scatt_int_woodbury}
        \doubleunderline{S}_{\mathrm{io}}(\omega) =& \doubleunderline{S}^{[\textrm{no nuclei}]}_{\mathrm{io}}(\omega_{\textrm{nuc}}) + 2\pi i \doubleunderline{\mathcal{W}}^\dagger \doubleunderline{\mathcal{D}}^{-1} \doubleunderline{g}^\dagger \nonumber
        \\
        & \cdot(\doubleunderline{\Lambda}^{-1}(\omega) + \doubleunderline{g}\,\,\doubleunderline{\mathcal{D}}^{-1}\doubleunderline{g}^\dagger)^{-1} \doubleunderline{g}\,\doubleunderline{\mathcal{D}}^{-1} \doubleunderline{\mathcal{W}}\,.
    \end{align}
    This formula shows the desired separation into the empty-cavity background \dl{\begin{align}\label{eq:S-io-nonuclei}
      \doubleunderline{S}^{[\textrm{no nuclei}]}_{\mathrm{io}}(\omega_{\textrm{nuc}}) =  \doubleunderline{\mathbb{I}} - 2\pi i \doubleunderline{\mathcal{W}}^\dagger(\omega) \doubleunderline{\mathcal{D}}^{-1}(\omega) \doubleunderline{\mathcal{W}}(\omega)
    \end{align}}
    and the additional nuclear contribution. The interpretation of the various contributions will become clearer in the following section discussing pertinent example cavities.
    
    \subsection{Example: Fano cavities\label{sec:fanocavity}}
    In this section, we illustrate Eq.~(\ref{eq::scatt_int_woodbury}) using standard cavity structures that are known to feature simple, yet relevant, nuclear line shapes in their frequency-dependent response. 
    The basic element contributing to the response can be understood using the single-cavity-mode approximation, in which the inverse $\doubleunderline{\mathcal{D}}^{-1}$ appearing in Eq.~\eqref{eq::scatt_int_woodbury} can be identified via  Eq.~\eqref{eq::adiabatic_spectralLevel2} as a Lorentzian.  In the general multi-mode case, since $\doubleunderline{g}\, \doubleunderline{\mathcal{D}}^{-1}\, \doubleunderline{g}^\dagger$ is a matrix coupling the different transitions, the nuclear contribution therefore is a superposition of coupled Lorentzian resonances interfering with themselves and the flat cavity background. As a result, the nuclear line shape can be rather complex for the general multi-mode multi-ensemble case.
    It is, however, well known that for certain cavity structures, the response reduces to a single Lorentzian interfering with the cavity background \cite{Heeg2015a,HeegPhD}, giving rise to Fano line shapes~\cite{Heeg2015a,Fano1961,Ott2013,Limonov2017}. In the following, we use such cases to illustrate Eq.~(\ref{eq::scatt_int_woodbury}).
    
    \subsubsection{Single nuclear ensemble}
    For a single nuclear ensemble, the index $l$ has only one value, such that $\doubleunderline{\Lambda}$ reduces to  a scalar. Similarly, $\doubleunderline{g}$ and, for single channel scattering, $\doubleunderline{\mathcal{W}}$ reduce to vectors with elements corresponding to the different cavity modes $\lambda$. As a result, for the reflection channel, $\doubleunderline{\mathcal{W}}^\dagger \doubleunderline{\mathcal{D}}^{-1} \doubleunderline{g}^\dagger$ and $\doubleunderline{g}\,\doubleunderline{\mathcal{D}}^{-1} \doubleunderline{\mathcal{W}}$ become complex numbers independent of $\omega$, thus characterizing the magnitude of the nuclear contribution as well as its relative phase to the empty-cavity contribution. 
   For the spectral shape, we are therefore left with analyzing
    \begin{align}
    &(\doubleunderline{\Lambda}^{-1}(\omega) + \doubleunderline{g}\,\doubleunderline{\mathcal{D}}^{-1}\doubleunderline{g}^\dagger)^{-1} \nonumber\\
    &\qquad = \frac{-1}{\omega - (\omega_{\textrm{nuc},l} + \Delta_\textrm{LS})  + i(\gamma + \Gamma_S)/2} \,,
    \end{align}
    where we have defined
    \begin{align}
        \Delta_\textrm{LS} - \frac{i}{2}\: \Gamma_S  = \sum_{\lambda, \lambda'} N\, g^{\mathstrut}_\lambda \: \mathcal{D}^{-1}_{\lambda \lambda'} \: g^*_{\lambda'} \,.
    \end{align}
    This is the Lorentzian line shape of a single nucleus, modified by the collective Lamb shift $\Delta_\textrm{LS}$ and the superradiance $\Gamma_S$ for the ensemble in the cavity. Due to the interference with the cavity background, this Lorentzian in general becomes a Fano line shape.
    
    Indeed, the above form of the nuclear line shape closely corresponds to what is obtained in the multi-mode phenomenological model (see in particular Eq.~(46) in \cite{Heeg2015c}). The main difference here is the form of the $\mathcal{D}$-matrix, which is diagonal in the phenomenological case \cite{Heeg2015c} and now includes the cross-mode coupling terms in the \textit{ab initio} generalization (see Table~\ref{tab::pheno_vs_aifmt}). In the special case of isolated cavity resonances \cite{Lentrodt2020}, the $\mathcal{D}$-matrix may well be approximated as diagonal, such that the phenomenological assumption can be well justified.
    
    We thus recover the corresponding predictions of the phenomenological model~\cite{Heeg2015a, Heeg2015c}, at the same time establishing an improved \textit{ab initio}  representation of the complex level shift $\Delta_\textrm{LS} - i\Gamma_S/2$.

    \subsubsection{Single cavity mode, uniform nuclear ensembles}
    Another special case where the nuclear contribution is of single Lorentzian form arises when there is only a single mode and, in addition, the $\omega_{\textrm{nuc},l}$ of all nuclear ensembles are equal \cite{Heeg2015a,HeegPhD}. 
    In this case, the index $\lambda$ is absent, $\doubleunderline{\mathcal{D}}$ reduces to a scalar, and $\doubleunderline{\mathcal{W}}$ and $\doubleunderline{g}$ become  column vectors with elements $\mathcal{W}_m$ corresponding to the external channels and $g_l$ to the layers.
    As a result, we obtain from Eq.~\eqref{eq::scatt_int} 
    \begin{align}\label{eq::scatt_int_special2}
    \doubleunderline{S}_{\mathrm{io}}(\omega) &= \doubleunderline{\mathbb{I}} + 2\pi i \underline{\mathcal{W}}^*\underline{\mathcal{W}}^T\left(\mathcal{D} -  \frac{\sum_l N^{\mathstrut}_l g^{*\mathstrut}_{l} g^{\mathstrut}_{l}}{\omega - \omega_{\textrm{nuc}} + i\frac{\gamma}{2}}\right)^{-1}\,.
    \end{align}
    Applying the scalar version of the Woodbury formula leads to
    \begin{align}\label{eq::scatt_int_special2_woodbury}
    \doubleunderline{S}_{\mathrm{io}}(\omega) =& \doubleunderline{S}^{[\textrm{no nuclei}]}_{\mathrm{io}}(\omega_{\textrm{nuc}})
    \\
    & - 2\pi i\: \frac{\underline{\mathcal{W}}^* \underline{\mathcal{W}}^T A}{\omega - \omega_{\textrm{nuc}} + i\gamma/2 - \mathcal{D}^{-1}A} \,, \nonumber
    \end{align}
    where $A= \sum_l \tilde{g}^{*\mathstrut}_{l} \tilde{g}^{\mathstrut}_{l} = \sum_l N_l g^{*\mathstrut}_{l} g^{\mathstrut}_{l}$. Again, the result indeed corresponds to a Fano line shape, recovering the phenomenological result \cite{Heeg2015c} with $A$ corresponding to $\mathcal{G}^2$ (Eq.~(48) in \cite{Heeg2015c}). We note again, that the result is fully expressible in terms of the rescaled couplings $\tilde{g}^{\mathstrut}_{l}$ (see also Appendix \ref{app::couplingsDef}).

    \subsection{Effective few-level scheme\\realized by the nuclei in the cavity}\label{sec::eff_scheme}
    As shown in \cite{Heeg2013b,Heeg2015c},  one can associate an effective nuclear level scheme to the nuclear cavity QED system, such that the spectroscopic response of the two agree in the low-excitation limit. This way, the spectra of the nuclear cavity QED system can be interpreted in terms of quantum optical phenomena. For instance, a spectral dip in certain two-ensemble x-ray cavities could be interpreted as an EIT phenomenon \cite{Rohlsberger2012,Heeg2015c}. 
    In~\cite{Heeg2013b,Heeg2015c}, the effective level scheme is derived using an adiabatic elimination of the cavity modes, giving rise to equations for the nuclear system alone with effective couplings between the ensembles. The parameters of this level scheme are then obtained using fits of the model to numerical calculations of the QED system's response or to experimental data. 

    In this section, we derive an effective few-level scheme from the \textit{ab initio} few-mode model given above and generalize it to arbitrary ensemble configurations.  While the adiabatic elimination employed here is not necessary to solve the linear scattering problem, it is necessary in order to obtain the effective level scheme. 
    Due to the \textit{ab initio}  character of the theory, this approach further allows us to express the quantum optical level parameters and couplings directly in terms of the cavity geometry. Alleviating the need for parameter fitting not only removes possible ambiguities due to overfitting or degeneracies, but also amplifies the interpretational power in terms of the relevant cavity modes and nuclear ensembles responsible for the observed effects.

    As a first step, we write Eq.~\eqref{eq::effFM} in terms of collective operators \cite{HeegPhD} $\hat{J}^{+/-/z}_l = \sum_n \hat{\sigma}^{+/-/z}_{ln}$, which gives
    \begin{subequations}\label{eq::effFM_collective}
        \begin{align}
        \frac{d}{dt}{\hat{J}}_l^-(t) &= -(i\omega_{\textrm{nuc}} + \frac{\gamma}{2})\hat{J}_l^-(t) + i \hat{J}_l^z(t) \underline{g}_l^{T} \underline{\hat{a}}^{\mathstrut}(t),
        \\
        \frac{d}{dt}{\hat{J}}_l^z(t) &= -2i \hat{J}_l^+(t) \underline{g}_l^{T} \underline{\hat{a}}^{\mathstrut}(t) + h.c. - \gamma(\hat{J}_l^z(t) + N^{\mathstrut}_l)\,. 
        \end{align}
    \end{subequations}
    As a next step, we adiabatically eliminate the cavity modes by substituting Eq.~\eqref{eq::eff_cavityEOM} into Eq.~\eqref{eq::effFM_collective} and employing the adiabatic approximation \cite{Heeg2013b,Heeg2015c} on the spectral level  Eqs.~(\ref{eq::adiabatic_spectralLevel1}, \ref{eq::adiabatic_spectralLevel2}).
    The adiabatic solution for the mode operator Eq.~\eqref{eq::eff_cavityEOM} can then be transformed into the time domain to give
    \begin{align}
	    \underline{\hat{a}}(t) = \doubleunderline{\mathcal{D}}^{-1} \big[&2\pi \doubleunderline{\mathcal{W}}\: \underline{\hat{b}}^{\mathrm{(in)}}(t) + \sum_l \underline{g}^*_l \hat{J}_l^-(t)\big] \,. \label{eq::eff_cavityEOM_adiabatic_timeDomain}
    \end{align}
    Substitution yields 
    \begin{subequations}\label{eq::effFM_modesEliminated}
    	\begin{align}
    	\frac{d}{dt}{\hat{J}}_l^-(t) =& -(i\omega_{\textrm{nuc},l} + \frac{\gamma}{2})\hat{J}_l^-(t) + i \hat{J}_l^z(t) \underline{\Omega}^T_l \underline{\hat{b}}^{\mathrm{(in)}}(t) \nonumber
    	\\
    	&+ i \hat{J}_l^z(t) \sum_{l'} G^{\mathstrut}_{ll'} \hat{J}_{l'}^-(t),
    	\\
    	\frac{d}{dt}{\hat{J}}_l^z(t) =& -2i \hat{J}_l^+(t) \underline{\Omega}^T_l \underline{\hat{b}}^{\mathrm{(in)}}(t) + h.c. \nonumber
    	\\
    	&-2i \hat{J}_l^+(t) \sum_{l'} G_{ll'} \hat{J}_{l'}^-(t) + h.c. \nonumber \\
    	&- \gamma(\underline{\hat{J}}_l^z(t) + N_l)\,,
    	\end{align}
    \end{subequations}
    where the \textit{effective drive coupling matrix} is given by
    \begin{align}
        \underline{\Omega}^T_l = \underline{\Omega}_l^T(\omega_{\textrm{nuc}}) = 2\pi \underline{g}_l^{T}\doubleunderline{\mathcal{D}}^{-1}(\omega_{\textrm{nuc}}) \doubleunderline{\mathcal{W}}(\omega_{\textrm{nuc}}) \,, \label{eq::aifmt_levScheme_driveVec}
    \end{align}
    and the \textit{effective level coupling matrix} is
    \begin{align}
        G_{ll'} = G_{ll'}(\omega_{\textrm{nuc}}) =  \underline{g}_l^{T}\doubleunderline{\mathcal{D}}^{-1}(\omega_{\textrm{nuc}}) \underline{g}_{l'}^* \,. \label{eq::aifmt_levScheme_coup}
    \end{align}
    These equations can be reformulated in terms of an effective Hamiltonian
    \begin{align}\label{eq::eff_H_levelScheme}
        \hat{H}_\textrm{eff} = \sum_l \frac{\omega_{\textrm{nuc},l}}{2}\hat{J}_l^z &+ \sum_{ll'} \hat{J}_l^+ \textrm{Re}[G_{ll'}] \hat{J}_{l'}^- \nonumber
        \\
        &+ \sum_l \hat{J}_l^+ \underline{\Omega}_l^T \underline{\hat{b}}^{\mathrm{(in)}}(t) + h.c. \,,
    \end{align}
    and the effective Lindblad operator
    \begin{align}\label{eq::effLevel_lind}
        \mathcal{L}_\textrm{eff}[\rho] =  & \sum_{ll'} -\textrm{Im}[G_{ll'}] (2\hat{J}_{l}^- \rho\hat{J}_{l'}^+ - \{\hat{J}_{l}^+\hat{J}_{l'}^-,\,\rho \}) \nonumber
        \\
        &+ \mathcal{L}_\textrm{SE}[\rho]\,,
    \end{align}
    where $\mathcal{L}_\textrm{SE}[\rho]$ is the local spontaneous emission term defined in Eq.~\eqref{eq::lindblad_SE_localDiss}. This effective level scheme again closely corresponds to the results in the phenomenological model (in particular Eqs.~(23-25) in \cite{Heeg2015c}), with the aforementioned differences in the \textit{ab initio} coupling parameters summarized in Sec.~\ref{sec::aifmt}.
    
\begin{figure}[t]
        \includegraphics[width=1.0\columnwidth]{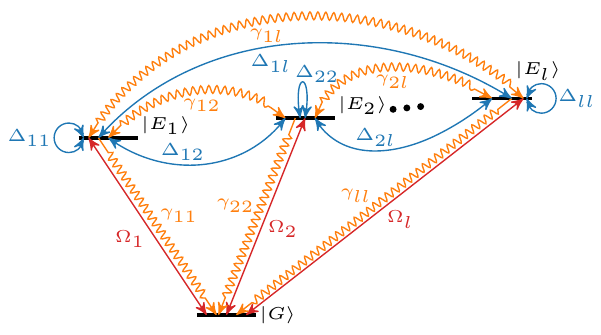}
        \caption{(Color online) Schematic illustration of effective level schemes equivalent to the total system of cavity and nuclei in the linear low-excitation regime. The structure of the level scheme is defined via the couplings given in Eqs.~\eqref{eq::eff_H_levelScheme} and \eqref{eq::effLevel_lind}, with $\Delta_{ll'} = \textrm{Re}[G_{ll'}]$ and $\gamma_{ll'} = -2\textrm{Im}[G_{ll}]$. Single nucleus decay terms are not depicted for clarity (see text).\label{fig::couplings_sketch}}
    \end{figure}

    Eqs.~\eqref{eq::eff_H_levelScheme} and \eqref{eq::effLevel_lind} can be interpreted as a system of interacting collective spins. Together, they define an effective level scheme in their low-excitation subspace. Starting from the ground state $|G\rangle$ defined as the state without photons in the cavity and with all nuclei in their ground state, the excited states of this level scheme are collective excitonic states of the many-body QED system, which can be defined as
    \begin{align}
      |E_l\rangle = \hat{J}_{l}^+\,|G\rangle\,.
    \end{align}
    The couplings within this effective level scheme are illustrated in Fig.~\ref{fig::couplings_sketch}. 
    For example, for a single ensemble $l$ the collective Lamb shift and the Purcell enhanced superradiance are given by
    $
        \Delta_{\textrm{LS}} = \textrm{Re}[G_{ll}]
    $
    and 
    $
        \gamma_{\textrm{S}} = -2\textrm{Im}[G_{ll}]
    $,
    respectively. With more than one ensemble, the coherent inter-ensemble couplings are given by
    $
        \Delta_{ll'} = \textrm{Re}[G_{ll'}]
    $
    and incoherent cross-ensemble decay terms with decay rate
    $
        \gamma_{ll'} = -2\textrm{Im}[G_{ll'}]
    $
    are also present. The latter can be shown to yield spontaneously generated coherences in certain cases \cite{Heeg2013a}. The various transitions are driven from an external channel $m$ with the effective Rabi frequency  $\Omega_{lm}$.

    The spontaneous-emission contribution in Eq.~\eqref{eq::effLevel_lind} requires additional care, because it {\it a priori} operates on the single-particle level, rather than on the level of the collective transition operators. But in the low-excitation or the linear regime, it simply adds to the collective decay rate \cite{Heeg2015c}, such that the system can fully be expressed in terms of collective operators as shown in Eqs.~\eqref{eq::effFM_modesEliminated}. 
    However, Eqs.~\eqref{eq::eff_H_levelScheme}, \eqref{eq::effLevel_lind} are also valid beyond the linear regime, where methods based on permutation symmetry \cite{Shammah2019,Kirton2017} can be employed to tackle the spontaneous emission term to extend the collective operator treatment in \cite{Heeg2016arxiv}. We note, however, that the non-linear dynamics may break the permutation symmetry by coupling to other parallel wave vector modes. 
    
    The effective many-body couplings contain the bare coupling constants $g_{l\lambda}$ instead of the rescaled coupling constants $\tilde{g}_{l\lambda} =\sqrt{N_l}\, g_{l\lambda}$, because the underlying equations contain collective operators defined as sums over single-particle operators. In the linear limit, the former can be replaced by effective single-particle operators, which leads to the appearance of the rescaled coupling constants~\cite{Heeg2015c,Heeg2016arxiv} (see also Sec.~\ref{sec::Green_lowExc_levScheme}). In this sense, the linear limit is independent of the number of coherently interacting nuclei, see also Appendix \ref{sec::Xray-AIFMT_complexRefractive}, as is also the case in the semi-classical layer formalism \cite{Rohlsberger2005}, where only the nuclear number density appears as a material parameter. However, beyond the linear excitation regime, the number of coherently interacting nuclei is crucial, since it determines the onset of non-linear effects \cite{Heeg2016arxiv}.
    
    We further note that by including the magnetic substructure of the nuclei, more versatile level schemes become accessible~\cite{Heeg2013a}.

    \subsection{Analytically solvable example system}\label{sec::aifmt_SimpleEx}
    In this section, we illustrate the \textit{ab initio} few-mode approach to nuclear many-body cavity QED using a specific example. We choose the simplest possible layer cavity geometry which features a resonance structure that resembles what is practically investigated in x-ray cavity QED~\cite{Rohlsberger2010,Rohlsberger2012,Heeg2013a,Heeg2015a,Heeg2015b,Haber2017}, and also has found applications in various contexts, see e.g.~\cite{PhysRevLett.89.237201,PhysRevB.58.8590,PhysRevB.50.10354,Rohlsberger1997,le_zak_2010}.  It furthermore has the advantage that it can be analytically solved within our framework.
    
    We present detailed comparisons to the semi-classical theory and to the phenomenological few-mode theory. In the process, we demonstrate the advantages and the potential of the method, showing differences to previous interpretations and completing the theoretical picture of the few-mode approach. As an application, we generalize the quantum optical model to cavities containing thick resonant layers. 
    
    \subsubsection{Cavity layout}

\begin{figure}
        \includegraphics[width=1.0\columnwidth]{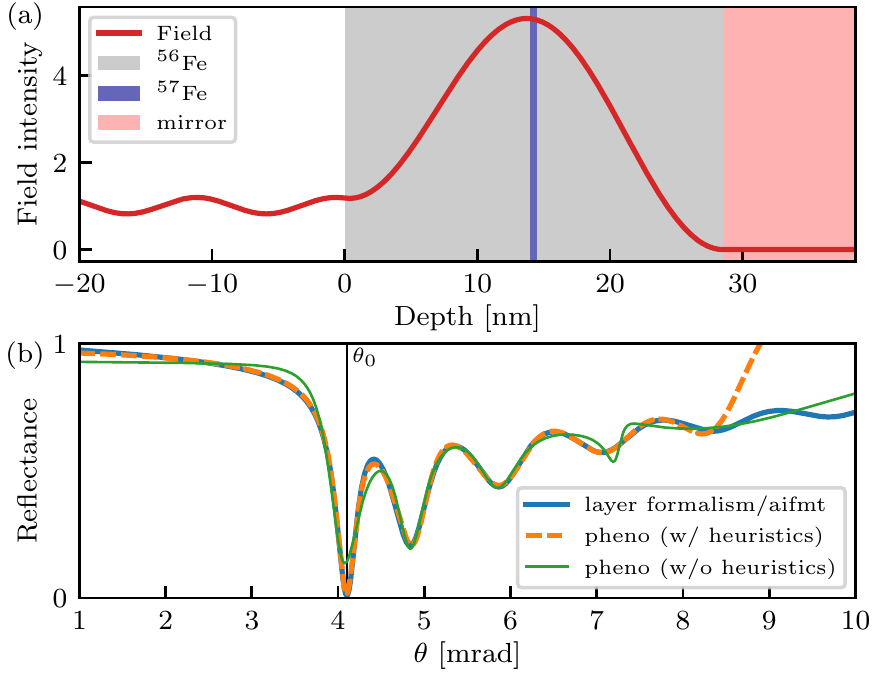}
        \caption{(Color online) Analytically solvable example within the \textit{ab initio} few-mode theory for x-ray cavities. 
        (a) The structure of the cavity under consideration (materials see legend) and its off-resonant field distribution (red line) at the incidence angle $\theta_0$ corresponding to the first guided mode. The cavity itself (gray ${}^{56}$Fe and blue ${}^{57}$Fe) has a total thickness of 28.5~nm, where the thickness of the resonant ${}^{57}$Fe dopant layer at the center is 0.5~nm (blue).
        (b) Rocking curve (off-resonant cavity reflectance   at $14.4$~keV in $\theta-2\theta$ geometry). The position $\theta_0$ of the first guided mode is indicated by the black dashed line at approximately 4~mrad. The panel compares \dl{$R^{[\textrm{no nuclei}]}$} for the {\it ab initio} formalism, which equivalently can be obtained using the layer formalism \cite{Rohlsberger2005} (solid blue), e.g., implemented in the software package \textsc{pynuss} \cite{pynuss}, to fits using the phenomenological model. A five-mode model with heuristic extensions (dashed yellow) is necessary for good agreement in the spectral range up to 8~mrad, reproducing the reflectance over the range of four modes. Without heuristic extensions (solid green), the fit is poor already in the first mode and beyond 6~mrad, properly describing only the second and third modes.  
        To fit a larger spectral range, the phenomenological model requires inclusion of more modes. This is a major drawback to the {\it ab initio} few-mode theory, where the off-resonant reflectance is always represented exactly for any mode number~\cite{Lentrodt2020}.\label{fig::benchmark1}}
    \end{figure}

    The cavity structure considered in the following is depicted in Fig.~\ref{fig::benchmark1}. We choose the resonant nucleus $^{57}$Fe as the archetype M\"ossbauer nucleus and practically most used species for nuclear x-ray cavity QED experiments, and consider a single unmagnetized thin resonant layer inside the cavity, where collective Lamb shifts \cite{Rohlsberger2010} and Fano line shapes \cite{Heeg2013a} can be observed.
    As the off-resonant dielectric material for the guiding layer, we choose the corresponding isomer $^{56}$Fe, which does not feature the nuclear resonance. Choosing two isomers has the advantage that the electronic contribution of the resonant layers is equal to the surrounding cavity material. Consequently, when the resonant layers are placed at varying positions, the electronic scattering properties of the cavity remain the same. This feature allows us to separate resonant effects, such as which modes the nuclear transitions couple to and which effective level scheme is realized, from cavity structure effects, such as which modal environment is realized in the cavity.  For the substrate, we choose an idealized material with a high refractive index, which essentially mimics a perfect mirror. This choice is motivated by the fact that a single layer of dielectric alone usually features weak resonances. A reflecting mirror substrate is the simplest way to realize a better resonance structure. 
    
    \subsubsection{Analytic solution}\label{sec::aifmt_SimpleEx_analyticSol}
    In order to employ the \textit{ab initio} few-mode theory, one first has to choose a few-mode basis. While the optimal basis may be difficult to find, one can use a constructive approach by choosing a basis that becomes locally complete in the region of the interaction if infinitely many discrete modes are considered \cite{Lentrodt2020}. Note that different choices for the basis vary in the number of modes required to achieve convergence of the results.
    
    Here, we choose Dirichlet modes, that is modes with hard wall boundary conditions at the top and bottom surface of the layer cavity, giving the transverse mode functions \cite{Domcke1983,Lentrodt2020,Viviescas2004}
    \begin{align}
    	\tilde{\chi}_\lambda(z) = \sqrt{\frac{2}{L}} \sin\left(\pi \lambda\frac{z}{L}\right)\,.
    \end{align}
    where $L$ is the cavity thickness. The few-mode basis is then constructed as a finite subset of these modes by choosing the relevant set of $\lambda$ indices. By observing the convergence of the spectral observables, we select the significantly contributing modes.

    The corresponding one-dimensional problem for this set of few-mode bases has been solved in \cite{Domcke1983,Lentrodt2020}, giving the system-bath couplings and few-mode propagator
        \begin{subequations}\label{eq::aifmt_singleLayer_sol_PropCoup}
            \begin{align}
            \mathcal{W}_\lambda(\omega, \theta) &= \frac{\pi \beta }{L} e^{-i\beta} \frac{1}{\alpha \cot(\alpha) - s - i\beta} \frac{\lambda (-1)^\lambda}{\sqrt{\omega_\lambda}} \,, \label{eq::aifmt_singleLayer_sol_PropCoupa}
            \\[2ex]
            \mathcal{W}^\dagger_\lambda(\omega, \theta) &= \frac{\pi \beta }{L} e^{+i\beta} \frac{1}{\alpha \cot(\alpha) - s + i\beta} \frac{\lambda (-1)^\lambda}{\sqrt{\omega_\lambda}}\,,
            \\[2ex]
            \mathcal{D}_{\lambda \lambda'}(\omega, \theta) &= \frac{\beta^2/L^2 - \omega_\lambda^2}{2\omega_\lambda} \delta_{\lambda \lambda'}\nonumber \\
            &\mkern-50mu + \frac{\pi^2}{L^2} \frac{1}{\alpha \cot(\alpha) - s - i\beta} \frac{\lambda (-1)^\lambda }{\sqrt{\omega_\lambda}} \frac{\lambda' (-1)^{\lambda'} }{\sqrt{\omega_{\lambda'}}}\,, \label{eq::aifmt_singleLayer_sol_PropCoupc}
            \\[2ex]
            \mathcal{D}^{-1}_{\lambda \lambda'}(\omega, \theta) &= \frac{2\omega_\lambda}{\beta^2/L^2 - \omega_\lambda^2} \delta_{\lambda \lambda'}- \frac{\pi^2}{L^2} \frac{1}{\alpha \cot(\alpha) - i\beta} \nonumber \\
            &\mkern-50mu \times \frac{2\lambda (-1)^\lambda \sqrt{\omega_\lambda}}{\beta^2/L^2 - \omega_\lambda^2} \frac{2\lambda' (-1)^{\lambda'} \sqrt{\omega_{\lambda'}}}{\beta^2/L^2 - \omega_{\lambda'}^2}\,, \label{eq::aifmt_singleLayer_sol_PropCoupd}
            \end{align}
        \end{subequations}
    where $L$ is the thickness of the cavity layer, $\omega_\lambda^2/2 = (\pi^2 \lambda^2)/(2 L^2) + \tilde{V}$ defines the system mode frequency, $\tilde{V}(\omega)=(1-n^2)\omega^2/2$ is the energy dependent potential, $n$ is the complex refractive index of the guiding layer material ($^{56}$Fe), $\beta = \omega L = kL$ is the scaled dimensionless momentum variable \cite{Domcke1983}, $k$ is the wave number in units of $\textrm{m}^{-1}$, 
    $\alpha=(\beta^2 - 2\tilde{V}L^2)^{\nicefrac{1}{2}}$, 
    $s = \sum_{\lambda \in \Lambda_Q} (2\lambda^2\pi^2)/(\alpha^2 - \lambda^2\pi^2)$, and $\lambda$ is the system-mode %
    index, which is summed over the chosen few-mode subset $\Lambda_Q$ of the locally complete Dirichlet basis. 
    Here, we have written the hermitian conjugate of the system-bath coupling $\mathcal{W}$ explicitly, in order to demonstrate which quantities should be excluded from the complex conjugation operation (see Sec.~\ref{sec::Xray-AIFMT_complexRefractive} for details). For example, $\alpha$ contains an imaginary part via the complex refractive index, but should not be conjugated within the non-hermitian Hamiltonian prescription.
    
    Due to the perfect mirror substrate, the system reduces to a single-channel problem with reflection only, such that that the system-bath couplings $\mathcal{W}_\lambda(\omega)$ do not depend on a channel index $m$. The background scattering matrix then reduces to a scalar given by \cite{Domcke1983}
    \begin{align}\label{eq::aifmt_Sbg}
    	S_\textrm{bg}(\omega, \theta) =& e^{-2i\beta}\: \frac{\alpha \cot(\alpha) + i\beta}{\alpha \cot(\alpha) - i\beta} \nonumber \\
    	& \qquad \times \frac{\alpha \cot(\alpha) - s - i\beta}{\alpha \cot(\alpha) - s + i\beta}\,.
    \end{align} 
    At a given incidence angle or parallel wave vector, this one-dimensional solution can be applied to our three-dimensional x-ray system using
    \begin{subequations}
        \begin{align}\label{eq::aifmt_3Dmap}
        n \rightarrow n(\theta) &= \frac{\sqrt{n^2-\cos^2(\theta)}}{\sin(\theta)} \,,
        \\
        \beta \rightarrow \beta(\theta) &= \omega L \sin(\theta) 
        \end{align}
    \end{subequations}
    as the index of refraction and scaled momentum variable, respectively, see Appendix~\ref{sec::classical_wave}.

    The interaction with the resonant nuclei is governed by the coupling constant
    \begin{align}\label{eq::aifmt_coup_analytic}
    g_{\lambda l} = -id\omega_{\textrm{nuc},l} \sqrt{\frac{f_\textrm{LM}\rho_N t_l}{2\omega_\lambda}} \tilde{\chi}_\lambda(z_l) \,,
    \end{align}
    where we have dropped the dependence on $\theta$ for brevity, $d$ is the nuclear dipole moment and $f_\textrm{LM}$ is the Lamb-M\"ossbauer factor. Details on the light-matter coupling constants are summarized in Appendix~\ref{sec::aifmt_DetailedDerivation_Dipole}.
    
    \dl{
    \subsubsection{A  practical guide to calculations in the \textit{ab initio} few-mode approach}\label{sec:aifmt_calc}
    
    After having derived general results for the \textit{ab initio} few-mode approach, we now show how to apply these results for practical calculations, that we will use to obtain the numerical results discussed in the following sections. For simplicity, we focus on the linear regime of the example geometry in Fig.~\ref{fig::benchmark1}(a), which is a single channel problem, since due to the perfect mirror substrate, only the reflection channel is open. The scattering ``matrices'' therefore reduce to scalar quantities corresponding to reflection coefficients as a function of incidence angle and frequency. According to Eq.~\eqref{eq::Sfull_product}, the full reflection coefficient is then given by the scalar product 
    \begin{align}
        r = {S} = {S}_{\mathrm{bg}}{S}_{\mathrm{io}}\,,     
    \end{align}
    with the reflection intensity, also known as reflectance, given by $R = |{r}|^2$.
    
    The first step of the calculation is to choose the set of mode indices $\{\lambda\}$ included in the few-mode analysis. The most straightforward approach not requiring any insight into the given problem is to start with the fundamental mode only and then to add higher-order modes successively until convergence of the observables is obtained. 
    
    After having chosen the few-mode basis, the coupling constants have to be determined.  For the example geometry, analytical results are provided in  Eqs.~\eqref{eq::aifmt_singleLayer_sol_PropCoup}-\eqref{eq::aifmt_coup_analytic}. For more general geometries, the coupling constants can be obtained by calculating matrix elements between the few-mode, the bath and the scattering states~\cite{Lentrodt2020}, using the separable expansion method \cite{Domcke1983}. 
    
    Next, the input-output reflection coefficient $r_{\mathrm{io}} = {S}_{\mathrm{io}}$ defined in Eq.~\eqref{eq::scatt_int} can be calculated by substituting the parameters Eqs.~\eqref{eq::aifmt_singleLayer_sol_PropCoup}, the nucleus-cavity coupling constant Eq.~\eqref{eq::aifmt_coup_analytic}, and the modified cavity propagator $\doubleunderline{\mathcal{D}}^{-1}_{(\textrm{int})}(\omega)$ containing the matrices in Eqs.~\eqref{eq:dmatrices} into Eq.~\eqref{eq:wood-decomposition}.  We note that in the  formulas for the parameters, the three-dimensional geometry substitutions Eqs.~\eqref{eq::aifmt_3Dmap} have to be used. 
    Alternatively, the input-output scattering matrix can be calculated in the form of Eq.~\eqref{eq::scatt_int_woodbury}, where the empty-cavity response is separated out.
    
    Analogously, the empty cavity reflectance $R^{[\textrm{no nuclei}]}$  evaluated using the scattering matrix ${S}^{[\textrm{no nuclei}]}_{\mathrm{io}}$ defined in Eq.~\eqref{eq:S-io-nonuclei} can readily be calculated. It corresponds to experimental situations in which the reflectance is recorded at x-ray energies detuned away from the nuclear resonance. For this, one may either set the nucleus-cavity coupling Eq.~\eqref{eq::aifmt_coup_analytic} in ${S}_{\mathrm{io}}$ to zero, or directly calculate it using  Eq.~\eqref{eq:S-io-nonuclei}.
    
    Similarly to the reflectances, the effective level schemes can be calculated. The effective couplings can be evaluated from Eqs.~(\ref{eq::aifmt_levScheme_driveVec}, \ref{eq::aifmt_levScheme_coup}), again with the help of the coupling constants already used for the reflectances. 
	}
    
    \subsubsection{Empty cavity}\label{sec::aifmt_SimpleEx_empty}

    \begin{figure}[t]
        \includegraphics[width=1.0\columnwidth]{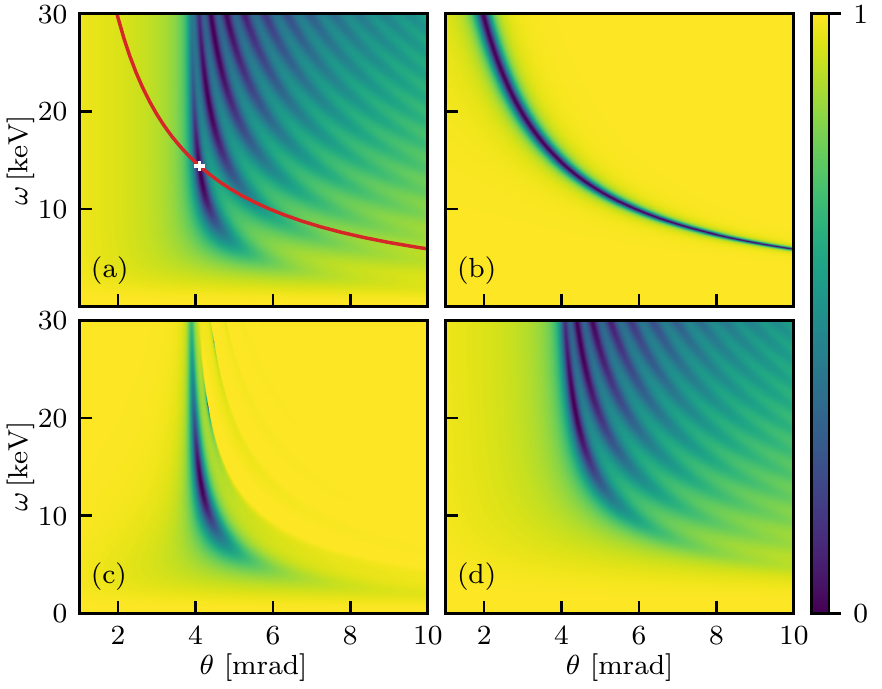}
        \caption{(Color online) Empty-cavity reflectance as a function of the incidence angle and the incident photon energy. 
        (a)  shows the full reflectance, which can be calculated via Parratt's formalism or equivalently  \dl{as $R^{[\textrm{no nuclei}]}$} via the \textit{ab initio}  few-mode theory. Note that the rocking curve in Fig.~\ref{fig::benchmark1} is a section through this figure at fixed energy. 
        The {\it ab initio} approach decomposes the total reflectance into an input-output \dl{($|S^{[\textrm{no nuclei}]}_\textrm{io}|^2$)} and a background \dl{reflectance ($|S_\textrm{bg}|^2$)}, which are shown for a single mode theory ($\lambda = 1$) in panels (c) and (d), respectively.
        As expected, the input-output contribution extracts the first resonance of the spectrum.
        The corresponding phenomenological single mode theory (b) does not correctly reproduce the first resonance. This is most clearly visible from the position of the resonance in the phenomenological model, which is shown as a red line in (a) and only coincides at the energy $14.4~$keV where the parameters of the model were fitted  (white cross). \label{fig::benchmark2_rocking2D}}
    \end{figure}
    
    \begin{figure}[t]
        \includegraphics[width=1.0\columnwidth]{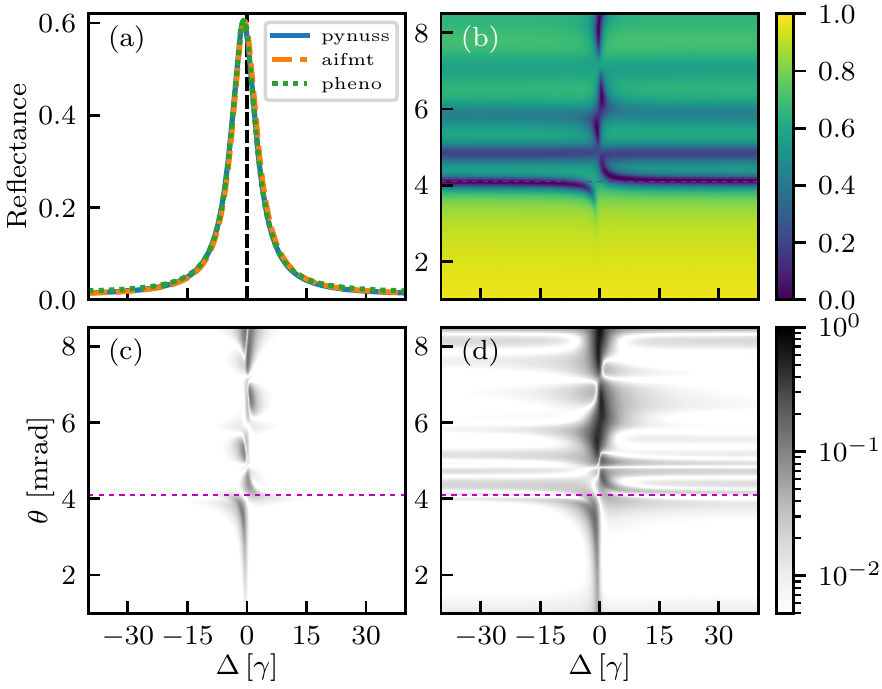}
        \caption{(Color online)
            Nuclear resonance spectra \dl{$R$} for the cavity in Fig.~\ref{fig::benchmark1} as a function of incidence angle and detuning. 
            Panel (a) shows spectra at the first cavity resonance 
            $\theta=\theta_0$ obtained from the exact layer formalism, the 5-mode \textit{ab initio} few-mode theory and a 5-mode phenomenological model, respectively.
            (b) shows the exact layer formalism results as a function of incidence angle and photon energy. (c) Residual deviation of the  5-mode \textit{ab initio} few-mode theory results to that of (b). (d) Corresponding results for the 5-mode phenomenological model. The magenta dashed line in (c,d) marks the section at $\theta_0$ shown in (a).
            On resonance, both the {\it ab initio} and the phenomenological approach yield excellent agreement, featuring deviations on the few-percent level and below. At higher incidence angles, however, the {\it ab initio} theory agrees significantly better. A more detailed comparison, including different approaches to fitting the phenomenological parameters, is presented in Appendix \ref{app:detailed_fitting}. \label{fig::singleLayer_nucSpectra}}
    \end{figure}

    We start by analyzing the off-resonant reflectance $R^{[\textrm{no nuclei}]}$ of the cavity, i.e., in the absence of the nuclear resonances, which represents a property of the empty cavity. In practice, this quantity as function of the incidence angle is also known as the rocking curve and can be measured in a cavity containing resonant nuclei by tuning the incident x-ray beam slightly away from the nuclear resonance, which does not have any influence on the much larger scales of the cavity linewidths. Here, we study the reflectance as a function of incidence angle and photon energy. We compare the results of the \textit{ab initio} framework to corresponding calculations using the phenomenological input-output theory, and further use the semi-classical layer formalism \cite{Rohlsberger2005} implemented in the software package \textsc{pynuss} \cite{pynuss} as a reference. The latter is a reimplementation of parts of the \textsc{conuss} software package~\cite{Sturhahn2000} with substantial extensions focused on applications in nuclear quantum optics.
     
    The lower panel of Fig.~\ref{fig::benchmark1} shows results for the rocking curve of the system, i.e.~the incidence-angle dependent reflectance observed with incidence angle equal to the exit angle ($\theta\,-\, 2\theta$ geometry). We find excellent agreement between the {\it ab initio} and the semi-classical approaches. Note that the different cavity mode resonances contributing to the rocking curve overlap strongly. Nevertheless, it is important to note that the \textit{ab initio} few-mode theory is equivalent to the layer formalism calculation for any set of chosen system modes, a feature which is based on the exact result shown in \cite{Lentrodt2020} and which can be checked by comparing the analytic formulas given in Sec.~\ref{sec::aifmt_SimpleEx_analyticSol} to the layer formalism result for the geometry \cite{Rohlsberger2005}. This has the practical advantage that the few-mode approximation only affects the interaction with the nuclei, but not the empty-cavity properties. This way, different approximations are disentangled, contributing to the understanding of the relevant processes. \dl{The analytical equivalence implies that the product of the background scattering matrix Eq.~\eqref{eq::aifmt_Sbg} and the empty cavity input-output result $\doubleunderline{S}^{[\textrm{no nuclei}]}_{\mathrm{io}}(\omega)$, given by the matrix contractions in Eq.~\eqref{eq::scatt_int} of the system-bath couplings Eqs.~\eqref{eq::aifmt_singleLayer_sol_PropCoup}, is identical to the Parratt calculation \cite{Rohlsberger2005,HeegPhD}. This mathematically surprising result can indeed be checked for the example geometry by substitution and simplification of the resulting sums in the few-mode case, similarly to what is shown for a Schr\"odinger equation example in \cite{Domcke1983}.}
    
    For the phenomenological few-mode theory on the other hand, no such equivalence is guaranteed. Instead, the model parameters are fitted for a chosen mode number such that the rocking curves agree in a certain angular range. Thus, the few-mode approximation affects the interaction with the nuclei and the empty-cavity response at the same time. The model fits in the bottom panel of Fig.~\ref{fig::benchmark1} show that for a five-mode model with heuristic extensions [dashed yellow] (see Sec.~\ref{sec::recap} for details), the agreement is good in the range of the four lowest resonance dips. Without heuristic extensions [solid green], however, the fit is poor already in the vicinity of the first cavity mode.
    
    These results demonstrate a key advantage of the \textit{ab initio}  few-mode model, that the off-resonant cavity scattering and hence the rocking curve is treated exactly. The model further does not require heuristic extensions and therefore gives a clear picture of the quantum mechanical interpretation of the cavity resonances.

    In order to investigate the origin of the difference between the phenomenological and the \textit{ab initio}  theory further, Fig.~\ref{fig::benchmark2_rocking2D} shows a two-dimensional generalization of the rocking curve, where the incidence angle and energy are varied. We note that these spectra are calculated at a fixed off-resonant material refractive index for simplicity. These two-dimensional empty cavity spectra give theoretical insight into the resonance structure of the cavity, which forms the electromagnetic environment for the nuclei. The layer formalism \cite{Rohlsberger2005} and the \textit{ab initio} few-mode theory are again numerically equivalent, only in the latter, the scattering matrix \dl{(panel (a), given by $|S|^2$ from Eq.~\eqref{eq::Sfull_product})} is decomposed into a mode-number dependent input-output \dl{(given by $|S^{[\textrm{no nuclei}]}_\textrm{io}|^2$ from Eq.~\eqref{eq:S-io-nonuclei})} and a background scattering part \dl{(given by $|S_\textrm{bg}|^2$ from Eq.~\eqref{eq::aifmt_Sbg})}. For the single mode results ($\lambda=1$) shown in the figure [panels (c,d)], we find that the input-output result captures the first resonance line across the whole two-dimensional space, including the total reflection cutoff at low incidence angles. The phenomenological single mode theory in (b), however, does not capture the resonance trajectory, except at $14.4~$keV, where the model parameters are fitted to the one-dimensional rocking curve. The reason for this difference is the angular and energy dependence of the quantum optical parameters, in particular $\Delta_C(\theta, \omega)$, which differs between the \textit{ab initio}  and the phenomenological model (see Sec.~\ref{sec::aifmt}). The phenomenological resonance trajectory can also be expressed analytically by solving $\Delta_{C,\lambda}(\theta, \omega)=0$ using the assumed form of the phenomenological cavity detuning Eq.~\eqref{eq::HE_CavDet_functionality} giving
    \begin{align}
    	\omega_\textrm{res,pheno}(\theta) = \omega_{\textrm{nuc}} \frac{\sin(\theta_0)}{\sin(\theta)}\,.
    \end{align}
    The comparison of this phenomenological resonance trajectory (shown as a red line in Fig.~\ref{fig::benchmark2_rocking2D}) to the actual spectral minimum of the first resonance reveals the difference between the two descriptions. 

	\subsubsection{Nuclear spectra and interacting systems}\label{sec::aifmt_SimpleEx_nucSpec}
    In Fig.~\ref{fig::singleLayer_nucSpectra}, we turn to the interacting scattering problem. We consider the nuclear resonance spectrum of the cavity as a function of detuning and incidence angle (panel b), which is accessible experimentally \cite{Heeg2015c,Heeg2015a,Rohlsberger2010,Rohlsberger2012,Heeg2013a}. Note that here, the scale of the detuning axis is on the order of neV ($\gamma_{^{57}\textrm{Fe}}\approx4.7~$neV), in contrast to the keV scale of the driving radiation, which determines the empty cavity properties investigated in Sec.~\ref{sec::aifmt_SimpleEx_empty}.
    
    The figure compares the nuclear spectra for the phenomenological and {\it ab initio} few-mode approaches to the semi-classical layer formalism \cite{Rohlsberger2005}, which serves as a well-understood reference in the linear scattering regime. In the phenomenological case, we use a five mode model to fit the nuclear spectra. For comparability, the {\it ab initio} few-mode result is also calculated using five cavity modes of the Dirichlet basis ($\lambda \in \{1,2,3,4,5\}$). \dl{The \textit{ab initio} spectra  $R$ are obtained as described in Sec.~\ref{sec:aifmt_calc}.}
    
    In the fitting procedure for the phenomenological model parameters, we first fit the empty cavity parameters to the rocking curve, yielding the yellow dashed line in Fig.~\ref{fig::benchmark1} as the best fit. This includes the heuristic dispersion phase, which is necessary to obtain a good result already in the empty cavity case (see Fig.~\ref{fig::benchmark1}). Due to the absence of a cladding, an envelope factor (see Sec.~\ref{sec::recap}) is not necessary for this cavity. The mode-ensemble interaction parameters are then determined by fitting the nuclear spectrum, using the two-dimensional layer formalism result in Fig.~\ref{fig::singleLayer_nucSpectra}(b) as the fit objective, without prioritizing a certain angular region. We note that other fitting procedures are possible, providing optimized representations for different spectral features. In Appendix \ref{app:detailed_fitting}, three options, some of which were already suggested in \cite{Heeg2015c}, are investigated and compared. The above procedure is found to provide the best description of the spectrum across the whole angular range up to 8.5~mrad. In contrast, for the {\it ab initio} method, no fit is required, and a single  model is adequate for the entire spectrum.
    
    Fig.~\ref{fig::singleLayer_nucSpectra}(c) and (d) show the residual deviations of the {\it ab initio} few-mode result and the phenomenological result obtained using above fitting procedure, both calculated for five cavity modes. The residual deviation is defined as $|R(\theta,\Delta)-R_\textrm{reference}(\theta,\Delta)|$ with the layer formalism result as the reference. We find that while both approaches provide good fits across the whole range, at higher incidence angles the {\it ab initio} theory progressively performs better than the phenomenological approach. The excellent agreement of both theories at resonance with the first cavity mode is illustrated by Fig.~\ref{fig::singleLayer_nucSpectra}(a), where the one-dimensional slice at $\theta=\theta_0$ is shown.
    
    We conclude that both approaches are well suited to model the given cavity, with the {\it ab initio} approach yielding a better global description. In comparison to the phenomenological model, it provides the main advantage that due to the absence of a fitting procedure, the quantum optical interpretation is unambiguous. In addition, the model captures the off-resonant properties exactly for any mode number and can be systematically brought to convergence by including larger mode numbers. These features are important in particular when going to more complex cavity structures and nuclear level schemes.
    
    Furthermore, increasing the number of modes provides a systematic way of improving the {\it ab initio} results. For 30 or more modes, the relative deviation is below 2\% in the entire Fig.~\ref{fig::singleLayer_nucSpectra}(c) and on this level is dominated by residual effects of the $0.5$~nm layer thickness (see also Sec.~\ref{sec::aifmt_ThickLayers}). This systematic improvement is not guaranteed in the phenomenological model (see also Appendix~\ref{app:detailed_fitting}).
    
    We note that these results do not invalidate the phenomenological approach, which may still provide numerical advantages for fitting unknown cavity structures, for example in the experiment, where the matrix elements required for the {\it ab initio} few-mode theory may be difficult to calculate (see, however, the Green's function approach in Sec.~\ref{sec::Green} for a numerically efficient alternative). Instead, the applicability of the phenomenological approach is clarified and an upgraded version can be employed when necessary. This progress in the understanding of the approach bears potential in particular for calculations beyond the linear regime \cite{Heeg2016arxiv}, where the theoretical models are required to be predictive and hence well understood.
    
    The investigation here is supplemented by a detailed comparison of the phenomenological and {\it ab initio} few-mode approaches in Appendix \ref{app:detailed_fitting}, where we also consider different fitting routines and the convergence at higher mode numbers.
    
    \subsubsection{Coupling constants and frequency dependence}
    \begin{figure}[t]
        \includegraphics[width=1.0\columnwidth]{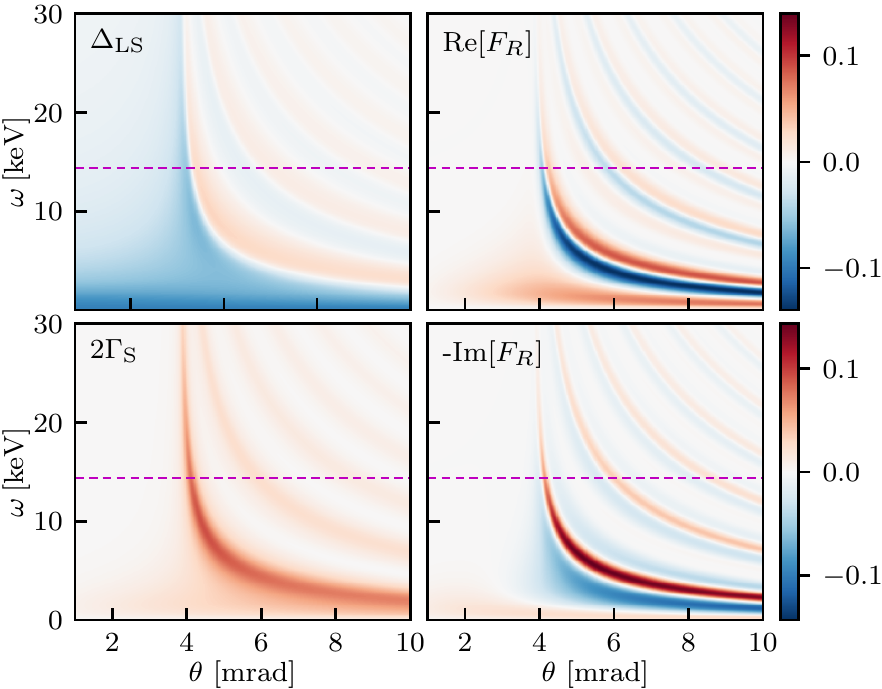}
        \caption{(Color online)
            Quantum optical parameters in the \textit{ab initio} few-mode theory for the example cavity with a single resonant thin-layer ensemble defined in Fig.~\ref{fig::benchmark1}. The results are calculated using the well-converged 20-mode theory and shown as a function of photon frequency and incidence angle. The left panels show the collective Lamb shift $\Delta_\textrm{LS}$ and superradiance $\Gamma_\textrm{S}$. The right panels show the real and imaginary parts of $F_R$ as defined in the main text. All parameters vary significantly both along the angular and frequency axis with non-trivial functional dependence. However, in the neV-$\mu$eV range around the nuclear resonant energy (magenta dashed line), which is relevant for the nuclear line shape, the variations of the parameters in frequency are negligible. \label{fig::couplings}}
    \end{figure}
    In Fig.~\ref{fig::couplings}, we investigate the parametric dependencies of the coupling constants in the \textit{ab initio} few-mode theory. In particular, we consider the nuclear complex level shift \cite{Domcke1983} given by $F= \Delta_\textrm{LS} - i\Gamma_\textrm{S}/2=\doubleunderline{g}\,\doubleunderline{\mathcal{D}}^{-1}\doubleunderline{g}^\dagger $ and the quantity $F_R = -2\pi i \doubleunderline{\mathcal{W}}^\dagger \doubleunderline{\mathcal{D}}^{-1} \doubleunderline{g}^\dagger\doubleunderline{g}\,\doubleunderline{\mathcal{D}}^{-1} \doubleunderline{\mathcal{W}}$, which quantifies the nuclear resonance peak height in analogy to $2\kappa_R$ for the cavity resonances \cite{Heeg2013b}. Together, these two quantities contain the information to calculate the nuclear spectrum in Fig.~\ref{fig::singleLayer_nucSpectra} according to Eq.~\eqref{eq::scatt_int_woodbury}. Since we consider a single nuclear ensemble in a cavity with only the reflection channel being open, both of these quantities are scalars \dl{and can be obtained using the substitution of the analytical formulas described in Sec.~\ref{sec:aifmt_calc}.}
    
    Results are shown in Fig.~\ref{fig::couplings} as a function of photon frequency and incidence angle over the range of the cavity spectrum in Fig.~\ref{fig::benchmark2_rocking2D}.  We note that all parameters  vary significantly in the frequency and angular direction, featuring non-trivial functional dependencies that give rise to the practical differences to the phenomenological model (see Sec.~\ref{sec::aifmt_SimpleEx_nucSpec}). However, while there are significant variations in the energy direction on the keV scale, the parameters are essentially constant in the neV-$\mu$eV range around the nuclear resonant energy. This confirms that the adiabatic approximation performed in Sec.~\ref{sec::adiabatic_lineShape} is valid for such cavities. The non-trivial angular dependence nevertheless persists. We note that since the cavity parameters are usually obtained by fitting to the rocking curve~\cite{Heeg2015c}, they are particularly susceptible to the differences in the phenomenological description.

    \subsection{Generalization to thick layers of resonant nuclei}\label{sec::aifmt_ThickLayers}
    In this section, we develop the quantum optical description for x-ray cavities with thick resonant layers. The case of thick resonant layers is practically relevant as it has been studied extensively in nuclear forward scattering experiments \cite{Helisto1991,Schindelmann2002,Smirnov2007,Shakhmuratov2013} and has been used in grazing incidence, for example, for novel spectroscopy techniques~\cite{Rohlsberger1997} or the study of magnetism in thin films, see, e.g.~\cite{PhysRevB.58.8590,le_zak_2010}. Placed inside cavities, thick resonant layers are particularly interesting from the theoretical perspective, since they are difficult to describe using a phenomenological approach. The reason is that in the latter, the mode functions are not known, such that one would have to fit a continuum of light-nucleus coupling parameters to describe the spatial variation throughout the thick layer.

    Within the \textit{ab initio} few-mode theory on the other hand, the description is rather straightforward and can be considered as the continuum limit of the general multi-layer solution derived in Sec.~\ref{sec::aifmt_GeneralSolution}. We further show that the thick layer system can be described as a self-coupled continuum ensemble, rather than the few-level schemes obtained in the thin-layer case.

    \subsubsection{Theory}
    \begin{figure}
        \includegraphics[width=1.0\columnwidth]{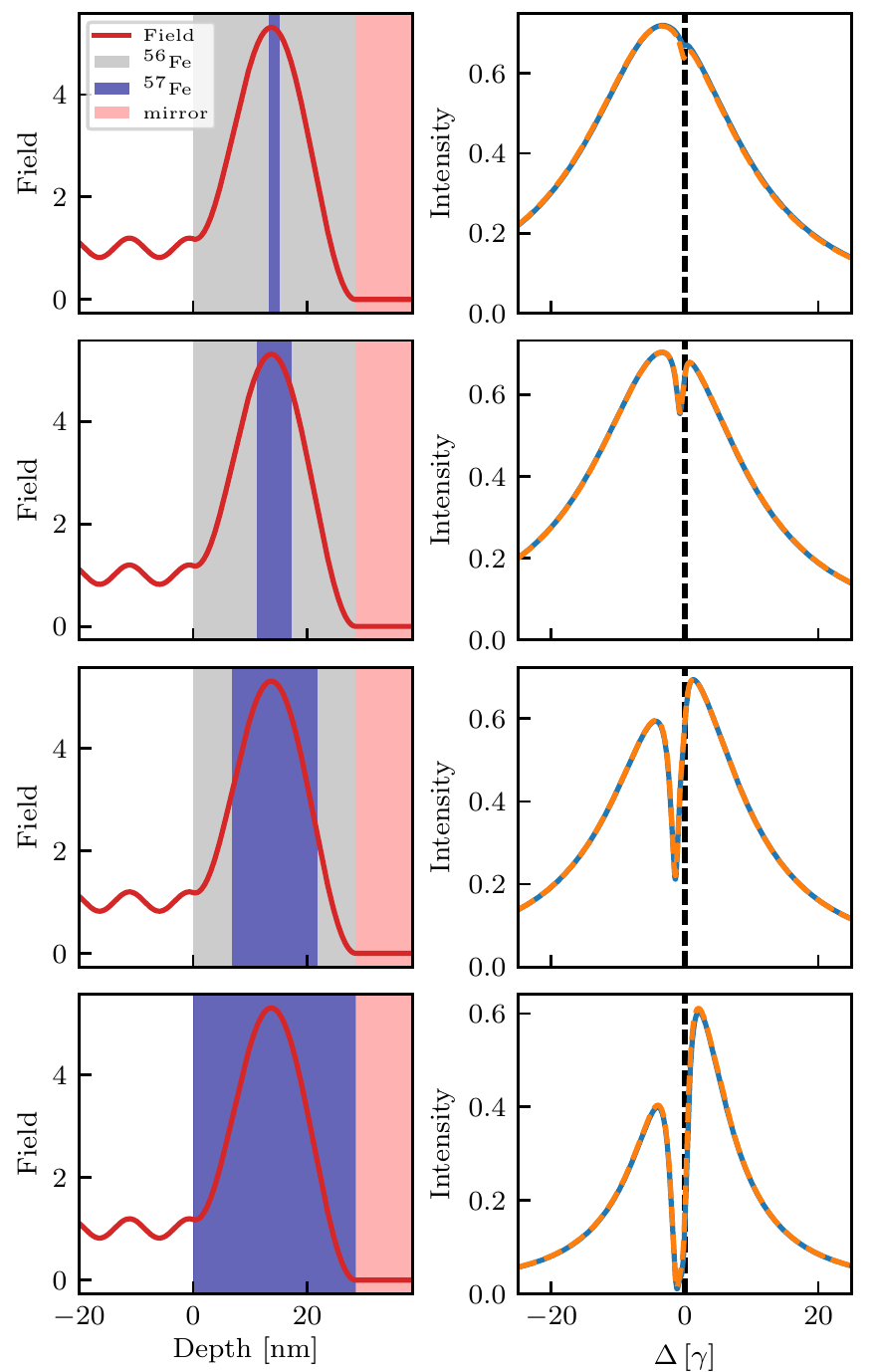}
        \caption{(Color online) Thick resonat layer effects in the nuclear x-ray cavity QED system. The four rows correspond to different thicknesses of the resonant layer. From top to bottom, the thickness of the resonant layer is increased ($2.0$~nm, $6.0$~nm, $15.0$, $28.4$~nm) while keeping the number of nuclei constant (with the usual enriched abundance of $0.95$ at the reference thickness $2.0$~nm). 
        The left column shows illustrations of the respective cavity structures. The right column shows corresponding nuclear spectra at the first rocking minima $\theta_0$.  
        We see that as the thickness increases, an EIT-like dip appears in the spectra. The continuum version of the \textit{ab initio} few-mode theory result (calculated at 15 modes, yellow dashed line) shows excellent agreement with the layer formalism result (solid blue line) in all cases.}
        \label{fig::thickLayer_nucSpectra}
    \end{figure}
    With the general solution of multi-mode multi-layer systems available, the case of thick layers can be solved simply by partitioning the layer into finely spaced sub-ensembles and taking the continuum limit of many such sub-ensembles.

    The main structure-dependent quantity to compute in order to obtain the scattering matrix is
    \begin{align}
    \Delta^{[\textrm{int}]}_{\lambda \lambda'}(\omega) = \sum_l \frac{N^{\mathstrut}_{l} g^{*\mathstrut}_{l\lambda} g^{\mathstrut}_{l\lambda'}}{\omega - \omega_{\textrm{nuc},l} + i\frac{\gamma}{2}} \,,
    \end{align}
    which modifies the intra-cavity propagator according to Eq.~\eqref{eq::mod_prop}.

    For a thick layer of resonant material, adjacent atomic layers of nuclei are usually separated by a distance much less than the variation in the mode profile. Dividing the thick layer into  thin sub-ensembles, we can then take the continuum limit of the sub-ensemble sum, such that
    \begin{align}\label{eq::cont_limit_int}
        \Delta^{[\textrm{int}]}_{\lambda \lambda'}(\omega) \approx \frac{|d|^2 f_\textrm{LM}}{2 \sqrt{\omega_\lambda \omega_{\lambda'}}} \int_{z_i}^{z_f} dz \rho_N(z)\frac{\omega_{\textrm{nuc}}^2(z)\tilde{\chi}^{*\mathstrut}_\lambda(z) \tilde{\chi}^{\mathstrut}_{\lambda'}(z)}{\omega - \omega_{\textrm{nuc}}(z) + i\frac{\gamma}{2}} \,,
    \end{align}
    where $\omega_{\textrm{nuc}}(z)$, $\rho_N(z)$ are the position dependent frequency and nuclear number density, respectively. For homogeneous resonant layers the $z$-dependence of the couplings is mainly determined by the mode profiles. $z_i$ [$z_f$] is the lower [upper] edge of the thick layer.

    Evaluating the integral above gives the necessary quantity to compute scattering observables. Similarly, one can compute a continuum version of the effective level scheme given by the continuum limit of Eqs.~(\ref{eq::eff_H_levelScheme}, \ref{eq::effLevel_lind})
    \begin{align}
    \hat{H}_\textrm{eff} =& \int dz \frac{\omega_{\textrm{nuc}}(z)}{2}\hat{J}^z(z) \nonumber
    \\
    &+ \int dz dz' \hat{J}^+(z) \textrm{Re}[G(z, z')] \hat{J}^-(z') \nonumber
    \\
    &+ \int dz \hat{J}^+(z) \underline{\Omega}^T(z) \underline{\hat{b}}^{\mathrm{(in)}}(t) + h.c. \,,
    \end{align}
    and the effective Lindblad term
    \begin{align}
    \mathcal{L}_\textrm{eff}[\rho] =  & -\int dzdz' \textrm{Im}[G(z,z')] (2\hat{J}^-(z) \nonumber \rho\hat{J}^+(z') \\
    &\qquad\qquad\qquad - \{\hat{J}^+(z)\hat{J}^-(z'),\,\rho \}) \nonumber
    \\
    &+ \mathcal{L}_\textrm{SE}[\rho]\,.
    \end{align}
	This yields a physical interpretation of the thick layer as a self-coupled continuum ensemble rather than an effective few-level scheme as in the thin layer case.

    \subsubsection{Thick resonant layer in the example cavity}
    As a practical example we again use the cavity from Fig.~\ref{fig::benchmark1} and consider a variable thickness of the resonant layer, while keeping it at the center of the cavity.

    Since the resonant layer material is homogeneous, the number density of resonant nuclei and the resonance frequency of the nuclear transition do not vary with $z$. In this case, denoting the thickness of the resonant layer by $t_\textrm{res}$, we can evaluate Eq.~\eqref{eq::cont_limit_int} to obtain
    \begin{align}
    \Delta^{[\textrm{int}]}_{\lambda \lambda'}(\omega) &\approx \frac{|d|^2 \omega_{\textrm{nuc}}^2 }{2 \sqrt{\omega_\lambda \omega_{\lambda'}}} \frac{f_\textrm{LM}\rho_N}{\omega - \omega_{\textrm{nuc}} + i\frac{\gamma}{2}}\nonumber \\
    & \qquad \times \int_{\frac{L-t_\textrm{res}}{2}}^{\frac{L+t_\textrm{res}}{2}} dz \tilde{\chi}^{*\mathstrut}_\lambda(z) \tilde{\chi}^{\mathstrut}_{\lambda'}(z) \nonumber
    \\[2ex]
    &= \frac{|d|^2 \omega_{\textrm{nuc}}^2 }{2 \sqrt{\omega_\lambda \omega_{\lambda'}}} \frac{f_\textrm{LM}\rho_N}{\omega - \omega_{\textrm{nuc}} + i\frac{\gamma}{2}} \Xi_{\lambda \lambda'}\,,
    \end{align}
    where the mode integral for the cavity under consideration is
    \begin{align}
    \Xi_{\lambda \lambda'} &= \frac{2}{L} \int_{\frac{L-t_\textrm{res}}{2}}^{\frac{L+t_\textrm{res}}{2}} dz \sin\left(\pi \lambda\frac{z}{L}\right) \sin\left(\pi \lambda'\frac{z}{L}\right)
    \,,
    \end{align}
    which can straightforwardly be evaluated analytically.

    Fig.~\ref{fig::thickLayer_nucSpectra} shows nuclear spectra of the example cavity at different layer thicknesses. The results from the continuum \textit{ab initio} few-mode theory show excellent agreement with the semi-classical \textsc{pynuss} \cite{pynuss} calculation, confirming the validity of our quantum model for cavities with thick resonant layers.
    
    Finally, to interpret the spectra, we note that  the self-coupled continuum ensemble features an EIT-like dip effect for thick layers. At layer thicknesses comparable to the cavity thickness, additional distortions are found, indicating the emergence of higher order spectral interferences.
    
    \section{\textit{Ab initio} Green's function approach to x-ray cavity QED}\label{sec::Green}
    In this section, we apply a well-known Green's function technique \cite{Gruner1996,Dung2002,Scheel2008,Asenjo-Garcia2017a} to the nuclear x-ray cavity QED system. This provides a second \textit{ab initio}  approach to the problem, which involves different approximations. The investigation completes the comparison of the three different \textit{ab initio}  approaches depicted in Fig.~\ref{fig::overview} and contributes as a numerically efficient method for calculating effective nuclear level schemes.
    
    As a theoretical result, we show that the layer formalism is essentially analogous to an effective propagation equation that can be derived from the Green's function quantization by linear dispersion theory, up to applicable but unnecessary approximations that can be removed (such as the rotating wave approximation). Consequently, in the linear regime, all three \textit{ab initio}  approaches (layer formalism, \textit{ab initio} few-mode theory, Green's function approach, see Fig.~\ref{fig::overview}) are essentially equivalent in that they can be used as alternative methods to calculate linear scattering spectra. However, each of these methods provides different advantages with regards to the interpretation of the quantum system or the mode structure of the cavity due to the different underlying approximations. In addition, the methods presented here go beyond the semi-classical linear dispersion theory in that they bear the potential for direct application in different sectors beyond the linear scattering and low excitation regimes.
   
    The Green's function technique presented in this section most importantly provides a numerically efficient method to calculate effective nuclear level schemes for arbitrary complex layer stacks. It also removes the need for a fitting procedure that is necessary in the phenomenological model in \cite{Heeg2013b,Heeg2015c}, with the main advantage over the \textit{ab initio} few-mode approach in Sec.~\ref{sec::eff_scheme} being the straightforward numerical implementation even for complex systems. On the other hand, the Green's function approach does not provide access to the dynamics of the cavity modes, which is naturally included in the  \textit{ab initio} few-mode approach. We demonstrate the validity and usefulness of the Green's function level scheme for practically relevant example systems, including the EIT and non-EIT cavities investigated in \cite{Rohlsberger2012,Heeg2015c}. In the process, we resolve an open question with regards to the quantum optical interpretation of this system that was raised in its phenomenological model description \cite{Heeg2015c}, where the main EIT feature was reproduced, but qualitative and quantitative differences of the two-dimensional nuclear spectra were found.
    
    \dl{For readers who are mainly interested in how to apply the method in practice, we refer to the corresponding guide in Sec.~\ref{sec::Green_calc}.}
    
    \subsection{\label{sec:macroQED}Green's function and macroscopic QED}
    The quantization of absorbing dielectrics (see \cite{Scheel2008} for a review) has been studied extensively in the literature (see e.g.~\cite{Huttner1992a,Gruner1996,Dung2002,Scheel2008,Buhmann2007,Buhmann2012}). A macroscropic prescription based on the Green's function \cite{Dung2002,Scheel2008} gives the Hamiltonian in the dipole approximation~\cite{Dung2002,Yao2009}
    \begin{align}\label{eq::hamgreen}
    \hat{H} =& \int d^3\mathbf{r} \int_{0}^{\infty} d\omega \hbar \omega \hat{\mathbf{f}}^\dagger(\mathbf{r}, \omega) \hat{\mathbf{f}}(\mathbf{r}, \omega) + \sum_{ln} \frac{\hbar \omega_{\textrm{nuc},l}}{2} \hat{\sigma}^z_{ln} \nonumber
    \\
    & - \sum_{ln} [\hat{\sigma}^+_{ln} \mathbf{d}^*_l + \hat{\sigma}^-_{ln} \mathbf{d}_l] \cdot \hat{\mathbf{E}}(\mathbf{r}_{ln}) \,,
    \end{align}
    where $\hat{\mathbf{f}}(\mathbf{r}, \omega)$ are bosonic operators, $\hat{\mathbf{E}}$ is the electric field operator and unlike in previous sections,  polarization is included. The bosonic operators appearing in the Hamiltonian are related to the field operator by \cite{Dung2002}
    \begin{align}
    \hat{\mathbf{E}}(\mathbf{r}) =& i \sqrt{\frac{\hbar}{\pi \varepsilon_0}}\int_0^\infty d\omega \int d^3\mathbf{r}' \sqrt{\mathrm{Im}[\varepsilon(\mathbf{r}')]} \nonumber \\[2ex]    
    & \qquad \times \mathbf{G}(\mathbf{r}, \mathbf{r}', \omega)\cdot \:\hat{\mathbf{f}}(\mathbf{r}', \omega) \,,
    \end{align}
    where the Green's tensor is defined via
    \begin{align}\label{eq:defgreen}
    [\nabla\times\nabla\times - \frac{\omega^2}{c^2} \varepsilon(\mathbf{r}, \omega)] \mathbf{G}(\mathbf{r}, \mathbf{r}', \omega) = \delta(\mathbf{r} - \mathbf{r}') \,,
    \end{align}
    and the dielectric permittivity is allowed to be frequency dependent. Approaches based on such Hamiltonians employing the classical electromagnetic Green's function of the system are known as macroscopic QED \cite{Scheel2008}, which has been used extensively, for example, for the study of dispersion forces and related phenomena \cite{Buhmann2012}, as well as recently for the description of atom-light scattering in nanostructures \cite{Asenjo-Garcia2017a,Asenjo-Garcia2017b} and atomic-waveguide QED \cite{Masson2019_preprint}, \dl{for the theory of Bose-Einstein condensation in multi-mode cavities \cite{Bennett2020a} and for inverse design of light-matter interactions in complex environments \cite{Bennett2019_arxiv}}. The Green's function quantization provides an alternative to the normal modes approach in Appendix~\ref{sec::cavQuant} that is used as the basis for the \textit{ab initio}  few-mode theory in Sec.~\ref{sec::aifmt}. Besides its numerical efficiency that will be demonstrated later, it has the advantage that absorption is described rigorously without the need for the non-Hermitian Hamiltonian prescription in Appendix~\ref{sec::Xray-AIFMT_complexRefractive}.
    
    As a drawback, the approach does not offer an interpretation in terms of the modes or resonances of the cavity structure, which are often a driving force for the design of novel and exotic cavity structures, both in the x-ray regime \cite{Heeg2015a,Heeg2016arxiv,Haber2017,Rohlsberger2012} and at lower wavelengths \cite{Rotter2017}. Instead, all the information about the cavity environment is contained in a single function. In this context, \dl{we note that effective modes for the macroscopic QED Hamiltonian have been introduced previously \cite{Buhmann2008} and have recently found applications for cavity interactions of multiple atoms \cite{Esfandiarpour2018,Esfandiarpour2019} and strongly coupled light-matter dynamics in complex environments \cite{SanchezBarquilla2020}}. Alternatively, the Green's function can in principle be approximated in terms of mode parameters \cite{Asenjo-Garcia2017a}. For overlapping modes structures, however, such an approximate treatment is non-trivial and results in similar problems as found in the phenomenological model. On the other hand, this provides a route towards direct numerical optimization of cavity structures via the Green's function approach (see e.g.~\cite{Dory2019,Bennett2019_arxiv}). These approaches could also be complemented by various decompositions of the Green's function known from resonance theory \cite{Krimer2014,Tureci2005,Tureci2006,Ching1998,Lalanne2018,Kristensen2014,Rotter2017}.
    
    We further note that the macroscopic QED framework, that is treating the material as a refractive index, also has its limitations. Complementary approaches connecting to electronic structure theory \cite{Ruggenthaler2014}, quantum chemistry \cite{Schaefer2018} and condensed matter physics \cite{Li2020} are available for a variety of parameter regimes and, while often being computationally demanding, allow to fully integrate associated effects.
    
    \subsection{Linear dispersion theory and relation to the layer formalism}\label{sec::Green_linDisp}
    The equations of motion for the Hamiltonian Eq.~(\ref{eq::hamgreen}) are all linear, except for the usual non-linear term featuring a product of $\hat{\sigma}^z$ and the field operators (see Eq.~\eqref{eq::effFM} for the few-mode counterpart of this contribution). In \cite{Yao2009}, it is shown that analogously to the linear calculation in Sec.~\ref{sec::aifmt_GeneralSolution} and the linear dispersion calculation in \cite{Zhu1990,Lentrodt2020}, these equations can be tackled by employing the low excitation approximation $\langle\hat{\sigma}^z(t)\rangle \approx -1$ and solving the resulting linearly coupled differential equations. The result can be expressed in frequency space as \cite{Yao2009}
    \begin{align}\label{eq::E_GreenSource}
    \langle\hat{\mathbf{E}}(\mathbf{r}, \omega)\rangle = \frac{1}{\varepsilon_0} \mathbf{G}(\mathbf{r}, \mathbf{r}_{ln}, \omega)\cdot \sum_{ln}[\mathbf{d}_l^*\langle\hat{\sigma}_{ln}^+(\omega)\rangle + \mathbf{d}_l\langle\sigma_{ln}^-(\omega)\rangle],
    \end{align}
    with \cite{Yao2009}
    \begin{align}
    	\langle\hat{\sigma}_{ln}^+(\omega)\rangle = \frac{\mathbf{d}_l\cdot\langle\hat{\mathbf{E}}(\mathbf{r}_{ln}, \omega)\rangle}{\hbar (\omega + \omega_{\textrm{nuc},l})} \,,
    \end{align}
    and
    \begin{align}
    	\langle\hat{\sigma}_{ln}^-(\omega)\rangle = -\frac{\mathbf{d}^*_l\cdot\langle\hat{\mathbf{E}}(\mathbf{r}_{ln}, \omega)\rangle}{\hbar (\omega - \omega_{\textrm{nuc},l})} \,.
    \end{align}
    Substitution into Eq.~(\ref{eq::E_GreenSource}) and using Eq.~(\ref{eq:defgreen}) shows that the electric field obeys the effective wave equation
    \begin{align}
    [\nabla\times\nabla\times - \frac{\omega^2}{c^2} \varepsilon(\mathbf{r}, \omega)] \langle\hat{\mathbf{E}}(\mathbf{r}, \omega)\rangle = \nonumber
    \\
    -\frac{1}{\varepsilon_0}   \sum_{ln}\delta(\mathbf{r} - \mathbf{r}_{ln})\frac{2\omega_{\textrm{nuc},l}\mathbf{d}^T_l\mathbf{d}_l}{\hbar (\omega^2 - \omega^2_{\textrm{nuc},l})}\langle\hat{\mathbf{E}}(\mathbf{r}, \omega)\rangle \,,
    \end{align}
    where we have assumed real $\mathbf{d}$ for simplicity. The effect of the coupling to the nuclear transitions can thus be interpreted as a modification of the frequency dependent refractive index \cite{Lentrodt2020,Malekakhlagh2016b}. We note that in contrast to the formula resulting from the simple model in \cite{Lentrodt2020}, the above approach includes polarization and a rigorous treatment of absorption. On the other hand, a factor of $\frac{\omega^2_{\textrm{nuc},l}}{\omega^2}$ is absent due to the $\mathbf{E} \cdot \mathbf{r}$ gauge that is adopted in \cite{Yao2009}, which is appropriate for our weak coupling scenario, but has to be adjusted at extreme coupling strengths \cite{Malekakhlagh2016b,Malekakhlagh2017,Schaefer2019}.
    
    We further note that by transforming the above effective wave equation into the time-domain, one can obtain a propagation equation that is of the form of Shvyd'ko's time and space wave equation \cite{Shvydko1999}, with the wave equation kernel expressed explicitly for the elastic scattering case considered here.
    
    When one is interested in steady-state scattering properties, the above equation can be solved directly in frequency space using transfer matrices or Parratt's method \cite{Parratt1954}, which has been generalized to the layer formalism in the nuclear resonance scattering community \cite{Rohlsberger2005,Sturhahn2000}. The above derivation thus unveils the connection to the semi-classical theories used in nuclear resonant scattering, clarifying its relation to the full quantum theory of the absorbing dielectric environment interacting with the nuclear transitions. The main insight is that the central approximation in these approaches is the low excitation approximation. In the linear excitation regime, where this approximation applies, the approaches are then analogous. We note that in practice, slight differences in the formulas can be found, since additional convenient approximations are made in the x-ray case, which are well applicable, but unnecessary from a formal perspective. Examples include the rotating wave approximation. We also refer to Appendix~\ref{app::aifmt_dipMom} for further context.
   
    \subsection{Nuclear Master equation~\label{sec:greenMaster}}
    
    Starting from the macroscopic QED Hamiltonian in the rotating-wave approximation, the Born-Markov approximation can be used to derive an effective Liouvillian for the nuclei interacting with each other via the electromagnetic field \cite{Dung2002,Asenjo-Garcia2017a}. Since nuclei and x-rays usually feature very weak coupling and the cavities in use are highly leaky, the Born-Markov approximation is applicable in most cases. Recently, systems featuring collective strong coupling have been reported \cite{Haber2016a}, where the Born-Markov approximation may break down and the \textit{ab initio}  few-mode theory presented in Sec.~\ref{sec::aifmt} may be advantageous. \dl{Alternatively, effective continuum modes \cite{Buhmann2008} based on the Green's function quantization may prove useful, which have been shown to be tractable via cumulant expansions \cite{SanchezBarquilla2020} in certain parameter regimes.}
    
    \begin{figure}[t]
    	\includegraphics[width=\columnwidth]{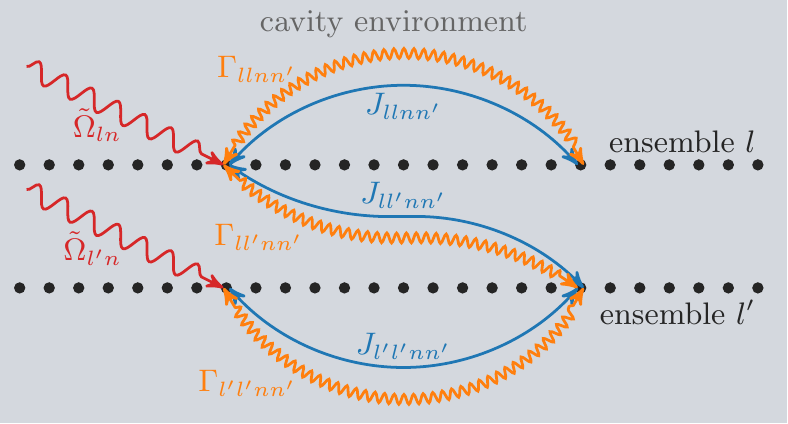}
    	\caption{(Color online) Schematic illustration of the effective nuclear Master equation obtained using the Green's function approach. $J_{ll'nn'}$ are couplings between the nuclei,  $\Gamma_{ll'nn'}$ decay constants, and  $\tilde{\Omega}_{ln}$ the effective driving strengths. The cavity does not appear explicitly, as it is treated as an environment in the Green's function approach.}
    	\label{fig::sketch-greenMB}
    \end{figure}
    
    Following the approach in \cite{Asenjo-Garcia2017a,Asenjo-Garcia2017b} and again excluding polarization effects for simplicity, we can write the effective Hamiltonian in our ensemble notation as
    \begin{align}\label{eff::Green_H}
    \hat{H}_\textrm{eff} =& \sum_{ln} \frac{\omega_{\textrm{nuc},l}}{2} \hat{\sigma}_{ln}^z - \sum_{ln} \sum_{l'n'} J_{ll'nn'} \hat{\sigma}_{ln}^+ \hat{\sigma}_{l'n'}^- \nonumber
    \\
    & - \sum_{ln} \left[ \textbf{d}_l^*\cdot\textbf{E}_\mathrm{in}(\mathbf{r}_{ln}) \hat{\sigma}_{ln}^+ + h.c.\right] \,,
    \end{align}
    and the Lindblad term as
    \begin{align}\label{eff::Green_Lind}
    \mathcal{L}_\textrm{eff}[\rho] = &\sum_{ln} \sum_{l'n'} \frac{\Gamma_{ll'nn'}}{2} (2\hat{\sigma}^-_{ln}\rho\sigma_{l'n'}^+ - \{\hat{\sigma}_{ln}^+\hat{\sigma}_{l'n'}^-,\,\rho\})\nonumber
    \\
    &+\mathcal{L}_\textrm{SE}[\rho]\,,
    \end{align}
    where $J_{ll'nn'}$ [$\Gamma_{ll'nn'}$] is the nucleus-nucleus coupling [decay] constant and $\textbf{E}_\mathrm{in}(\mathbf{r}_{ln}) = \langle \textbf{E}_\mathrm{in}(\mathbf{r}_{ln}) \rangle$ is the driving field \cite{Asenjo-Garcia2017a} providing a driving strength $\tilde{\Omega}_{ln}=\textbf{d}^*\cdot\textbf{E}_\mathrm{in}(\mathbf{r}_{ln})$ for the nucleus $n$ in ensemble $l$. Within the Born-Markov approximation, the nuclei are thus described as driven by the incident cavity field without resonant modification.
   
    The couplings and decay constants can be obtained from the Green's function of the system \cite{Asenjo-Garcia2017a} by
    \begin{align}
    J_{ll'nn'} &= \frac{\mu_0 \omega^2_{\textrm{nuc},l}}{\hbar} \textbf{d}_l^*\cdot \mathrm{Re}[\textbf{G}(\mathbf{r}_{ln}, \mathbf{r}_{l'n'}, \omega_{\textrm{nuc},l})]\cdot \textbf{d}_{l'}\,,\\
    \Gamma_{ll'nn'} &= 2\frac{\mu_0 \omega^2_{\textrm{nuc},l}}{\hbar} \textbf{d}_l^*\cdot \mathrm{Im}[\textbf{G}(\mathbf{r}_{ln}, \mathbf{r}_{l'n'}, \omega_{\textrm{nuc},l})]\cdot \textbf{d}_{l'} \,.
    \end{align}
    In the above form, the effective nuclear Hamiltonian includes intra-layer couplings between individual nuclei, constituting an effective level scheme beyond the single parallel wave vector approximation employed in Sec.~\ref{sec::eff_scheme}. The resulting many-body coupling scheme is illustrated in Fig.~\ref{fig::sketch-greenMB}.
    
    In order to obtain an ensemble picture as in Fig.~\ref{fig::couplings_sketch}, one can introduce spin-wave operators \cite{Masson2019_preprint,Asenjo-Garcia2017b,Zhou2007} $\hat{\sigma}^\pm_l(\textbf{k}_\parallel) = \sum_n \hat{\sigma}^\pm_{ln} e^{\pm i \textbf{k}_\parallel \cdot \textbf{r}_{\parallel,n}}$, which under the translational invariance assumption diagonalize the effective Hamiltonian \cite{Asenjo-Garcia2017b}. Here, since we are mainly interested in the linear sector, where the parallel wave vector $\textbf{k}_\parallel$ of a given excitation is conserved, we can then again derive an effective Hamiltonian for the subspace at a single parallel wave vector (see also Appendix~\ref{sec::aifmt_DetailedDerivation_eff1D}), as we show in the following. Such an effective level scheme is of interest, since it corresponds directly to the effective level schemes derived in Sec.~\ref{sec::eff_scheme} from the {\it ab initio} few-mode approach and to the corresponding interpretation of recent experiments \cite{Rohlsberger2010,Rohlsberger2012,Heeg2013a,Heeg2015a,Heeg2015b,Heeg2015c,Haber2017,Haber2019}. The Green's function approach provides an alternative {\it ab initio} method to calculate these level schemes, with different approximations involved bearing the potential for different generalizations. The main advantage of the Green's function approach over the \textit{ab initio} few-mode theory is its numerically efficiency for calculating effective level schemes in the case of the layered cavity geometry, as we demonstrate in the following sections.
    
    \subsection{Effective nuclear level scheme in the low excitation subspace}\label{sec::Green_lowExc_levScheme}
    In the low excitation regime, where $\langle\hat{\sigma}^z_{ln}\rangle\approx-1$, the equation of motion for the lowering operator resulting from the Born-Markov Master equation reads
    \begin{align}
    	\frac{d}{dt}\hat{\sigma}^-_{ln}  = &-i(\omega_{\textrm{nuc},l} + i\frac{\gamma}{2})\hat{\sigma}^-_{ln} + i \sum_{l'n'} \mathcal{G}(\mathbf{r}_{ln}, \mathbf{r}_{l'n'}) \hat{\sigma}^-_{l'n'} \nonumber
    	& \\
    	&+i\textbf{d}_l^*\cdot\textbf{E}_\mathrm{in}(\mathbf{r}_{ln}) \,,
    \end{align}
    where 
    \begin{align}
    	\mathcal{G}(\mathbf{r}_{ln}, \mathbf{r}_{l'n'}) =& J_{ll'nn'} + i\frac{\Gamma_{ll'nn'}}{2}\nonumber
    	\\
    	=&\frac{\mu_0 \omega^2_{\textrm{nuc},l}}{\hbar} \textbf{d}_l^*\cdot \textbf{G}(\mathbf{r}_{ln}, \mathbf{r}_{l'n'}, \omega_{\textrm{nuc},l})\cdot \textbf{d}_{l'} \,,
    \end{align}
    and we have dropped the expectation value brackets for brevity. Using the approximate translational invariance of the x-ray cavity, we can write the Green's function as~\cite{Tomas1995}
    \begin{align}\label{green-approx}
    	\textbf{G}(\mathbf{r}_{ln}, \mathbf{r}_{l'n'}, \omega) = \int \frac{d^2\mathbf{k}_\parallel}{(2\pi)^2} \textbf{G}(z_l, z_{l'}, \textbf{k}_\parallel, \omega) e^{i\mathbf{k}_\parallel \cdot (\mathbf{r}_{\parallel,n} - \mathbf{r}_{\parallel,n'})} \,,
    \end{align}
    where $\textbf{G}(z_l, z_{l'}, \textbf{k}_\parallel, \omega)$ is the Fourier transformed Green's function and lattice offsets between the ensembles of less than a lattice constant are neglected. Similarly, we define 
    \begin{align}\label{green-curly-g}
        \mathcal{G}(z_l, z_{l'}, \textbf{k}_\parallel)=\frac{\mu_0 \omega^2_{\textrm{nuc},l}}{\hbar} \textbf{d}_l^*\cdot \textbf{G}(z_l, z_{l'}, \textbf{k}_\parallel, \omega_{\textrm{nuc},l})\cdot \textbf{d}_{l'}\,.
    \end{align}
    
    Transferring to the spin-wave basis \cite{Masson2019_preprint}, the equations of motion simplify to
    \begin{align}\label{equ::eff_lev_scheme_linEOM}
    	\frac{d}{dt} \hat{\sigma}^-_{l}(\textbf{k}_\parallel) =& -i(\omega_{\textrm{nuc},l} + i\frac{\gamma}{2}) \hat{\sigma}^-_{l}(\textbf{k}_\parallel) \nonumber
    	\\
    	& + i\sum_{l'}\frac{N}{A_\parallel}\mathcal{G}(z_l, z_{l'}, \textbf{k}_\parallel) \hat{\sigma}^-_{l'}(\textbf{k}_\parallel) \nonumber
    	\\
    	&+ i\frac{N}{A_\parallel} \mathbf{d}_l^* \cdot \textbf{E}_\mathrm{in}(z_{l}, \textbf{k}_\parallel) \,,
    \end{align}
    where we have used $\sum_n e^{i(\textbf{k}_\parallel - \textbf{k}'_\parallel)\cdot\textbf{r}_{\parallel,n}} = \frac{(2\pi)^2 N}{A_\parallel}\delta(\textbf{k}_\parallel - \textbf{k}'_\parallel)$, defined
    \begin{align}
    	\textbf{E}_\mathrm{in}(z_{l}, \textbf{k}_\parallel) = \int d^2\mathbf{r}_{\parallel,n} \textbf{E}_\mathrm{in}(\mathbf{r}_{ln})e^{-i\textbf{k}_\parallel\cdot\textbf{r}_{\parallel,n}} \,,
    \end{align} 
    and used the approximation $\sum_n\textbf{E}_\mathrm{in}(\mathbf{r}_{ln})e^{-i\textbf{k}_\parallel\cdot\textbf{r}_{\parallel,ln}} \approx\frac{N}{A_\parallel}\int d^2\mathbf{r}_{\parallel,n} \textbf{E}_\mathrm{in}(\mathbf{r}_{ln})e^{-i\textbf{k}_\parallel\cdot\textbf{r}_{\parallel,n}} $, which is valid for grazing incidence illumination, where the phase variation of the illuminating beam is small over the length scale of the lattice parameter, and which effectively neglects Bragg scattering.
    
    Importantly, the driving field and inter-ensemble couplings above are also approximated as independent of $n$ within one ensemble $l$. If all nuclei are located at the same $z_l$, this is an exact representation of the ensemble in the considered geometry. Practically, however, one often wishes to interpret a resonant layer of finite thickness as a single ensemble \cite{Heeg2015c,Rohlsberger2012,Rohlsberger2010}. The ensemble quantities are then approximated as constant over the layer thickness and taken equal to their central value at $z_l$, which we denote as the thin-layer approximation. We refer to Sec.~\ref{sec::Green_applications} for a practical discussion of this approximation and its consequences, as well as to the previous discussion of thick layers in the context of the {\it ab initio} few-mode theory in Sec.~\ref{sec::aifmt_ThickLayers}.
    
    We see that Eq.~\eqref{equ::eff_lev_scheme_linEOM} provides a closed set of operator equations at a given parallel wave vector. The effective subspace Liouvillian corresponding to this linear equation of motion is given by
    \begin{align}\label{eff::Green_H_effLin}
    	\hat{H}_\textrm{eff}(\textbf{k}_\parallel) =& \sum_{l} \frac{\omega_{\textrm{nuc},l}}{2} \hat{\sigma}_{l}^z(\textbf{k}_\parallel) + \sum_{ll'} \Delta_{ll'}(\textbf{k}_\parallel) \hat{\sigma}_{l}^+(\textbf{k}_\parallel) \hat{\sigma}_{l'}^-(\textbf{k}_\parallel) \nonumber
   		 \\
    	& + \sum_{l} \left[ \textbf{d}_l^*\cdot\textbf{E}_\mathrm{in}(z_l, \textbf{k}_\parallel) \hat{\sigma}_{l}^+(\textbf{k}_\parallel) + h.c.\right] \,
    \end{align}
    and the Lindblad term as
    \begin{align}\label{eff::Green_Lind_effLin}
    	\mathcal{L}_\textrm{eff}[\rho](\textbf{k}_\parallel) = &\sum_{ll'} \frac{\gamma_{ll'}(\textbf{k}_\parallel)}{2} \biggl(2\hat{\sigma}^-_{l}(\textbf{k}_\parallel)\rho\sigma_{l'}^+(\textbf{k}_\parallel) \nonumber
    	\\
    	&\qquad - \{\hat{\sigma}_{l}^+(\textbf{k}_\parallel)\hat{\sigma}_{l'}^-(\textbf{k}_\parallel),\,\rho\}\biggr)\nonumber
    	\\[2ex]
    	&+\mathcal{L}_\textrm{SE}[\rho]\,,
    \end{align}
    where
    \begin{align}
    	\Delta_{ll'} &= \frac{N}{A_\parallel} \mathrm{Im}[\dl{\mathcal{G}(z_l, z_{l'}, \textbf{k}_\parallel)}] \,, \label{eq::Green_couplings_lin}
    	\\ 
    	\gamma_{ll'} &= 2 \frac{N}{A_\parallel} \mathrm{Im}[\dl{\mathcal{G}(z_l, z_{l'}, \textbf{k}_\parallel)}] \,. \label{eq::Green_decay_lin}
    \end{align}
    We see that this effective level scheme corresponds in close analogy to the one derived from the few-mode approach in Sec.~\ref{sec::eff_scheme} (as depicted in Fig.~\ref{fig::couplings_sketch}), with the driving field being expressed as a channel mode expansion in the few-mode case $\textbf{d}_l^*\cdot\textbf{E}_\mathrm{in}(z_l, \textbf{k}_\parallel) = \underline{\Omega}_l^T \underline{\hat{b}}^{\mathrm{(in)}}(t)$.
    
    In summary, in the linear regime, the nuclear dynamics for an incident field of a defined parallel wave vector are given by the effective level scheme Eqs.~\eqref{eff::Green_H_effLin}, \eqref{eff::Green_Lind_effLin}. For an incident field containing multiple parallel wave vectors, the superposition principle applies in the linear regime.
    
    We note that as also observed in the \textit{ab initio}  few-mode theory  (see Appendix \ref{sec::aifmt_DetailedDerivation_Dipole}), the effective linear level scheme parameters depend on the dipole moment scaled by $\sqrt{N/A_\parallel}$, such that only the nuclear number density is relevant for linear observables.
    
    \subsection{Linear solution in frequency space\label{sec:solgreen}}
    Eq.~\eqref{equ::eff_lev_scheme_linEOM} can be solved in frequency space to give \cite{Asenjo-Garcia2017a,Lentrodt2020}
    \begin{align}\label{eq::Green_linSol}
    	\hat{\sigma}^-_{l}(\textbf{k}_\parallel, \omega) = -\sum_{l'} (\mathcal{M}^{-1})_{ll'} \tilde{\Omega}_{l'}\,,
    \end{align}
    where
    \begin{align}
    	\mathcal{M}_{ll'} = -i(-\Delta_{\textrm{nuc},l} + i\frac{\gamma}{2})\delta_{ll'} - i\frac{N}{A_\parallel}\mathcal{G}(z_l, z_{l'}, \textbf{k}_\parallel) \,,
    \end{align}
    and
    \begin{align}
    	\tilde{\Omega}_l = i\frac{N}{A_\parallel} \mathbf{d}_l^* \cdot \textbf{E}_\mathrm{in}(z_{l}, \textbf{k}_\parallel, \omega) \,.
    \end{align}

    \subsection{Reconstructing spectral observables}\label{sec::Green_reconstructScatt}
    The Born-Markov Master equation or the effective nuclear level scheme given above define the dynamics of nuclear observables for a given driving field, which can be solved to obtain, for example, the linear regime solution Eq.~\eqref{eq::Green_linSol}. Scattering observables can be reconstructed using a generalized input-output equation \cite{Asenjo-Garcia2017a}, which is also valid beyond the linear sector. At a given parallel wave vector and in frequency space, it reads
    \begin{align}
    &\hat{\textbf{E}}(z, \textbf{k}_\parallel, \omega) = \hat{\textbf{E}}_\textrm{in}(z, \textbf{k}_\parallel, \omega) \nonumber
    \\
    &\qquad + \mu_0 \sum_{l} \omega^2_{\textrm{nuc},l}\textbf{G}(z, z_l, \textbf{k}_\parallel, \omega)\dl{\cdot \mathbf{d}_l}  \: \hat{\sigma}^-_l(\textbf{k}_\parallel, \omega)\,. \label{eq::Green_io-rel}
    \end{align}
    Substituting the linear frequency space solution for the lowering operator yields
    \begin{align}
    &\hat{\textbf{E}}(z, \textbf{k}_\parallel, \omega) = \hat{\textbf{E}}_\textrm{in}(z, \textbf{k}_\parallel, \omega) \nonumber
    \\
    & \qquad - \mu_0 \sum_{ll'} \omega^2_{\textrm{nuc},l} \textbf{G}(z, z_l, \textbf{k}_\parallel, \omega)\dl{\cdot \mathbf{d}_l} \: (\mathcal{M}^{-1})_{ll'} \tilde{\Omega}_{l'}\,. \label{eq::Green_scattSol}
    \end{align}
    This formula describes the field at position $z$ for a given parallel wave vector $\textbf{k}_\parallel$ and frequency $\omega$, including the nuclear response. It is thus to be interpreted as a steady state solution for a driving field at frequency $\omega$. $\hat{\textbf{E}}_\textrm{in}(z, \textbf{k}_\parallel, \omega)$ is the corresponding field in the absence of the nuclei. It therefore is proportional to the mode profile and a driving amplitude factor \dl{$\alpha^q_\textrm{in}$} for each polarization, resulting in
    \begin{align}\label{eq::Green_drive}
    	\hat{\textbf{E}}_\textrm{in}(z, \textbf{k}_\parallel, \omega) = \dl{\sum_q \alpha^q_\textrm{in} }\:\mathbfcal{E}^{0(n)}_q(z, \mathbf{k}_\parallel, \omega)\,,
    \end{align}
    where our notation is chosen close to \cite{Tomas1995}. The function $\mathbfcal{E}^{0(n)}_q(z, \mathbf{k}_\parallel, \omega)$ is the mode profile that is normalized to the surface of the cavity from where the radiation is incident \cite{Tomas1995}, with $0$ [$n$] indicating top [bottom] illumination. The polarization index $q$ accounts for $s$- and $p$-polarization. For details on the mode functions used to expand the Green's function \cite{Tomas1995} we refer to Appendix~\ref{app::tomas}.

    In grazing incidence experiments at synchrotron facilities \cite{Rohlsberger2005}, a common observable is the reflection spectrum as investigated in Sec.~\ref{sec::aifmt_SimpleEx}. In the Green's function approach, such spectra can be obtained by evaluating the total field at the surface of the cavity. \dl{For example, for $s$-polarized incident light and $s$-polarized detection},
    \begin{align}
    	r_\textrm{green}(\textbf{k}_\parallel, \omega) = \dl{\frac{\mathbf{e}_s \cdot \hat{\textbf{E}}(\tilde{z}_0, \textbf{k}_\parallel, \omega)}{\alpha^s_\textrm{in}}} - 1, \label{Green_r-coeff}
    \end{align}
    \dl{where $\mathbf{e}_s$ is the unit vector in $s$-direction.} Similarly, the transmission can be obtained by
    \begin{align}
	    t_\textrm{green}(\textbf{k}_\parallel, \omega) =\dl{ \frac{\mathbf{e}_s \cdot\hat{\textbf{E}}(\tilde{z}_n, \textbf{k}_\parallel, \omega)}{\alpha^s_\textrm{in}} }\,, \label{Green_t-coeff}
    \end{align}
    where $\tilde{z}_0$ [$\tilde{z}_n$] is the position of the surface boundary of the uppermost [substrate] layer (see Fig.~\ref{fig::app_GF_illu}). For substrates with an absorptive character via a non-zero imaginary part of the refractive index, the wave amplitude decays upon propagation through the medium, such that the transmission coefficient characterizes the amplitude ratio at the last layer surface \cite{Tomas1995}.

      \begin{figure}[t]
    	\includegraphics[width=0.95\columnwidth]{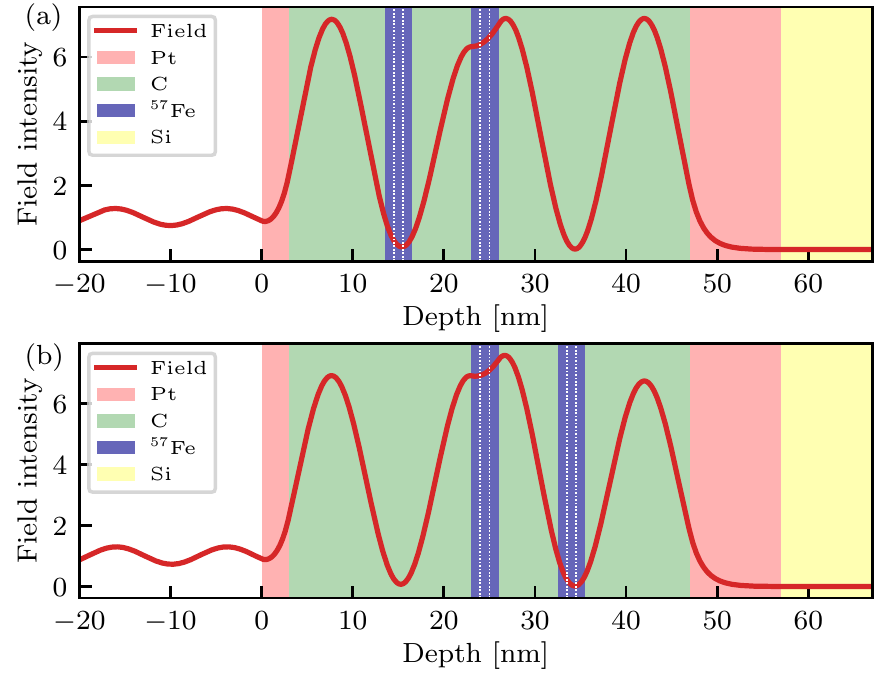}
    	\caption{(Color online) x-ray cavities with nuclei in EIT (a) and non-EIT (b) configuration, which were investigated experimentally in \cite{Rohlsberger2012} and modeled theoretically in \cite{Heeg2015c}. 
    	(a) and (b) show the cavity structure (for materials, see legend) and off-resonant field distribution for the EIT (a, cavity 1) and non-EIT (b, cavity 2) case, respectively. We note that over the width of the blue resonant layers, the field distribution shows visible variations. The white dashed lines indicate the boundaries of sub-ensembles (see main text) over which the field distribution does not vary significantly.
    	}
    	\label{fig::GF_EIT_1Dbenchmark-upper}
    \end{figure}

      \begin{figure}[t]
    	\includegraphics[width=0.95\columnwidth]{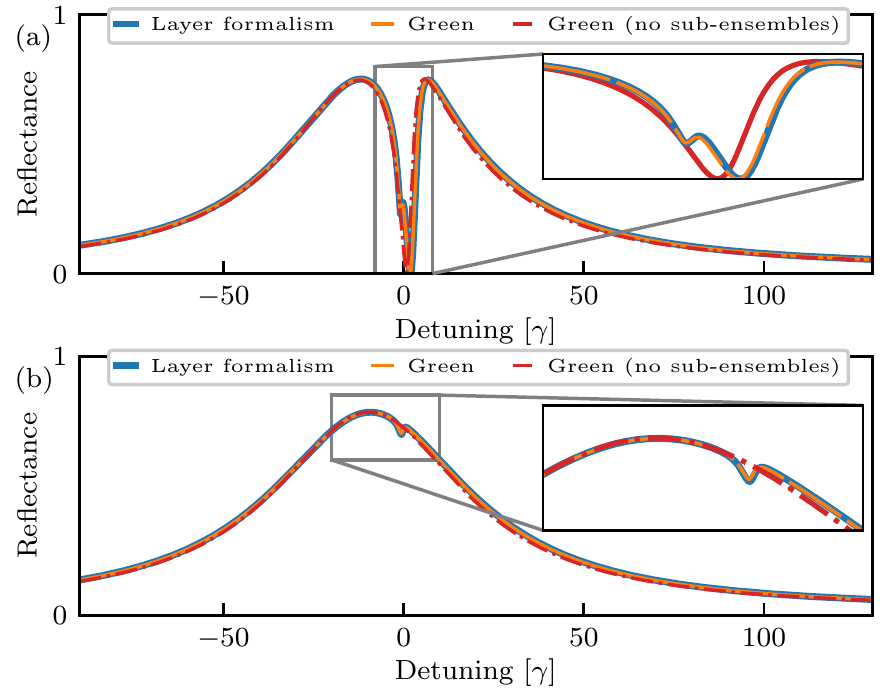}
    	\caption{(Color online)  Nuclear spectra for the EIT (a) and non-EIT (b) cavities shown in Fig.~\ref{fig::GF_EIT_1Dbenchmark-upper}. The spectra are calculated using the Green's function approach for illumination at the third cavity mode, and the well-known layer formalism (blue) is compared to the Green's function approach presented in this paper.
    	Without dividing the two nuclear ensembles into sub-ensembles (red), the spectra fit well and reproduce the main spectral features, but fail to reproduce more subtle additional features (see insets). These are captured if each resonant layer is divided into three sub-ensembles (yellow), and therefore can be attributed to the layer thickness resulting in a field gradient across the layer (see also Fig.~\ref{fig::GF_EIT_2Dbenchmark}).}
    	\label{fig::GF_EIT_1Dbenchmark-lower}
    \end{figure}
    
    \subsection{Numerical efficiency for the layer geometry\label{sec:efficiency}}
    Conveniently, the layered x-ray cavity geometry is one of the few cases \cite{Asenjo-Garcia2017a,Buhmann2012} where the Green's function for the cavity is known analytically \cite{Tomas1995}. In particular, $\textbf{G}(z, z', \textbf{k}_\parallel, \omega)$ can be expressed algebraically via an analytic recursion formula \cite{Tomas1995} similar to Parratt's formalism \cite{Parratt1954}. This feature makes the above approach highly numerically efficient. Compared to the {\it ab initio} few-mode approach to calculating effective nuclear level schemes presented in Sec.~\ref{sec::Green_lowExc_levScheme}, this feature poses a major advantage when one is interested in the calculation of effective quantum optical parameters as depicted in Fig.~\ref{fig::couplings_sketch}, and, for example, opens optimization opportunities.
    
    The formula for the Green's function and its practical evaluation, as presented in \cite{Tomas1995}, are summarized in the following section and in Appendix~\ref{app::tomas}. \dl{In subsequent sections}, we employ a numerical implementation thereof in order to benchmark the approach, and to demonstrate its usefulness. As a main result, we provide an \textit{ab initio} nuclear level scheme for the EIT cavity configuration investigated experimentally in \cite{Rohlsberger2012}, resolving previous discrepancies in the quantum optical description of the system \cite{Heeg2015c}.
    
    \dl{
    \subsection{A practical guide to calculations in the Green's function \textit{ab initio} approach}\label{sec::Green_calc}
    Before we present our results for concrete systems, we show how common observables are calculated in the Green's function formalism in practice. This section serves as a recipe to reproduce the calculations discussed in the following sections and to provide clarity on the approach from an algorithmic perspective, in analogy to Sec.~\ref{sec:aifmt_calc} for the few-mode approach.
    
    The first step is to calculate the parallel Fourier transform of the Green's function $\textbf{G}(z, z_l, \textbf{k}_\parallel, \omega)$ for the cavity structure under study, which is the basic quantity appearing in the equations of the quantum theory. This is achieved by using the refractive indices of the cavity layers to compute the Fresnel coefficients for neighboring layers given by Eqs.~\eqref{eq::app_Fresnel_neighbor}. The reflection and transmission coefficients for multi-layer stacks of the empty cavity, that is disregarding the nuclear resonances, can then be obtained from the recursion formulas Eqs.~\eqref{eq::app_Fresnel_recursion}. Substituting these coefficients into  Eqs.~\eqref{eq::modeProfiles} yields the field distributions, which in turn directly give the Green's function using Eq.~\eqref{eq::app_GF_parallel}. We note that the $\delta$-function term in  Eq.~\eqref{eq::app_GF_parallel} can be disregarded for calculations of the effective level scheme, since it is identical to the free space term that renormalizes the transition frequency \cite{Masson2019_preprint}. Since the latter is used as a parameter corresponding to the experimentally observed value, it is already accounted for.
    
    To obtain a nuclear level scheme, one has to specify the nuclear ensembles used in the calculation, corresponding to the summation index $l$ in the formulas. For thin  resonant layers with no magnetic or other splittings, it is natural to treat each layer as one ensemble. For thicker layers, the spatial variation of the cavity field across the layer requires a splitting of the thick layer into multiple sub-ensembles for a more accurate treatment, as discussed in more detail below.
    
    The coupling constants in the effective low-excitation level scheme can be obtained by evaluating Eq.~\eqref{green-curly-g} using the effective dipole moments of the nuclear ensembles. The coupling constants then follow by substitution of the latter result and the number density of the resonant material into Eqs.~(\ref{eq::Green_couplings_lin}, \ref{eq::Green_decay_lin}).
    The effective level scheme Hamiltonian Eq.~\eqref{eff::Green_H_effLin} and Lindblad term Eq.~\eqref{eff::Green_Lind_effLin} are then fully determined after the driving term is calculated from the field distributions, where one has to manually specify the polarization state of the incoming field according to Eq.~\eqref{eq::Green_drive}.

    Spectral observables can be computed using the linear scattering solution for the nuclear operators Eq.~\eqref{eq::Green_linSol}. These can be substituted  into the input-output relation Eq.~\eqref{eq::Green_io-rel} to obtain the combined field distribution including the nuclear resonance contribution in frequency space Eq.~\eqref{eq::Green_scattSol}. For a given external driving field, this quantity encodes the entire output field and therefore the scattering information. However, in the linear regime, the spectral observables can also be computed directly using Parratts formalism, which does not make use of the ensemble interpretation in the effective level scheme. A comparison of the two approaches allows one to evaluate how accurately the effective level scheme describes the scattering process.
    
    Finally, common observables such as reflection coefficients and transmission coefficients can then be computed for defined polarization directions as exemplified by Eqs.~(\ref{Green_r-coeff}, \ref{Green_t-coeff}).  
    
    This algorithm is used to obtain the linear scattering results discussed in the following.
	}
    
    \subsection{Application: Electromagnetically induced transparency cavities}\label{sec::Green_applications}
    The interesting cavity geometry mentioned above, which has notably been studied experimentally in \cite{Rohlsberger2012}, features a sharp spectral dip in the spectrum. The latter has been shown to correspond to an EIT phenomenon, and a corresponding effective level scheme has been derived~\cite{Rohlsberger2012,Heeg2015c}.
    The system can be described by the phenomenological few-mode theory and the comparison to spectral observables from semi-classical calculations showed that the main EIT feature can be reproduced. However, there are unexplained quantitative and qualitative disagreements \cite{Heeg2015c}. In addition, it is unclear to what extent the heuristic extensions of the model (see Sec.~\ref{sec::recap}) that were found to be necessary \cite{Heeg2015c} influence the interpretation in terms of an effective level scheme.
  
    In this section, we apply the Green's function approach developed above to the two cavities studied in \cite{Rohlsberger2005,Heeg2015c}, resolving these open questions. In particular, we show that the qualitative disagreements in the spectra arise due the relatively thick resonant layers that were used in these cavities, causing the formation of cavity mode field gradients across the layers. The new approach can incorporate such gradients, and thus provides excellent quantitative agreement, improving the previous phenomenological description of the system significantly. In addition, we unambiguously calculate the effective nuclear level schemes using the \textit{ab initio}  method and investigate its quantum optical parameter trends.
    
    \begin{figure}[t]
		\includegraphics[width=1.0\columnwidth]{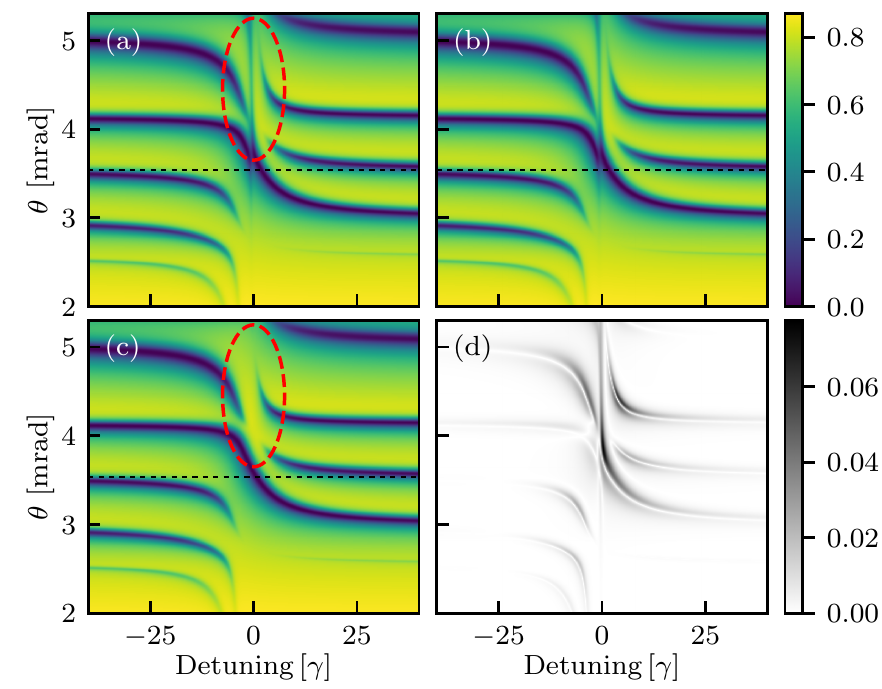}
		\caption{Nuclear spectra of the EIT cavity (cavity 1) as a function of incidence angle. The panels show the layer formalism result (a) and the Green's function result with (b) and without (c) sub-ensembles. (d) shows the deviation $|$(a)$-$(b)$|$. The dashed black line indicates the third mode minimum $\theta_3$. The dashed red ellipse marks a region where (a) and (c) differ significantly, while (a) and (b) agree well, indicating that the layer thickness plays a role in particular in the higher modes due to more rapidly varying field distributions.}
		\label{fig::GF_EIT_2Dbenchmark}
	\end{figure}
	
	\begin{figure}[t]
		\includegraphics[width=1.0\columnwidth]{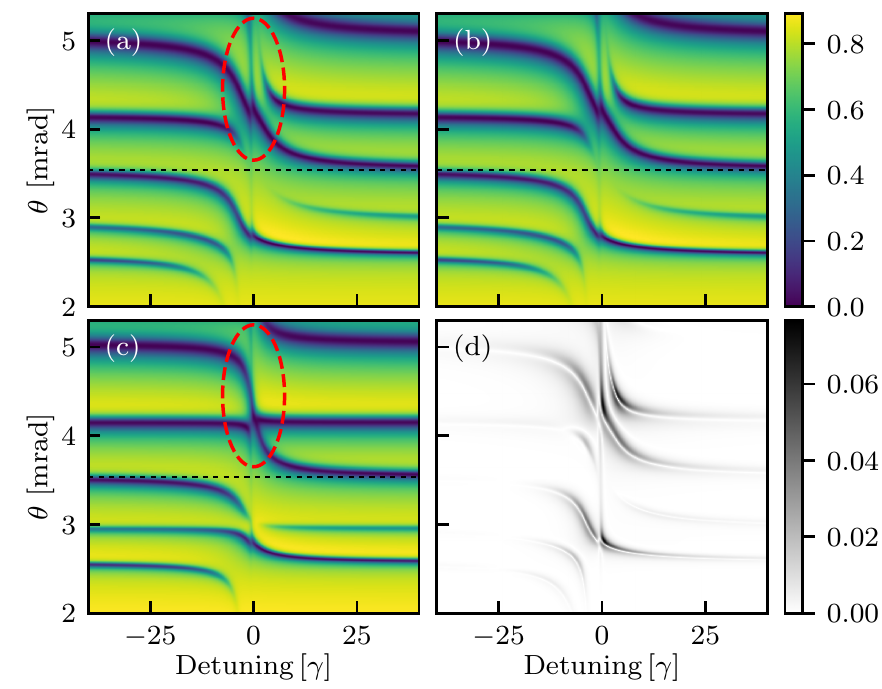}
		\caption{Nuclear spectra of the non-EIT cavity (cavity 2), analogous to Fig.~\ref{fig::GF_EIT_2Dbenchmark} for the EIT cavity. The additional spectral feature inside the red dashed ellipse is accompanied by changes in the surrounding mode structure, with an avoided crossing in (a) turning into a merging point in (c).}
		\label{fig::GF_nonEIT_2Dbenchmark}
	\end{figure}

    \subsubsection{Double resonant layer cavities and spectral benchmarks}
    
    The two cavity layer stacks under consideration are depicted in Fig.~\ref{fig::GF_EIT_1Dbenchmark-upper} and are identical to the geometries investigated in \cite{Heeg2015c} to understand the EIT effect observed in \cite{Rohlsberger2012}. They consist of platinum cladding layers enclosing a carbon guiding layer that is doped with $3~$nm thick resonant $^{57}$Fe layers at certain positions. In both cases (cavity 1 and cavity 2), one of the resonant layers is located at the cavity center, where the field distribution in the third cavity mode at incidence angle $\theta=\theta_3$ features an anti-node. As shown in Fig.~\ref{fig::GF_EIT_1Dbenchmark-upper}, in the EIT cavity (cavity 1, panel a) the second resonant layer is placed in the adjacent field distribution node closer to the cavity surface, while in the non-EIT cavity (cavity 2, panel b), it is placed in the node closer to the bottom of the cavity. 
    
    We refer to the two structures as ``EIT'' (``non-EIT'') cavities, because they feature (do not feature) a pronounced spectral dip in the respective nuclear spectra shown in~Fig.~\ref{fig::GF_EIT_1Dbenchmark-lower}~\cite{Rohlsberger2012}. The two panels compare the spectra obtained using the Green's function approach \dl{(computed using Eq.~\eqref{Green_r-coeff}, see Sec.~\ref{sec::Green_calc} for a description of the algorithm)} to reference spectra calculated using the semi-classical layer formalism \cite{Rohlsberger2005,Sturhahn2000} implemented in the software package \textsc{pynuss} \cite{pynuss}. For the Green's function method, two different curves are shown in each panel. The red line corresponds to a model in which each of the two layers is treated as a single nuclear ensemble, neglecting possible field gradients across the layer. While the qualitative agreement is good and the major spectral features are reproduced, there are quantitative differences, most visible in additional small spectral features shown in the insets. 
    To capture these features, we divide each layer into three sub-layers in our theoretical model (see white dashed lines in Fig.~\ref{fig::GF_EIT_1Dbenchmark-upper}), to better account for the variation of the field intensity across the layers. The result is shown as the solid yellow line, which essentially agrees perfectly with the semi-classical calculation. The improvement can be understood since in Fig.~\ref{fig::GF_EIT_1Dbenchmark-upper} it is clear that the field distribution varies visibly over the thickness of each resonant layer, while it is constant to a good approximation over the thickness of each of the sub-ensembles.
    
    We further note that the agreement between the spectra is much better than in the case of the phenomenological few-mode fits shown in \cite{Heeg2015c}, even when the sub-ensemble partition is not considered (red line in Fig.~\ref{fig::GF_EIT_1Dbenchmark-lower}). This general quantitative improvement is due to the absence of heuristic extensions and problems related to the fitting procedure  \cite{Heeg2015c}, which are not needed in the \textit{ab initio}  theories reported here.
    
    \subsubsection{The EIT effect and thick layer sub-ensembles}
    
	Next, we extend the discussion to the two-dimensional spectra as a function of detuning and incidence angle, which were also investigated for the two cavities in \cite{Heeg2015c} in the context of the phenomenological few-mode model. In this reference, it was found that in addition to the quantitative differences in the one-dimensional spectra (see Fig.~\ref{fig::GF_EIT_1Dbenchmark-lower} and the discussion above), there are qualitative features that are not captured by the phenomenologically fitted model even with the heuristic extensions included.
	
	Results are shown in Figs.~\ref{fig::GF_EIT_2Dbenchmark} and \ref{fig::GF_nonEIT_2Dbenchmark} for the EIT- and the non-EIT cavity, respectively. In each figure, panel (a) corresponds to the layer formalism calculation which again serves as a benchmark. Panels (b) and (c) show the Green's function results with and without sub-ensemble partition of the resonant layers, respectively \dl{(spectra are computed using Eq.~\eqref{Green_r-coeff}, see Sec.~\ref{sec::Green_calc} for a description of the algorithm)}. Finally, panel (d) shows the absolute value of the difference between the results in panels (a) and (b).
		
	We see that while  (d) demonstrates excellent agreement of the  Green's function description with sub-ensembles to the layer formalism, the corresponding results without sub-ensembles (panel c) are missing a spectral feature as indicated by the red dashed ellipse. Its absence shows that approximating thicker resonant layers as a single thin layer can lead to qualitative differences in the theoretical description. 
	
	For the EIT case in Fig.~\ref{fig::GF_EIT_2Dbenchmark}, we further find that the agreement of the model without sub-ensembles is still rather good close to the third mode minimum $\theta=\theta_3$, justifying the thin-layer approximation  at this incidence angle. However, the differences become sizable in the region of the red dashed ellipse, which can be understood by noting that the field distributions vary more rapidly as function of position in the cavity at higher incidence angles. Fig.~\ref{fig::GF_nonEIT_2Dbenchmark} shows analogous results for the non-EIT cavity (cavity 2). In this case, the absence of the spectral feature in the two-ensemble model also causes the surrounding spectral structure to change, with an avoided mode crossing turning into a merging point (see region inside the red dashed ellipse).
	
	\subsubsection{Effective nuclear level schemes}
	\begin{figure}[t]
		\includegraphics[width=0.8\columnwidth]{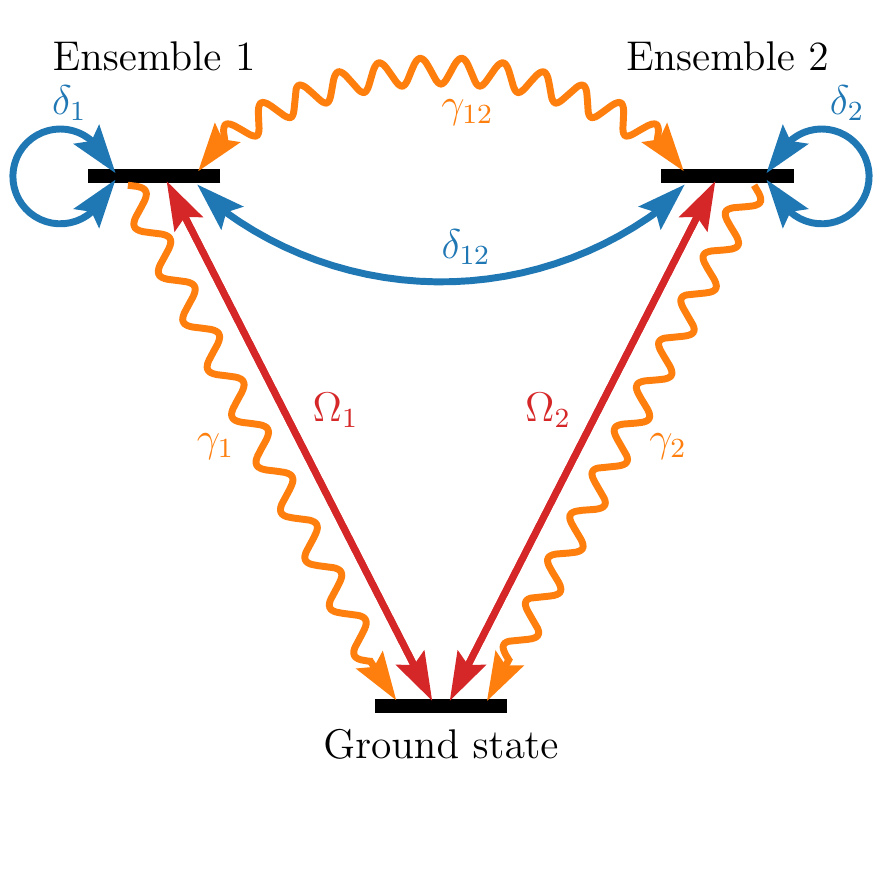}
		\begin{center}
			\begin{tabular}{ c | c | c }
				\hline
				System & Coupling matrix [$\gamma$] & Drive vector \\
				\hline
				&&
				\\
				General & $\begin{pmatrix} \color{pltC0}\delta_1\color{black} - i\color{pltC1}\gamma_1\color{black}/2 & \color{pltC0}\delta_{12}\color{black}-i\color{pltC1}\gamma_{12}\color{black}/2 \\ \color{pltC0}\delta_{12}\color{black}-i\color{pltC1}\gamma_{12}\color{black}/2 & \color{pltC0}\delta_2\color{black}-i\color{pltC1}\gamma_2\color{black}/2  \end{pmatrix} $  & $\begin{pmatrix} \color{pltC3}{\tilde{\Omega}}_1\color{black} \\ \color{pltC3}{\tilde{\Omega}}_2\color{black}  \end{pmatrix}$
				\\
				&&
				\\
				Cavity 1 & $\begin{pmatrix} -0.15-0.94i & 6.21+2.85i \\ 6.21+2.85i & -6.47-28.28i  \end{pmatrix}$  & $\begin{pmatrix} 0.01+0.14i\\ -0.85-0.51i  \end{pmatrix}$
				\\
				&&
				\\
				Cavity 2 & $\begin{pmatrix} -9.46-31.73i & 0.98-2.88i \\ 0.98-2.88i & -0.45-0.64i  \end{pmatrix} $ & $\begin{pmatrix} -0.82-0.57i \\ -0.03-0.09i  \end{pmatrix}$
			\end{tabular}
		\end{center}
		\caption{(Color online) Nuclear level scheme in the Green's-function approach. The figure shows the two-ensemble description of the cavities in Fig.~\ref{fig::GF_EIT_levelSchemes} as an example. The coupling constants can be calculated directly using the Green's function approach and are shown in the table in matrix form for the EIT (cavity 1) and non-EIT (cavity 2) case. As before, we consider excitation in the third cavity mode ($\theta=\theta_3$). The drive vector has been normalized, since it is proportional to the applied x-ray field amplitude. The corresponding quantities as a function of incidence angle are shown in Fig.~\ref{fig::GF_EIT_levelSchemes_angleTrends}. We note that the layer ensembles are labeled by their position in the cavity from top to bottom, such that the central layer in the anti-node corresponds to index 1 in cavity 1 and to index 2 in cavity 2.}
		\label{fig::GF_EIT_levelSchemes}
	\end{figure}
	\begin{figure}[t]
		\includegraphics[width=1.0\columnwidth]{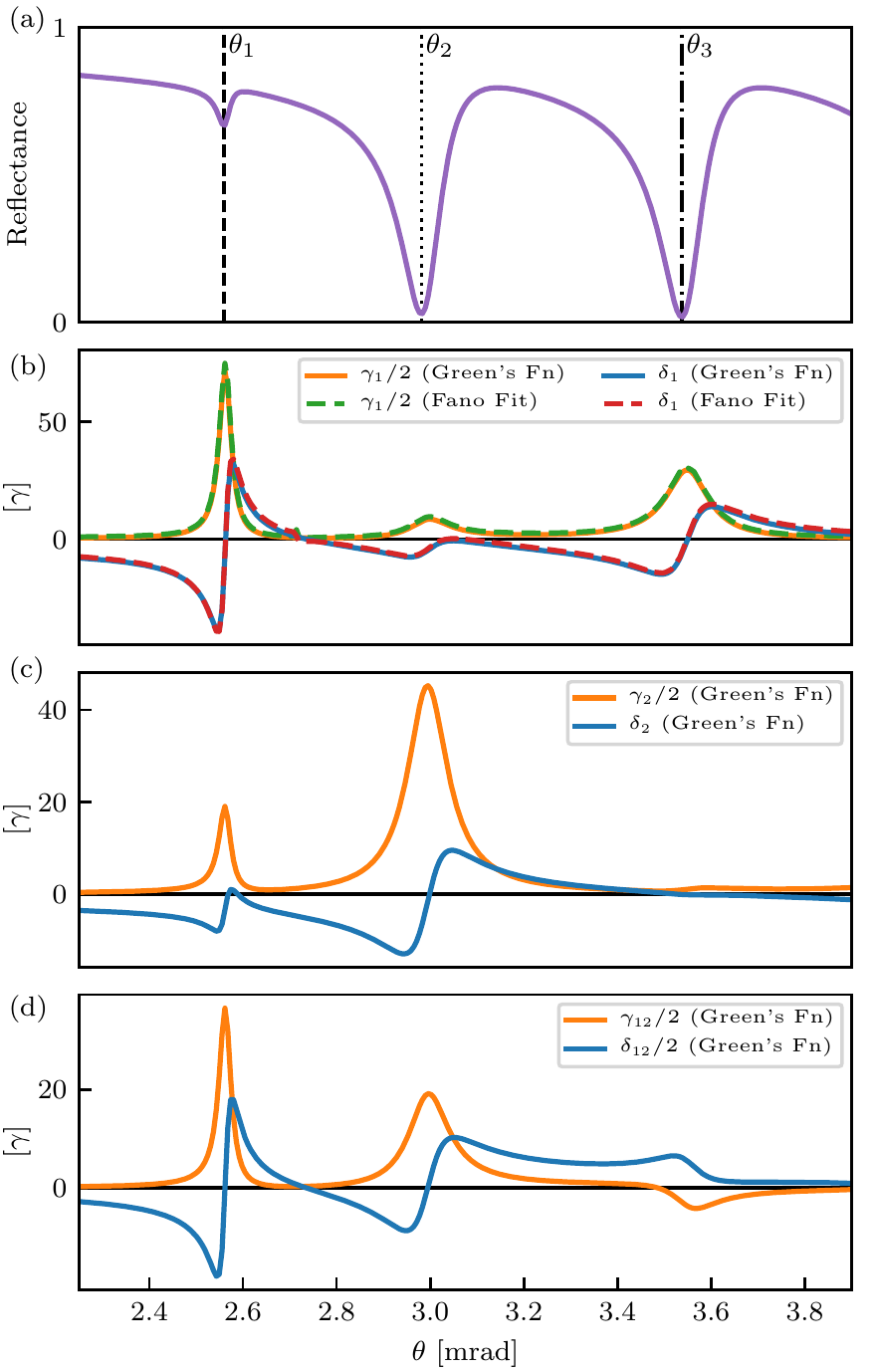}
		\caption{Rocking curve (a), collective Lamb shifts and superradiant decay rates of ensemble 1 (b) and ensemble 2 (c), and level couplings (d) in the EIT cavity as a function of incidence angle (solid lines, see legend). The meaning of the quantum optical couplings is illustrated in the effective level scheme Fig.~\ref{fig::GF_EIT_levelSchemes} and the shown result is calculated using the Green's function approach. As an additional benchmark (see also spectral benchmarks in previous figures), panel (a) shows the superradiance (dashed green) and collective Lamb shift (dashed red) extracted from a Fano fit (details see main text).}
		\label{fig::GF_EIT_levelSchemes_angleTrends}
	\end{figure}
	
	Finally, we discuss the effective level scheme resulting from the Green's function approach to x-ray cavity QED. Fig.~\ref{fig::GF_EIT_levelSchemes} shows the effective nuclear level scheme and the quantum optical coupling constants for the EIT- and non-EIT cavities in Fig.~\ref{fig::GF_EIT_1Dbenchmark-upper} \dl{(the couplings are obtained from Eqs.~(\ref{eq::Green_couplings_lin}, \ref{eq::Green_decay_lin}), see Sec.~\ref{sec::Green_calc} for a description of the algorithm used to obtain the level scheme)}. For simplicity, the case without sub-ensemble partitioning is shown, with the notation adopted from \cite{Heeg2015c}. 
	Within the Green's function approach, the couplings and other quantum optical parameters  can be calculated directly and unambiguously from the cavity geometry. The resulting parameters at the incidence angle $\theta=\theta_3$ are tabulated in Fig.~\ref{fig::GF_EIT_levelSchemes}.
	
	In Fig.~\ref{fig::GF_EIT_levelSchemes_angleTrends}, we show the quantum optical parameters as a function of incidence angle, revealing the changes of the three level system over the different modes. In particular, the collective Lamb shifts, the superradiant decay rate enhancements and the level couplings between the two ensembles in the EIT cavity (cavity 1)  are depicted. Around each mode minimum ($\theta_1, \theta_2, \theta_3$), the collective Lamb shift and superradiance roughly show the typical behavior of real and imaginary parts of a complex Lorentzian \cite{Longo2016,Heeg2015a}. For the coupling parameters $\delta_{12}$ and $\gamma_{12}$, similar structures are found. Only in the third mode, where the EIT phenomenon is observed, a different functional dependence and a negative $\gamma_{12}$ is found.
	
	These results show how complex cavity structures can be unambiguously interpreted in terms of quantum optical models, which paves the way for designing effective nuclear level schemes via tailored mode environments.
	
	As a last consistency check, we show that the collective Lamb shift and superradiance calculated here indeed correspond to what these quantities have been associated with in the nuclear cavity QED literature so far~\cite{Heeg2015a,Rohlsberger2010}. To this end, we replace the second resonant $^{57}$Fe-layer by its off-resonant $^{56}$Fe counterpart and fit a generic Fano model \cite{Heeg2015a,Ott2013,Barnthaler2010,Zhou2008} to the resulting line shapes at each incidence angle. The Fano profile fit function for the reflectance spectrum is given by \cite{Heeg2015a}
	\begin{align}
		R(\Delta) = |r(\Delta)|^2 = \sigma_0 \frac{|q + \epsilon(\Delta)|^2}{1 + \epsilon^2(\Delta)} \,,
	\end{align}
	where $\epsilon(\Delta)=(\Delta-\delta_1)/(\gamma_1/2)$. 
	The scale factor $\sigma_0$, the complex Fano parameter $q$ and $\delta_1, \gamma_1$ are the fit parameters. This fit provides an alternative way to obtain the collective Lamb shift $\delta_1$ and the superradiance $\gamma_1$ from the spectra, which can be compared to the {\it ab initio} predictions. Indeed, the fit results (dashed lines in Fig.~\ref{fig::GF_EIT_levelSchemes_angleTrends}b) show excellent agreement with the Green's function calculation, confirming the interpretation of the effective nuclear level scheme derived from the Born-Markov Master equation in Sec.~\ref{sec::Green_lowExc_levScheme}.

    \section{Discussion and Summary}\label{sec::conclusion}
    
    In summary, we have presented two \textit{ab initio}  approaches to describe thin-film x-ray cavities doped with narrow resonances such as those provided by M\"ossbauer nuclei on a quantum mechanical level. Our results improve the previously introduced phenomenological input-output model \cite{Heeg2013b,Heeg2015c} for such systems and resolve multiple open questions in the theory, which up to now have hindered extensions towards new parameter regimes and cast doubts on the models' predictive capabilities at higher intensities. This progress to an {\it ab initio} and, in the linear regime, essentially exact theory provides qualitatively new value to quantum optical interpretations used in the field, in the same spirit as recent developments connecting other sectors of quantum optics and \textit{ab initio} theory \cite{Ruggenthaler2014,Schaefer2019,Cerjan2016,Tureci2006}.
    
    The two \textit{ab initio} methods presented here are directly applicable to model linear scattering experiments in grazing incidence, which have been performed extensively in the platform of hard x-ray cavity QED with nuclei \cite{Rohlsberger2010,Rohlsberger2012,Heeg2013a,Heeg2015a,Heeg2015b,Haber2016a,Haber2017} and electronic resonances \cite{Haber2019}, and to interpret them quantum optically in terms of an \textit{ab initio}  effective nuclear level scheme. The {\it ab initio} perspective and the exact treatment of the linear sector put the quantum optical interpretation on firm grounds, establishing it as a tool in the quantum theory of nuclear resonance scattering. Beyond that, we provide clear theoretical connections between the existing approaches and outline the approximations involved in each case, paving the way for generalizations. As a consequence, one can now seamlessly switch between the different theories, depending on which formulation is most favorable for the respective purpose or regime. In this context, we emphasize that a main advantage of our method over the existing theories is their applicability in the fully quantum mechanical sector. That is, the approaches bear the potential to describe non-linear and correlated quantum dynamical effects in hard x-ray cavity QED, going beyond linear, low-excitation, semi-classical or mean-field phenomena.
    
    Our first approach is based on a few-mode description of the system, and promotes the established phenomenological  models~\cite{Heeg2013b,Heeg2015c} to an {\it ab initio} theory without the need for heuristic extensions, thereby  resolving previously not understood discrepancies in the modeling. The {\it ab intio} theory features modified angular and frequency dependencies of the quantum optical parameters, which can now be calculated directly from the cavity geometry. We further presented a closed form solution to the resulting equations of motion for general multi-mode multi-ensemble systems in the low excitation regime, including the derivation of scattering observables and the effective nuclear level scheme.
    
    For an analytically solvable example cavity that features strongly overlapping modes, we demonstrate the advantages of our theory over the previous phenomenological models, and show that it provides quantitative agreement with semi-classical approaches. We investigate the functional dependencies of the quantum optical parameters, showing where the improvements provided by our theory originate from and finding non-trivial behavior as a function of incidence angle. As an application, we for the first time extend the quantum optical modeling to the realistic case of resonant layers with  arbitrary thickness. We find that the resulting effects can straightforwardly be captured in the \textit{ab initio} few-mode theory, a task which is difficult in phenomenological approaches due to the large set of fitting parameters required. Our results show that thick layers lead to coupled continuum ensemble descriptions, a feature which may lead to the x-ray implementation of qualitatively different quantum optical model systems than the few-level systems realized so far \cite{Heeg2013a,Rohlsberger2012,Haber2017,Haber2016a,Heeg2015b,Rohlsberger2010,Heeg2015a,Haber2019}.
    
    As our second approach, we develop an alternative method based on well-known Green's function techniques \cite{Knoll1991,Gruner1996,Dung2002,Scheel2008,Asenjo-Garcia2017a,Buhmann2012}. The main motivation for this second approach is its numerical efficiency for calculating effective nuclear level schemes. We again demonstrate the theory's connection to the existing semi-classical nuclear resonant scattering literature \cite{Rohlsberger1999,Rohlsberger2005}, deriving an effective wave equation including the nuclear resonance dynamics in the linear regime. We then show how a quantum optical description analogous to the previously investigated effective nuclear level schemes can be obtained and demonstrate its numerical efficiency, which is ensured by an analytic solution for the Green's function of the relevant geometry \cite{Tomas1995}.
    
    To showcase the power of this approach, we apply the Green's function method to the case of the EIT cavities investigated in \cite{Heeg2015c,Rohlsberger2012}, where qualitative and quantitative discrepancies of linear spectra between the phenomenological few-mode model and semi-classical theory were found \cite{Heeg2015c}. Our approach is able to calculate the quantum optical description of the system in terms of an effective nuclear level scheme without the need for heuristic model extensions or a fitting procedure, which were necessary before \cite{Heeg2015c}. It further reveals the importance of sub-ensembles of the relatively thick resonant layers in the system which are responsible for the aforementioned qualitative deviations. We show that quantitatively, the approach yields essentially perfect agreement if the resonant layers are divided into a sufficient number of sub-ensembles. We further present  the first \textit{ab initio} calculation of the resulting effective nuclear level schemes and investigate trends of the quantum optical parameters. These results pave the way for tailoring mode environments in x-ray cavities to design nuclear quantum systems. 
    
    Beyond the layer geometry, the Green's function formalism is easily adaptable to alternative photonic environments for the nuclei. Such structures may become accessible with improving fabrication techniques and already reported examples include other dimensional waveguides \cite{Pfeiffer2002}, curved channels \cite{Salditt2015}, nanowires \cite{Chen2002} and periodically structured surfaces such as nanodots or nanodiscs \cite{Ellrich2012}. While unlike in the layer geometry and some others \cite{Buhmann2007,Asenjo-Garcia2017a,Scheel2008}, the Green's function is usually not known analytically in these cases, various numerical and modeling approaches exists to tackle such geometries \cite{Asenjo-Garcia2017a,Buhmann2012,NovotnyHecht2006}.
    
    In this context, we also note that in comparison to the {\it ab initio} few-mode approach, in the presented Green's function formulation the mode structure of the cavity is hidden and cannot be included in the quantum optical modeling. It remains to be seen if the Green's function can also be modeled in terms of the modes or resonances \cite{Asenjo-Garcia2017a} to address the relevant case of overlapping modes characteristic of x-ray cavities, e.g., using momentum space representations of the Green's function.
    
    As a whole, our results provide a comprehensive solution for the linear scattering regime of thin-film x-ray cavity QED with  M\"ossbauer nuclei or ultra-narrow resonances, resolving previous discrepancies and revealing the connection between different existing theoretical approaches. As an outlook, we expect this progress in the understanding of the low-excitation sector to provide a solid theoretical foundation for describing phenomena in the non-linear and correlated quantum dynamics regime of this platform, which may be accessible at current and upcoming x-ray facilities \cite{Heeg2016arxiv,Adams2019} such as x-ray free electron lasers, where the first experiment on nuclear quantum optics has recently been performed \cite{Chumakov2018}.
    
    \section{Acknowledgements}
    We gratefully acknowledge fruitful discussions with R.~Bennett, L.~Bocklage, S.~Bragin, S.~Y.~Buhmann, H.~S.~Dhar, J.~Haber, O.~Leupold, M.~Mycroft, R.~R\"ohlsberger, S.~Rotter, M. Stobi\'nska, T.~J.~Sturges, S.~Velten and M. Zens. We thank S.~Y.~Buhmann for reading of the manuscript and pointing out pertinent references. This work is part of and supported by the Deutsche Forschungsgemeinschaft Collaborative Research Centre ``SFB 1225 (ISOQUANT)''.

    \appendix
    \section{Detailed derivation of the \textit{ab initio} few-mode theory}\label{sec::aifmt_DetailedDerivation}
    In Sec.~\ref{sec::aifmt}, we explained how the phenomenological few-mode model for thin-film x-ray cavities with M\"ossbauer nuclei \cite{Heeg2013b,Heeg2015c} can be modified to comply with \textit{ab initio}  theory. We focused on differences to the established model \cite{Heeg2013b,Heeg2015c}, which has been used extensively for interpreting experiments \cite{Heeg2013a,Heeg2015a,Heeg2015b,Heeg2015c,Haber2017,Haber2019}, and on what advantages the {\it ab initio} version can provide practically, which we illustrated extensively for example systems.
    
    In this appendix, we provide a detailed derivation of the improved {\it ab initio} model. We outline which approximations are necessary to obtain such a description and provide a clear physical interpretation of the model parameters and operator degrees of freedom.
    
    \subsection{Classical wave equation for the layered geometry}\label{sec::classical_wave}
    The thin-film x-ray cavities under consideration are made up of a dieletric cavity structure (see Fig.~\ref{fig::sketch} for an illustration), which acts as an off-resonant background confining the electromagnetic field. Before quantizing the system in order to describe the field-nucleus interaction, we consider the classical wave propagation in this geometry in the absence of the resonant nuclei, that is for the ``empty cavity''. The system can then be described by Maxwell's equations with a spatially varying dielectric permittivity \cite{Rohlsberger2005,Rotter2017}.
    
    Again omitting polarization degrees of freedom, the scalar Maxwell mode equation for a homogeneous isotropic medium \cite{Scheel2008,Buhmann2012} in frequency space reads \cite{Rotter2017,Glauber1991}
    \begin{equation}
    	\nabla^2 f_m(\textbf{r}, \omega) + \varepsilon(\textbf{r}) \omega^2 f_m(\textbf{r}, \omega) = 0 \,, \label{eq::app_wave}
    \end{equation}
    where $f_m(\mathbf{r}, \omega)$ are the normal modes for the scattering problem. The index $m$ encodes all additional necessary degrees of freedom. We note that we neglect the frequency dependence of the dielectric permittivity $\varepsilon(\mathbf{r})$ of the off-resonant cavity material here, which is typically irrelevant for the nuclear dynamics due to the narrow resonances, but can be important if electronic resonances \cite{Haber2019} or collective strong coupling \cite{Haber2016a} are considered.
    
    In the case of thin-film x-ray cavities, the system is approximately translation invariant in two directions, such that the dielectric index only depends on the transverse coordinate, $\varepsilon(\textbf{r})=\varepsilon(z)$. As can be seen by a product ansatz, the normal modes then take the form
    \begin{equation}
    f_m(\textbf{r}, \omega, \textbf{k}_\parallel) = C\, e^{i\textbf{k}_\parallel \cdot \textbf{r}_\parallel} \tilde{f}_m(z, k_\perp, \theta) \,,
    \end{equation}
    where $\textbf{r}_\parallel$ is a displacement vector in the layer plane, $\textbf{k}_\parallel$ is the corresponding parallel wave vector, $C$ is a normalization factor for the parallel mode component, and $\tilde{f}_m(z, k_\perp, \theta)$ is an effective one-dimensional mode function at incidence angle $\theta$. The latter fulfills the effective one-dimensional mode equation
    \begin{align}
    \frac{d^2}{dz^2} \tilde{f}_m(z, k_\perp, \theta) + \tilde{\varepsilon}(z, \theta) k^2_\perp \tilde{f}_m(z, k_\perp, \theta) = 0 \,,
    \end{align}
    where
    \begin{align}
    \tilde{\varepsilon}(z, \theta) = \frac{\varepsilon(z) - \cos^2(\theta)}{\sin^2(\theta)} \,.
    \end{align}
    The perpendicular wave component can be obtained from the mode energy via $
    k_\perp = \omega \sin(\theta)$ and is to be understood as a scattering boundary condition, that is the incoming perpendicular wave vector in the far field (see Fig.~\ref{fig::sketch} for an illustration).
    
    In summary, the classical three dimensional problem for the layer geometry can be reduced to an effective one-dimensional problem with an effective dielectric constant
    \begin{align}
    \varepsilon(r) \rightarrow \tilde{\varepsilon}(r, \theta)= \frac{\varepsilon(r) - \cos^2(\theta)}{\sin^2(\theta)} \,,
    \end{align}
    and wave energy
    \begin{align}
    \omega \rightarrow k_\perp(\omega, \theta) = \omega \sin(\theta) \,.
    \end{align}
    These substitutions are employed in the analytical calculation in Sec.~\ref{sec::aifmt_SimpleEx}.
    
    \dl{\subsection{Recap of the \textit{ab initio} few-mode construction}\label{app::recab_aifmt}
  	Before we turn to applying the effective few-mode theory to the three-dimensional wave equation above, we first briefly summarize its concept as presented in \cite{Lentrodt2020}. For brevity, we focus on the basic principles that are necessary to understand the more detailed derivation for the layered geometry in the following section.
  	
  	The starting point of the few-mode scheme is a continuum theory, that is an open or scattering problem with a continuum Hamiltonian, which can, for example, be obtained from canonically quantizing a wave equation such as Eq.~\eqref{eq::app_wave}. In quantum optical scenarios, the wave equation can, for example, describe the electromagnetic field inside and around a cavity. The considered Hamiltonian is typically of the form
  	\begin{align}
	  	H_\textrm{field} &= \sum_{m} \int d\omega  \omega \hat{b}_m^{\dagger}(\omega)\hat{b}_m^{\mathstrut}(\omega)\,.
  	\end{align}
  	In three dimensions, the index $m$ can also include continuous parameters such as the propagation direction of a wave.
  	
  	The central feature of this Hamiltonian is that it comprises a continuum of frequencies, as is typical for an open quantum system. When coupling the field to a two-level system within the dipole approximation, one obtains a field-matter Hamiltonian that is well-known in the theory of light-matter interactions in free space \cite{Scully1997} or cavities \cite{Glauber1991,Krimer2014,Malekakhlagh2017}.
  	
  	However, the continuum Hamiltonian conceals the resonance structure of the cavity, which is encoded in the frequency-dependence of the light-matter coupling \cite{Krimer2014,Malekakhlagh2017,Breuer2002_BOOK}. For this reason, phenomenological few-mode Hamiltonians and input-output theory \cite{Gardiner1985,Gardiner2004} are common tools to describe light-matter interaction in structured environments such as cavities. The few-mode model for thin-film X-ray cavity QED with M\"ossbauer nuclei by Heeg \& Evers \cite{Heeg2013b,Heeg2015c} is based on the latter approach.
  	
  	The \textit{ab initio} few-mode theory \cite{Lentrodt2020} provides a connection between the two sides and allows to systematically construct few-mode Hamiltonians from the continuum description. Most importantly, the \textit{ab initio} construction allows to generalize the approach such that overlapping modes and bad cavities can be described accurately. It is the latter feature which is particularly valuable in the X-ray case, where the cavities in use are highly leaky, feature overlapping modes and are doped with nuclear resonances boasting high spectral resolution.
  	
  	In the \textit{ab initio} few-mode theory, the first step is a basis transformation, splitting the continuum modes into a discrete part (the ``system'') and an interacting continuum (the ``bath''). The few modes that are included in the system part can be chosen arbitrarily and the resulting theory for the free field remains exact. In particular, an exact version of the input-output formalism can be obtained. For scattering observables, the latter allows to compute an input-output scattering matrix $S_\textrm{io}(\omega)$ analogously to the phenomenological case, which describes the scattering between bath modes. The full scattering matrix is then obtained by translating the bath modes into the asymptotically free modes \cite{Glauber1991,Domcke1983} via
  	\begin{align}
  		S(\omega) = S_\textrm{bg}(\omega)S_\textrm{io}(\omega) \,,
  	\end{align}
	 with the so-called background scattering matrix $S_\textrm{bg}(\omega)$.
  	
  	While in the free theory, the few-mode construction is exact independently of the choice of modes, for the interacting case it is useful to choose the few-mode basis such that the relevant resonances of the system are well approximated \cite{Lentrodt2020}. The advantage is that if a sufficient number of modes are included in the few-mode basis, the so called few-mode approximation can be applied: the direct interaction between the external bath modes and the atom or matter degrees of freedom can be neglected, such that the latter only couple directly to the few system modes. This feature, in turn, allows for various existing solution methods to be applied \cite{Gardiner2004,Breuer2016,DeVega2017,Carmichael1993}. For example, a Markovian Master equation can often be derived for a strongly coupled light-matter system if the strongly coupled modes are included in the system and only the remaining weakly coupled bath modes are traced out, which is reminiscent of an input-output version of the pseudo-modes theory \cite{Garraway1997b,Tamascelli2018}.
  	
  	We note that the few-mode construction results in a converging expansion scheme in the mode number \cite{Lentrodt2020}, such that the validity of the few-mode approximation can be ensured by including more modes.
  	
  	The form of the resulting \textit{ab initio} Hamiltonian \cite{Viviescas2003,Lentrodt2020} is very close to the one used in the phenomenological model for thin-film X-ray cavity QED \cite{Heeg2013b,Heeg2015c}, such that the \textit{ab initio} few-mode theory can provide new insights into the system by directly connecting to and extending an existing and well established model. In \cite{Lentrodt2020}, only one-dimensional special cases are considered. The thin-film geometry and the extreme parameter regimes encountered in the hard X-ray and nuclear resonance case require additional consideration. The precise derivation of the relevant few-mode Hamiltonian and an outline of its limitations are given in the following section.
  	
  	As a last remark on the basics of the \textit{ab initio} few-mode theory, we note that from a computational perspective, the main step to obtain the few-mode Hamiltonian and its coupling constants is to calculate matrix elements between the few-mode, bath and scattering states \cite{Lentrodt2020,Domcke1983}. While the resulting formulas in \cite{Lentrodt2020} are general, the explicit computation for the one-dimensional example systems relies on separable expansions of the cavity structure \cite{Domcke1983}. In this paper, the same approach is applied to simple cases of the three-dimensional layered geometry. More general cavity structures may become accessible using related methods from quantum chemistry~\cite{Berman1984,Domcke1983}}.

    \subsection{Application to the layered geometry}\label{sec::cavQuant}
    The wave equation Eq.~\eqref{eq::app_wave} can be quantized canonically \cite{Glauber1991}, resulting in the Hamiltonian
    \begin{align}\label{eq::norm_modes_H}
    H_\textrm{cav} =& \sum_{m} \int d^2\mathbf{k}_\parallel dk_\perp
    \omega(\mathbf{k}_\parallel, k_\perp) \hat{c}_m^\dagger(\mathbf{k}_\parallel, k_\perp) \hat{c}_m(\mathbf{k}_\parallel, k_\perp),
    \end{align}
    where $\hat{c}_m(\mathbf{k}_\parallel, k_\perp)$ are bosonic operators at a given parallel and perpendicular wave vector, with $\omega(\mathbf{k}_\parallel, k_\perp)$ representing their corresponding frequency. The operator equations of motion for this Hamiltonian are equivalent to the Maxwell wave equation \cite{Glauber1991}. The advantage of the quantized formalism is that we can treat interactions with resonant nuclei (see Sec.~\ref{sec::aifmt_nuc}) beyond mean-field Maxwell-Bloch \cite{Shvydko1999,Liao2012} or semi-classical scattering treatments \cite{Rohlsberger1999,Rohlsberger2005}. We note that the canonical normal modes quantization is valid for real refractive indices. To account for complex refractive indices, which are relevant in the hard x-ray regime, we employ a non-hermitian Hamiltonian prescription, which allows the above Hamiltonian to be used directly (see Appendix \ref{sec::Xray-AIFMT_complexRefractive} for details). For a complete treatment of absorptive processes within the framework of macroscopic QED see Sec.~\ref{sec::Green}, where an alternative approach is developed.
    
    The Hamiltonian Eq.~\eqref{eq::norm_modes_H} features the typical normal mode continuum of open scattering systems \cite{Glauber1991,Lentrodt2020}. In order to connect to the successful phenomenological model \cite{Heeg2013b,Heeg2015c} for thin-film x-ray cavities with M\"ossbauer nuclei, the notion of a few resonant modes has to be introduced. To perform this step systematically without a model or fitting prescription, we employ the \textit{ab initio} few-mode theory \cite{Lentrodt2020}, \dl{whose concept is summarized in the previous section.}
    
    Following this approach, we can partition the cavity Hamiltonian given above into a few-mode part and an external bath. If we choose the few mode basis to respect the translation symmetry, the full Hamiltonian can be written as
    \begin{align}
    H_\textrm{cav} = H_\textrm{few} +  H_\textrm{ext}\,,
    \end{align}
    where
    \begin{align}
    H_\textrm{few} = \int d^2\mathbf{k}_\parallel \sum_{\lambda \in \textrm{modes}} \omega^{\mathstrut}_\lambda(\mathbf{k}_\parallel) \hat{a}_\lambda^\dagger(\mathbf{k}_\parallel) \hat{a}_\lambda^{\mathstrut}(\mathbf{k}_\parallel)\,,
    \end{align}
    \begin{align}
    H_\textrm{ext} &= \sum_{m \in \textrm{channels}} \int d\omega d^2\mathbf{k}_\parallel \tilde{\omega}(\omega, \mathbf{k}_\parallel) \hat{b}_m^{\dagger}(\omega, \mathbf{k}_\parallel)\hat{b}_m^{\mathstrut}(\omega, \mathbf{k}_\parallel) \nonumber
    \\
    +& \sum_{\substack{m \in \textrm{channels}\\\lambda \in \textrm{modes}}} \int d\omega d^2\mathbf{k}_\parallel  \mathcal{W}^{\mathstrut}_{\lambda m}(\omega, \mathbf{k}_\parallel) \hat{b}_m^{\mathstrut}(\omega, \mathbf{k}_\parallel)\hat{a}_\lambda^{\dagger}(\mathbf{k}_\parallel) \nonumber
    \\
    &+ h.c. \,,
    \end{align}
    where we have employed the relabeling from Appendix~\ref{sec::classical_wave} and $\hat{a}_\lambda^{\mathstrut}(\mathbf{k}_\parallel)$ are bosonic few-mode operators at each parallel wave vector, which have frequency $\omega^{\mathstrut}_\lambda(\mathbf{k}_\parallel)$. Similarly, $\hat{b}_m^{\mathstrut}(\omega, \mathbf{k}_\parallel)$ are the external bath operators, which couple to the few-mode operators with coupling strength $\mathcal{W}^{\mathstrut}_{\lambda m}(\omega, \mathbf{k}_\parallel)$.
    
    We see that the Hamiltonian is a linear combination of one-dimensional few-mode terms at each parallel wave vector $\mathbf{k}_\parallel$. The parallel direction thus still features a continuum. Since there is no confinement or resonance structure in the parallel direction, this continuum can not easily be removed by another few-mode projection.
    
    \subsection{Nuclear resonant interaction}\label{sec::aifmt_nuc}
    The interaction with nuclear transitions can be described within the dipole and rotating wave approximation by the Hamiltonian
    \begin{align}\label{eq::H_full_FM_int}
    H = H_\textrm{cav} + H_\textrm{nuc} +  H_\textrm{int}\,,
    \end{align}
    where
    \begin{align}
    H_\textrm{nuc} = \sum_{\substack{l \in \textrm{ensembles}\\n \in 1, 2\dots N_l}} \frac{\omega_{\textrm{nuc},l}}{2} \hat{\sigma}^z_{ln}\,,
    \end{align}
    \begin{align}\label{eq::aifmt_h_int_3D_fm}
    H_\textrm{int} &= \int d^2\mathbf{k}_\parallel \sum_{\substack{\lambda \in \textrm{modes}\\l \in \textrm{ensembles}\\n \in 1, 2\dots N_l}} g^{\mathstrut}_{\lambda l n}(\mathbf{k}_\parallel) \hat{a}^{\mathstrut}_\lambda(\mathbf{k}_\parallel) \hat{\sigma}^+_{ln} + h.c. \nonumber
    \\
    + \int &d\omega d^2\mathbf{k}_\parallel \sum_{\substack{m \in \textrm{channels}\\l \in \textrm{ensembles}\\n \in 1, 2\dots N_l}} \tilde{g}^{\mathstrut}_{m l n}(\omega, \mathbf{k}_\parallel) \hat{b}_m(\omega, \mathbf{k}_\parallel) \hat{\sigma}^+_{ln} + h.c. \,,
    \end{align}
    and $\hat{\sigma}^{+,-,z}$ are the Pauli operators for the nuclear transitions. We have included the effect of multiple ensembles (indexed by $l$), each of which contains $N_l$ individual nuclei (indexed by $n$) and couples to the system modes (indexed by $\lambda$) with coupling constant $g^{\mathstrut}_{\lambda l n}(\mathbf{k}_\parallel)$ as well as to the bath modes with coupling constant $\tilde{g}^{\mathstrut}_{m l n}(\omega, \mathbf{k}_\parallel)$. These coupling constants can be calculated from nuclear transition and material parameters (see Appendix \ref{sec::aifmt_DetailedDerivation_Dipole} for details), which are also used in the layer formalism \cite{Rohlsberger2005}.
    
    We note that if a sufficient number of system modes is chosen, the few-mode approximation can be performed \cite{Lentrodt2020}, that is the direct interaction of the nuclei with the bath modes can be neglected\dl{, as explained in Appendix \ref{app::recab_aifmt}}.
    
    We further note that in addition to these Hamiltonian terms which arise from the nucleus-light coupling, nuclear transitions can feature additional incoherent decay channels, such as internal conversion \cite{Rohlsberger2005,Sturhahn2004,Hannon1999}. For example for $^{57}$Fe where the $\alpha$-factor is $8.56$ \cite{Rohlsberger2005}, internal conversion makes up \dl{the majority} of the incoherent decay rate (see Appendix \ref{app::aifmt_dipMom} for details). In our prescription, this effect can be accounted for by a Lindblad term \cite{Heeg2013a}
    \begin{equation}\label{eq::lindblad_IC_localDiss}
    \mathcal{L}_\textrm{IC}[\rho] = \sum_{\substack{l \in \textrm{ensembles}\\n \in 1, 2\dots N_l}} \frac{\gamma_\textrm{IC}}{2} (2\hat{\sigma}^-_{ln}\rho\hat{\sigma}_{ln}^+ - \{\hat{\sigma}_{ln}^+\hat{\sigma}_{ln}^-,\,\rho \}) \,,
    \end{equation}
    where $\gamma_\textrm{IC}$ is the internal conversion decay constant.
    
    We note that since the field continuum is still present in the theory, radiative incoherent losses are already included in the light-nucleus interaction Hamiltonian and should not be added as an additional incoherent Lindblad term. If the few-mode approximation is performed, a small residual decay contribution is added to the incoherent decay rate to account for the weak interaction with the removed bath modes, which tends to zero at large system mode numbers.
    
    Within the dipole, rotating-wave and non-hermitian field Hamiltonian approximations (see also Appendices \ref{sec::Xray-AIFMT_complexRefractive}, \ref{app::aifmt_dipMom}), the Hamiltonian derived above is a fully general description in few-mode form, with the translational invariance of the system explicitly implemented.
    
    \subsection{Effective one-dimensional problem}\label{sec::aifmt_DetailedDerivation_eff1D}
    We see that the derived few-mode Hamiltonian is already very similar to the phenomenological model \cite{Heeg2013b,Heeg2015c}, with the main difference being the continuum of parallel wave vectors. In order to complete the connection, we derive an effective one-dimensional description in this section, which is well applicable in the linear excitation regime.
    
    We first note that in practice, the nuclear cavity QED system is often studied spectroscopically. That is the system is excited by a collimated and highly monochromatic\footnote{On the spectral scale of the cavity, the beams are monochromatic. On the scale of the nuclei on the other hand, the exciting radiation has a broad spectrum.} beam from a modern low-emittance x-ray facility such as a synchrotron  or an x-ray free electron laser in conjunction with a high-resolution monochromator. At anticipated light sources such as x-ray free electron laser oscillators \cite{Adams2019}, similar setups are to be expected.
    
    In such a setup, the exciting light defines a narrow range of incidence angles and consequently a narrow range of parallel wave vectors. Due to the assumed translational invariance of the system in the layer plane, the parallel wave vector is a conserved quantity in the low-excitation regime (see Sec.~\ref{sec::Green_lowExc_levScheme} for details, and~\cite{PhysRevA.90.063834} for a related discussion of higher excitations), such that each parallel wave vector forms an isolated subspace of the dynamics. Therein, we can then obtain the effective one-dimensional Hamiltonian
    \begin{align}\label{eq::aifmt::eff_1D_H}
    H^{\textrm{1D}}(\mathbf{k}^{\textrm{(in)}}_\parallel) =& H_\textrm{cav}^{\textrm{1D}}(\mathbf{k}^{\textrm{(in)}}_\parallel) + H_\textrm{nuc} + H_\textrm{int}^{\textrm{1D}}(\mathbf{k}^{\textrm{(in)}}_\parallel) \,,
    \end{align}
    where
    \begin{align}
    H_\textrm{cav}^{\textrm{1D}}(\mathbf{k}^{\textrm{(in)}}_\parallel) = H_\textrm{few}^{\textrm{1D}}(\mathbf{k}^{\textrm{(in)}}_\parallel) + H_\textrm{ext}^{\textrm{1D}}(\mathbf{k}^{\textrm{(in)}}_\parallel) \,,
    \end{align}
    with
    \begin{align}
    H_\textrm{few}^{\textrm{1D}}(\mathbf{k}^{\textrm{(in)}}_\parallel) = \sum_{\lambda \in \textrm{modes}} \omega^{\mathstrut}_\lambda(\mathbf{k}^{\textrm{(in)}}_\parallel) \hat{a}_\lambda^\dagger(\mathbf{k}^{\textrm{(in)}}_\parallel) \hat{a}_\lambda^{\mathstrut}(\mathbf{k}^{\textrm{(in)}}_\parallel)\,,
    \end{align}
    \begin{align}
    H_\textrm{ext}^{\textrm{1D}}&(\mathbf{k}^{\textrm{(in)}}_\parallel) = \sum_{m \in \textrm{channels}} \int d\omega \tilde{\omega}(\omega, \theta) \hat{b}_m^{\dagger}(\omega, \mathbf{k}_\parallel^{\textrm{(in)}})\hat{b}_m^{\mathstrut}(\omega, \mathbf{k}_\parallel^{\textrm{(in)}}) \nonumber
    \\
    +& \sum_{\substack{m \in \textrm{channels}\\\lambda \in \textrm{modes}}} \int d\omega \mathcal{W}^{\mathstrut}_{\lambda m}(\omega, \mathbf{k}_\parallel^{\textrm{(in)}}) \hat{b}_m^{\mathstrut}(\omega, \mathbf{k}_\parallel^{\textrm{(in)}})\hat{a}_\lambda^{\dagger}(\mathbf{k}_\parallel^{\textrm{(in)}}) \nonumber
    \\
    &+ h.c. \,,
    \end{align}
    and
    \begin{align}
    H_\textrm{int}^{\textrm{1D}}(\mathbf{k}^{\textrm{(in)}}_\parallel) = \sum_{\substack{\lambda \in \textrm{modes}\\l \in \textrm{ensembles}\\n \in 1, 2\dots N_l}} g^{\mathstrut}_{\lambda l n}(\mathbf{k}_\parallel^{\textrm{(in)}}) \hat{a}^{\mathstrut}_\lambda(\mathbf{k}_\parallel^{\textrm{(in)}}) \hat{\sigma}^+_{ln} + h.c.\,.
    \end{align}
    In addition to the internal conversion Lindblad term, we obtain a radiative contribution to the incoherent decay
    \begin{equation}\label{eq::lindblad_rad_localDiss}
    \mathcal{L}_\textrm{rad}[\rho] = \sum_{\substack{l \in \textrm{ensembles}\\n \in 1, 2\dots N_l}} \frac{\gamma_\textrm{rad}}{2} (2\hat{\sigma}^-_{ln}\rho\sigma_{ln}^+ - \{\hat{\sigma}_{ln}^+\hat{\sigma}_{ln}^-,\,\dl{{\rho}} \}) \,,
    \end{equation}
    where $\gamma_\textrm{rad}$ is the spontaneous emission rate of the nuclear transition in the cavity environment. The two incoherent terms can be combined to
    \begin{equation}\label{eq::lindblad_SE_localDiss}
    \mathcal{L}_\textrm{SE}[\rho] = \sum_{\substack{l \in \textrm{ensembles}\\n \in 1, 2\dots N_l}} \frac{\dl{\gamma}}{2} (2\hat{\sigma}^-_{ln}\rho\sigma_{ln}^+ - \{\hat{\sigma}_{ln}^+\hat{\sigma}_{ln}^-,\,\dl{{\rho}} \}) \,,
    \end{equation}
    which is the single-transition version of the term used in \cite{Heeg2013b}. \dl{We note that $\gamma = \gamma_\textrm{IC} + \gamma_\textrm{rad}$ can in principle differ slightly from the natural linewidth since the cavity modes are retained in the Hamiltonian, but is essentially equal for all practical purposes.}
    
    Noting that $\mathbf{k}_\parallel^{\textrm{(in)}}$ appears only as a parametric dependence now, we see that the above description provides the improved input-output model summarized in Sec.~\ref{sec::aifmt}, where the parallel wave vector dependence is rewritten in terms of the incidence angle and the frequency of the transition energy.
    
    We have thus provided an \textit{ab initio} generalization of the successful phenomenological model \cite{Heeg2013b,Heeg2015c} for thin-film x-ray cavities with M\"ossbauer nuclei and clarified its origin as well as the involved approximations. The basic structure and its relation to the original phenomenological version \cite{Heeg2013b,Heeg2015c} are summarized in Sec.~\ref{sec::aifmt}.
    
    In the main text, we show that the improved model provides a number of qualitative advantages, allowing the quantum optical description to be applied to new systems and featuring essentially exact predictions in the linear regime. The {\it ab initio} character of the theory further provides a solid foundation, which, as a main motivation beyond the quantum interpretation of linear scattering experiments, will allow the method to be applied as a predictive tool in the non-linear and correlated quantum dynamics regime.
    
    \subsection{Complex refractive index in the \textit{ab initio}  few-mode theory}\label{sec::Xray-AIFMT_complexRefractive}
    In the treatment presented in Appendix \ref{sec::cavQuant} and also in the original development of the \textit{ab initio} few-mode theory \cite{Lentrodt2020}, a real refractive index is considered. X-ray cavities, however, feature significant material absorption and as a result a complex refractive index should be accounted for.
    
    A rigorous treatment of material absorption and the resulting effective quantum theory can be obtained in various ways (see \cite{Scheel2008,Buhmann2012} for a review as well as \cite{Franke2019} for a recent advance). In general, the resulting light-matter Hamiltonian is highly complex even if the resonant quantum interaction is not included.
    
    Here, we resort to a simple approach, which is also employed in the standard nuclear resonant scattering literature including the semi-classical scattering theory \cite{Rohlsberger1999,Rohlsberger2005}. In the latter approach, the Maxwell wave equation is directly coupled to the resonant nuclei or atoms, while neglecting quantization effects of the light field. For a real refractive index, the generalization to the quantum level is given by the canonical quantization scheme (see Appendix~\ref{sec::cavQuant}). We then include the absorptive character of the material by using the complex refractive index to obtain a non-Hermitian Hamiltonian. Since this approximation is also included in the standard semi-classical x-ray scattering theory \cite{Rohlsberger1999,Rohlsberger2005}, previous experiments substantiate its validity at least for intensity observables at low driving fields. The approach can be complemented by the quantum jump formalism \cite{Minganti2019,Carmichael1999,Carmichael2008}, similarly to recent work using Green's function techniques \cite{Masson2019_preprint}, where the absorptive bath is treated rigorously from the outset (see Sec.~\ref{sec::Green} for a detailed comparison of the approaches).
    
    Practically, the non-Hermitian few-mode Hamiltonian approach is implemented by performing calculations as for a real refractive index, and then transferring to the non-Hermitian theory by substitution of the complex refractive index. We note that for numerical implementations, one has to ensure that complex conjugation operations, as they are used for example in the quantum scattering theory \cite{Domcke1983} underlying the projection scheme, are not applied to the refractive index.
    
    \subsection{Coupling constant in the \textit{ab initio}  few-mode theory}\label{sec::aifmt_DetailedDerivation_Dipole}
    In this appendix, we derive the mode-nucleus coupling in terms of known nuclear resonance and material parameters.
    
    \subsubsection{Mode coupling and quantization area}\label{app::couplingsDef}
    The few-mode coupling in our approach can be written as \cite{Lentrodt2020}
    \begin{align}
    g_{\lambda l n}(\mathbf{k}_\parallel) = - \frac{id\omega_{\textrm{nuc},l}}{\sqrt{2\omega_\lambda(\mathbf{k}_\parallel)}} \chi_\lambda(z_{l}, \mathbf{r}_\parallel, \mathbf{k}_\parallel) \,,
    \end{align}
    where $\chi_\lambda(z, \mathbf{r}_\parallel, \mathbf{k}_\parallel)$ is the three-dimensional system mode from Eq.~\ref{eq::aifmt_h_int_3D_fm} and $z_l$ the vertical position of the nuclei, which is by construction independent of $n$ within the thin-layer approximation. If $z_l$ varies within one ensemble, the latter should be divided into multiple sub-ensembles (see also the thick layer treatment in Sec.~\ref{sec::aifmt_ThickLayers}). Since the system modes themselves also fulfill the translational invariance, we have
    \begin{equation}
    \chi_\lambda(z_l, \mathbf{r}_\parallel, \mathbf{k}_\parallel) = C \,e^{i\textbf{k}_\parallel \cdot \textbf{r}_\parallel} \tilde{\chi}_\lambda(z_l) \,,
    \end{equation}
    where $\tilde{\chi}_\lambda(z_l)$ is the effective one-dimensional mode function evaluated at the nuclear layer position $z_l$ and $C$ is a normalization factor. We therefore see that the mode normalization constant is important to determine the value of the coupling. In order to obtain the constant, we resort to the usual box quantization procedure \cite{Scully1997} for the parallel direction. We require that
    \begin{align}
    \int_{A_\parallel} d^2\mathbf{k}_\parallel \int dz\chi_\lambda(z, \mathbf{r}_\parallel, \mathbf{k}_\parallel) = 1 \,,
    \end{align}
    where $A_\parallel$ is a parallel quantization area, such that we obtain
    \begin{equation}
    C = \frac{1}{\sqrt{A_\parallel}} \,.
    \end{equation}
    Consequently, the coupling constant is given by
    \begin{align}\label{eq::Xray-AIFMT_coupling_unsimplified}
    g_{\lambda l n}(\mathbf{k}_\parallel) = - \frac{id\omega_{\textrm{nuc},l}}{\sqrt{2A_\parallel\omega_\lambda(\mathbf{k}_\parallel)}} e^{i\textbf{k}_\parallel \cdot \textbf{r}_{\parallel,ln}} \tilde{\chi}_\lambda(z_l) \,.
    \end{align}
    The phase factor $e^{i\textbf{k}_{\parallel} \cdot \textbf{r}_{\parallel,ln}}$ can be absorbed into an algebra preserving redefinition of the $\hat{\sigma}$-operators in the effective one-dimensional Hamiltonian \cite{HeegPhD}, amounting to a basis change that affects the nuclear dipole moment by a phase. The coupling is thus independent of the nuclear index $n$ within one ensemble, such that we have
    \begin{align}\label{eq::Xray-AIFMT_coupling}
    g_{\lambda l}(\mathbf{k}_\parallel) = - \frac{id\omega_{\textrm{nuc},l}}{\sqrt{2A_\parallel\omega_\lambda(\mathbf{k}_\parallel)}} \tilde{\chi}_\lambda(z_l) \,.
    \end{align}
    We see that the coupling remains dependent on the quantization area $A_\parallel$. While this feature may initially seem unphysical, it is to be expected, since the number of nuclei $N_l$ participating in the dynamics also depends on this area. As is also noted in the phenomenological model \cite{Heeg2013b}, in the linear regime the collective coupling constant
    \begin{align}
    \tilde{g}_{l\lambda} =\sqrt{N_l} g_{l\lambda} &= - \frac{id\omega_{\textrm{nuc},l}}{\sqrt{2\omega_\lambda}} \sqrt{\frac{N_l}{A_\parallel}} \tilde{\chi}_\lambda(z_l)
    \\
    &= -id\omega_{\textrm{nuc},l} \sqrt{\frac{f_\textrm{LM}\rho_N t_l}{2\omega_\lambda}} \tilde{\chi}_\lambda(z_l)
    \end{align}
    is the relevant quantity, where $\rho_N$ is the number density of the resonant nuclei and $f_\textrm{LM}$ is the Lamb-M\"ossbauer factor encoding the fraction of nuclei which participate in the recoil free scattering or decay process \cite{Rohlsberger2005,Hannon1999}. $t_l$ is the thickness of the ensemble layer $l$ and we assume the thin layer limit for a single ensemble by taking $z_l$ to be independent of the nuclear index $n$. We further drop the parametric parallel wave vector dependence for brevity. In the linear limit, the properties of the system thus only depend on the number density and not on the absolute number of nuclei participating in the dynamics, which is consistent with the semi-classical layer formalism \cite{Rohlsberger2005,HeegPhD,Hannon1999}.
    
    Beyond the linear limit, however, the absolute number of nuclei and the quantization area become important and do not reduce to the number density in the equations. This feature can be understood by recognizing that in a fully translation invariant system, the number of nuclei is necessarily infinite. Physically, however, only a finite number participates in the dynamics due to a limited coherence volume and a limited size of the excitation beam. The quantization area $A_\parallel$ should therefore be chosen to capture these physical features. We further note that the translationally invariant description essentially neglects finite excitation size effects and excitation spreading at the edge of the excitation region. A phenomenological argument to include such contributions is given in \cite{Heeg2016arxiv}.
    
    \subsubsection{Effective nuclear dipole moment}\label{app::aifmt_dipMom}
    A remaining question is how to compute the effective single nucleus transition matrix element $d$ from tabulated resonance parameters. The connection to quantities that are conventionally used for example in the layer formalism \cite{Rohlsberger2005} can be obtained by a simple comparison of physical scattering observables.
    
    The formula for the resonant contribution to the refractive index of a resonant nuclear medium is \cite{HeegPhD}
    \begin{align}\label{eq::refractive_nucResScatt}
    n_{^{57}\textrm{Fe}}-n_{^{56}\textrm{Fe}} = -2\pi \frac{\rho_N}{k_0^3} \frac{f_\textrm{LM}}{2(1+\alpha)} \frac{2I_e + 1}{2I_g + 1} \frac{1}{2\Delta/\gamma + i} \,,
    \end{align}
    where we have taken $^{57}$Fe as an example, and ``magnetic splitting and polarization dependence'' \cite{HeegPhD} have been neglected \cite{HeegPhD,Siddons1999}. The quantities here are defined as in \cite{HeegPhD}, including the Lamb-M\"ossbauer factor $f_\textrm{LM}$, the resonant wave number $k_0$, the spins for the ground (excited) state $I_g$ ($I_e$) and the internal conversion factor $\alpha$.
    
    A formula for the resonant contribution to the refractive index in the linear regime can alternatively be obtained from our effective transition theory, for example via a linear dispersion theory \cite{Lentrodt2020} calculation, which gives
    \begin{equation}
    \varepsilon_\textrm{res}(r) = - \frac{|d|^2}{\omega - \omega_{\textrm{nuc}} + i\frac{\gamma}{2}} f_\textrm{LM}\rho_N \,.
    \end{equation}
    where we have neglected the $A^2$-term contribution, assumed a dense lattice, which for example results in Bragg scattering being neglected, and performed the rotating wave approximation by setting $\omega^2-\omega_{\textrm{nuc}}^2 \approx 2\omega_{\textrm{nuc}}(\omega-\omega_{\textrm{nuc}})$ as well as $\frac{\omega_{\textrm{nuc}}^2}{\omega^2} \approx 1$ \cite{Lentrodt2020}. We note that all of these approximations are already implicit in the nuclear refractive index formula Eq.~\eqref{eq::refractive_nucResScatt}, with more general treatments including lattice effects and other scattering processes being available \cite{Hannon1999}. The Lamb-M\"ossbauer factor $f_\textrm{LM}$ encodes the fraction of nuclei that participate in the recoil free scattering process \cite{Rohlsberger2005,Hannon1999} and therefore modifies the nuclear number density.
    
    A further approximation which is commonly performed in nuclear resonant scattering and which is implicit in Eq.~\eqref{eq::refractive_nucResScatt} is the small response approximation $n = \sqrt{\varepsilon} = \sqrt{1 + \chi} \approx 1 + \chi/2$. Performing this approximation also in the linear dispersion theory case we obtain
    \begin{equation}\label{eq::refractive_linDisp}
    n - n_\textrm{electronic} \approx - \frac{|d|^2}{\omega - \omega_{\textrm{nuc}} + i\frac{\gamma}{2}} \frac{f_\textrm{LM}\rho_N}{2} \,.
    \end{equation}
    In order to obtain the effective nuclear transition matrix element $d$, we can therefore straightforwardly compare Eq.~\eqref{eq::refractive_linDisp} to Eq.~\eqref{eq::refractive_nucResScatt}, which gives
    \begin{align}\label{eq::eff_dipole}
    |d_{^{57}\textrm{Fe}}|^2 \approx \frac{2\pi \gamma}{k_0^3} \frac{1}{2(1+\alpha)} \frac{2I_e + 1}{2I_g + 1} \,.
    \end{align}
    We recall that as our previous calculations this formula is given in natural units with $\hbar = c = \varepsilon_0 = 1$.
    
    This calculation also clarifies the connection to the semi-classical layer formalism \cite{Rohlsberger2005}, which is a standard tool for describing resonant x-ray scattering experiments \cite{Rohlsberger2005,Sturhahn2000} and can be used to include various experimental imperfections in the description \cite{Rohlsberger2005,Sturhahn2000}. In this context, we note that our quantum optical approach is based on nuclear \textit{transitions}, whose properties are assumed to be known, as the starting point. In the established perturbative scattering theory \cite{Hannon1999,Siddons1999}, the transition structure of the nuclei is investigated in detail and effects such as interaction with the lattice are included. Related approaches such as Shvyd'ko's time and space picture \cite{Shvydko1999} allow for the inclusion of additional effects such as inelastic scattering \cite{Shvydko2000}. For further details on the relation between the formalisms refer to Sec.~\ref{sec::Green_linDisp} and Fig.~\ref{fig::overview}.
    
    \section{Detailed comparison of the \textit{ab initio} and phenomenological few-mode approaches}\label{app:detailed_fitting}
    \begin{figure*}[t] 
    	\includegraphics[width=0.7\textwidth]{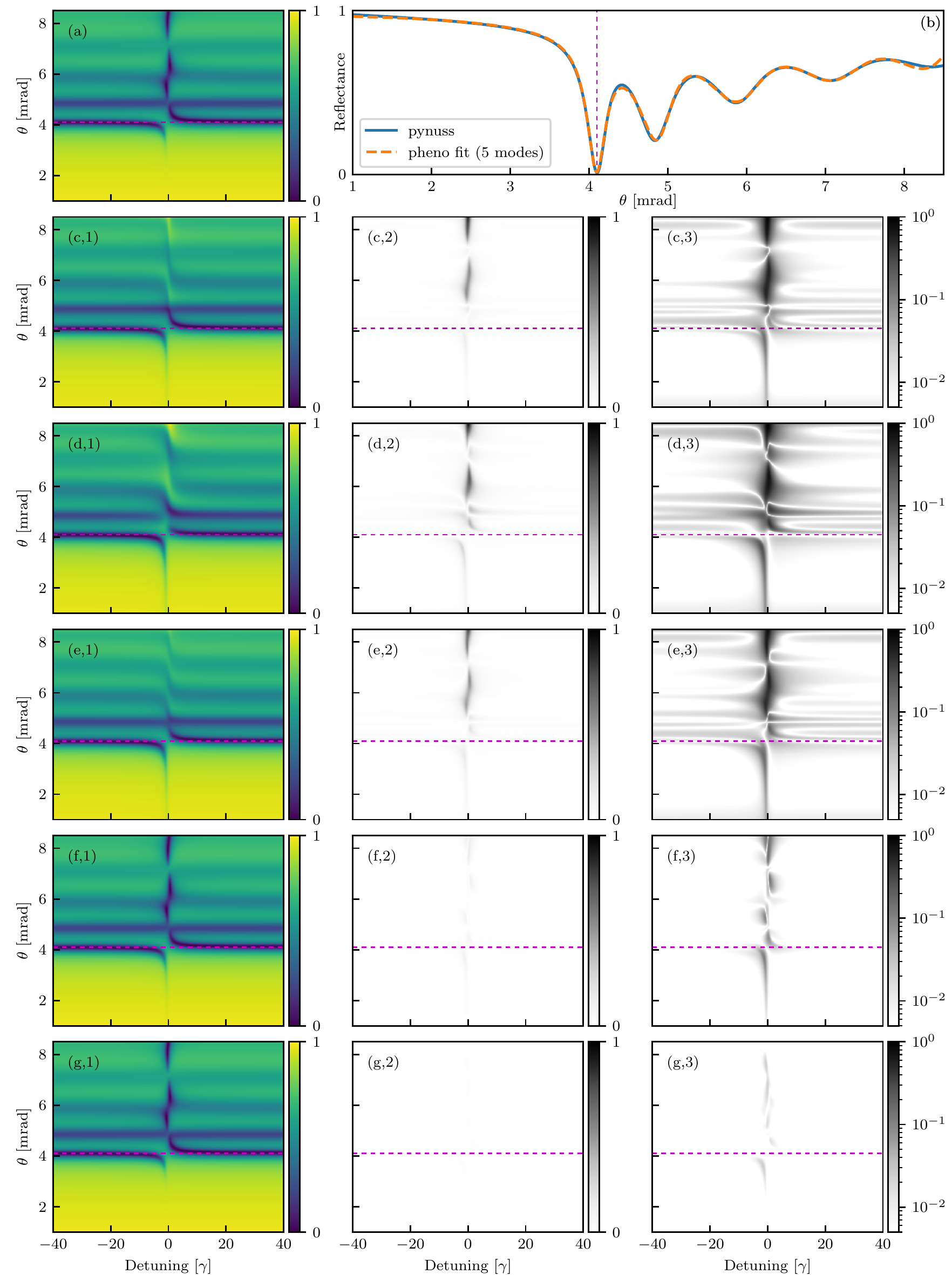}
    	\caption{(Color online)
    		Comparison of the {\it ab initio} few-mode model with corresponding phenomenological approaches. (a) Reference spectrum obtained using a semi-classical calculation. Panel (b) shows the best phenomenological model fit (five modes, yellow dashed) of the rocking curve (solid blue), which is used to constrain the cavity parameters ($\kappa_\lambda, \kappa_{R,\lambda}, \theta_\lambda$) in the phenomenological models. Panels (c)-(g) show results for different few-mode approaches, with (c)-(e) comprising the phenomenological model with different parameter fitting procedures (fit method 1, 2 and 3, respectively, definition see text) and (f), (g) being the {\it ab initio} few-mode theory with 5 and 20 modes, respectively. Panels (c-g,1) show the model spectrum in each case, while panels (c-g,2) and (c-g,3) show the residual deviation on a linear and logarithmic scale, respectively.\label{fig::detailed_fitting}}
    \end{figure*}
    \begin{figure}[t]
    	\includegraphics[width=1.0\columnwidth]{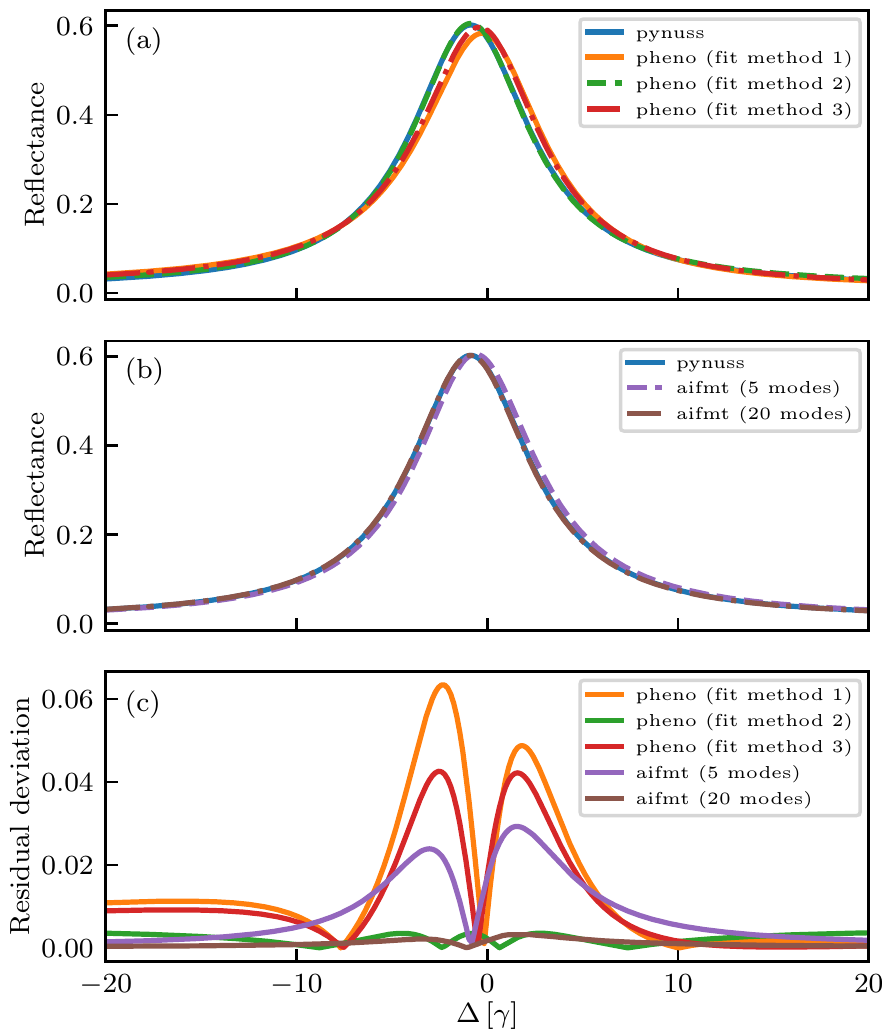}
    	\caption{(Color online) One dimensional slices at the first cavity resonance $\theta=\theta_0$ of the spectra in Fig.~\ref{fig::detailed_fitting} resulting from the different models. Panel (a) compares the various phenomenological fits (see legend) to the semi-classical reference (solid blue). Panel (b) compares the {\it ab initio} few-mode result for 5 and 20 modes (see legend). In panel (c), the residual deviations are shown.\label{fig::detailed_fitting2}}
    \end{figure}
    
    In Sec.~\ref{sec::aifmt_SimpleEx_nucSpec}, we presented a comparison of the phenomenological and {\it ab initio} few-mode approaches with regards to their capability to model nuclear spectra of the example cavity. 
    However, this comparison is not unique, because the phenomenological models are based on fits of their parameters to predictions from semi-classical theories, and the fits can be obtained using different fit objectives, i.e., different approaches to quantify the differences between the model and the references. Note that these differences are often not of practical relevance close to the resonance at which the parameters are fitted, but deviations between different parameter sets are expected further away from the resonance. In contrast, the parameters of the {\it ab initio} model are unique, as they are calculated from the cavity structure without a fitting procedure.
    
    Because of the ambiguity in the phenomenological parameters, in this appendix, we provide additional detail on this comparison to support our conclusions.  In particular, we consider multiple ways of fitting the phenomenological model parameters.   Fig.~\ref{fig::detailed_fitting}(a) shows the layer formalism spectrum used as a reference for this analysis. In order to obtain such a spectrum from the phenomenological model,  
    in all cases, the empty cavity parameters are first fitted to the rocking curve as described in Sec.~\ref{sec::aifmt_SimpleEx_nucSpec}, yielding the best fit shown in Fig.~\ref{fig::detailed_fitting}(b).
    Next, we employ three fitting procedures for the mode-ensemble interaction parameters.
    
    \paragraph{Fit method 1.}
    In this method, we only fit one global interaction parameter. This method was already employed successfully in \cite{Heeg2015c} for the EIT-scenario and uses an extraction of the individual mode parameters. The latter is achieved by decomposing the coupling constants as \cite{Heeg2015c}
    \begin{align}
    	\sqrt{N_l}g_{\lambda l} = \tilde{\mathcal{E}}_{\lambda l} (\sqrt{N_l}g_{l}) \,,
    \end{align}
    where $\tilde{\mathcal{E}}_{\lambda l}$ is taken as the mode amplitude at the location of ensemble $l$ when illuminated at the rocking minimum corresponding to mode $\lambda$ \cite{Heeg2015c}. For our single ensemble case, $\sqrt{N_l}g_{l}$ is then a single global scale that is fitted to the two-dimensional spectrum.
    
    \paragraph{Fit method 2.}
    In this approach we replace the mode parameter extraction prescription of fit method 1 and instead fit the $\sqrt{N_l}g_{\lambda l}$ parameters for each mode. As a fit objective, we use the one-dimensional spectrum at resonance with the first cavity mode, that is at $\theta=\theta_0$.
    
    \paragraph{Fit method 3.}
    Here, we proceed as in fit method 2, only that agreement to the full two-dimensional spectrum (see Fig.~\ref{fig::detailed_fitting}) is used as a fit objective.
    
    The results from the three methods are shown in Fig.~\ref{fig::detailed_fitting}. Panels (c,1)-(g,1) show the two-dimensional spectra calculated using the phenomenological model with the model parameters extracted from the respective fitting methods. The remaining panels show the corresponding residual deviation to the layer formalism result in panel (a), both on a linear scale in panels (c,2)-(g,2) and on a log scale in panels (c,3)-(g,3).
    
    We find that all models capture the behavior well, in particular at low incidence angles. At higher incidence angles, that is going towards the edge of the fitting range for the cavity parameters, the deviations become more significant. The agreement at higher incidence angles can be improved by including a larger set of modes, resulting in more fitting parameters. We further see that the various fitting procedures result in deviations due to the emphasis on different regions of the spectrum, but yield the same qualitative spectral features and the same overall quantitative level of agreement.
    
    In comparison, Fig.~\ref{fig::detailed_fitting}(f) shows analogous results for the {\it ab initio} few-mode theory with 5 cavity modes. We see that the performance is similar to the phenomenological cases around the first resonance, but better at higher incidence angles. Panel (g) shows the corresponding 20 modes result, illustrating the systematic convergence when including more cavity modes. In the latter case, the residual deviations are mainly due to the thin layer approximation, with higher modes only having a tiny contribution.
    
    In Fig.~\ref{fig::detailed_fitting2}, we focus on the slice at the first mode resonance ($\theta=\theta_0$). To illustrate the excellent level of agreement of all approaches, the nuclear spectra are plotted on top of the semi-classical reference (panels a,b). The residual deviations are shown in panel (c). We find that the {\it ab initio} result at 20 modes is well converged. Indeed, the phenomenological result gives a comparable level of agreement with only 5 modes for the case of fit method 2, illustrating that the phenomenological model can yield very good agreement if one focuses on a particular spectral region. However, it is not known {\it a priori} which phenomenological fit works best for a given problem. The remaining fitting methods and the 5-mode {\it ab initio} theory show similar deviations, which are structurally comparable, with already a good quantitative agreement on the level of a few percent deviation.
    
    \section{Analytic formula for the layer stack Green's function}\label{app::tomas}
    In this appendix, we summarize the analytical form of the Green's function and its efficient numerical calculation as presented in \cite{Tomas1995}. The appendix is structured to allow for a convenient numerical implementation of the formulas. 
    
\begin{figure}[t]
    	\includegraphics[width=1.0\columnwidth]{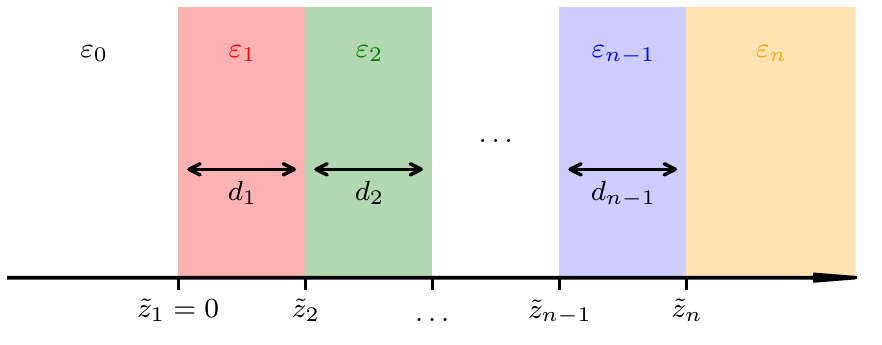}
    	\caption{(Color online) Setup of the cavity layer stack and notation for the analytical Green's function formula, adapted from \cite{Tomas1995}.  The first (last) layer with $j=0$ ($j=n$) extends infinitely to the left (right).}
    	\label{fig::app_GF_illu}
    \end{figure}

    As shown in \cite{Tomas1995}, the in-plane Fourier transformed Green's function is given by
    \begin{align}
	    \textbf{G}&(z \in j, z' \in j', \textbf{k}_\parallel, \omega) = -\frac{4\pi}{k_j^2} \hat{\mathbf{z}}\hat{\mathbf{z}} \delta(z-z') + \frac{2\pi i}{\beta_{j_n}} \nonumber
	    \\
	    &\times\sum_{q \in p,s} \frac{\xi_q}{t^q_{0/n}} \biggl[\mathbfcal{E}^{0}_q(z, \mathbf{k}_\parallel, \omega)\mathbfcal{E}^{n}_q(z', -\mathbf{k}_\parallel, \omega) \Theta(z-z')  
	    \nonumber \\
	    &  + \mathbfcal{E}^{n}_q(z, \mathbf{k}_\parallel, \omega)\mathbfcal{E}^{0}_q(z', -\mathbf{k}_\parallel, \omega)\Theta(z'-z)\biggr] \,, \label{eq::app_GF_parallel}
    \end{align}
    where $\Theta$ is the step function, $\xi_{p(s)}=+(-)1$, $\hat{\mathbf{z}}$ is the unit vector in $z$-direction and $z\in j$ denotes that the position in $z$-direction lies in the $j$th layer of the cavity stack (see Fig.~\ref{fig::app_GF_illu} for details on the notation, which is adapted from \cite{Tomas1995}). $k_j=\sqrt{\varepsilon_j(\omega)}\omega/c$ is the in-medium wave number and $\beta_j = \sqrt{k_j^2 - k_\parallel^2}$ is its $z$-component \cite{Tomas1995}. 
    
    The information about the spatial dependence of the Green's function is given by the mode profiles, which are given explicitly by \cite{Tomas1995}
    \begin{subequations}\label{eq::modeProfiles}
    	\begin{align}
    	\mathbfcal{E}^{n(0)}_p(z &\in j, \mathbf{k}_\parallel, \omega) = \frac{t^p_{n(0)/j} e^{i\beta_j d_j}}{D_{pj}} \nonumber
    	\\
    	&\times \biggl[\pm \frac{\beta_j}{k_j} (e^{-i\beta_jz^\mp} - r^p_{j/0(n)}e^{i\beta_jz^\mp})\hat{\mathbf{k}}
    	\\
    	& + \frac{k_\parallel}{k_j} (e^{-i\beta_jz^\mp} + r^p_{j/0(n)}e^{i\beta_jz^\mp})\hat{\mathbf{z}}\biggr]\,,
    	\\
    	\mathbfcal{E}^{n(0)}_s(z &\in j, \mathbf{k}_\parallel, \omega) = \frac{t^s_{n(0)/j} e^{i\beta_j d_j}}{D_{pj}} \nonumber
    	\\
    	&\times\biggl[(e^{-i\beta_jz^\mp} + r^s_{j/0(n)}e^{i\beta_jz^\mp})\hat{\mathbf{k}}\times\hat{\mathbf{z}}\biggr]\,,
    	\end{align}
    \end{subequations}
    where $D_{qj}=1 - r^q_{j/0}r^q_{j/n}e ^{2i\beta_j d_j}$, $z^+ = d_j - (z-\tilde{z}_j)$ and $z^-=z-\tilde{z}_j$ with $\tilde{z}_j$ being the position of the surface of layer $j$ and $d_j$ its thickness as shown in Fig.~\ref{fig::app_GF_illu}. We note that in the case of grazing incidence, the $p$- and $s$-polarization mode profiles have approximately identical magnitude \cite{Parratt1954,HeegPhD}.
    
    The result is now expressed in terms of the coefficients $t_{i/j}$ and $r_{i/j}$, which are defined as the transmission and reflection coefficients, respectively, from layer $i$ into layer $j$ \cite{Tomas1995}. They can be calculated via a recursion formula \cite{Tomas1995} analogous to Parratt's formalism \cite{Parratt1954}. Specifically, the $i/k$ coefficients can be expressed in terms of $i/j$, $j/k$ coefficients for any in-between $j$ by \cite{Tomas1995}
    \begin{subequations}
    	\label{eq::app_Fresnel_recursion}
    	\begin{align}
	    	r^q_{i/k} &= r^q_{i/j/k}\nonumber \\[2ex]
	    	&= \frac{r^q_{i/j} + (t_{i/j}^q t^q_{j/i} - r^q_{i/j}r^q_{j/k})r^q_{j/k}e^{2i\beta_j d_j}}{D^{(ik)}_{qj}}\,,
	    	\\
    		t^q_{i/k} &= t^q_{i/j/k} = \frac{1}{D^{(ik)}_{qj}}t_{i/j}^q t^q_{j/k}e^{i\beta_j d_j} \,.
    	\end{align}
    \end{subequations}
    If the coefficient $i/k$ is required, the conversion is conveniently started by choosing $j=i+1$ for $i<k$ or $j=i-1$ for $i>k$, and terminated by the Fresnel coefficients for adjacent layers \cite{Tomas1995}
    \begin{subequations}
    	\label{eq::app_Fresnel_neighbor}
    	\begin{align}
	    	r^q_{ij} &= \frac{\beta_i - \gamma_{ij}^q \beta_j}{\beta_j + \gamma_{ij}^q\beta_j} \,,\\[2ex]
		    t^q_{ij} &= \sqrt{\gamma_{ij}^q}(1+r^q_{ij}) \,,
	    \end{align}
    \end{subequations}
    where the absence of a slash indicates that the layers are adjacent and $\gamma^p_{ij}=\varepsilon_i(\omega)/\varepsilon_j(\omega)$ as well as $\gamma^s_{ij}=1$.
    
    Together, the results from \cite{Tomas1995} summarized here provide a recursively analytic and hence numerically efficient way to calculate the Green's function at a given parallel wave vector for the layer cavities, which is the central quantity appearing in the effective nuclear level scheme described in the main text.

    \bibliographystyle{myprsty}
    \bibliography{library}
\end{document}